\documentclass[
journal=jpcafh,
manuscript=article,
maxauthors=10,
etalmode=truncate
]{achemso}

\usepackage{chemmacros}   
\usepackage[T1]{fontenc}
\usepackage{easyReview}
\usepackage{multicol}     
\usepackage{multirow}     
\usepackage{appendix}

\usepackage{amsmath}      
\usepackage{cases}        

\usepackage[pdftex]{hyperref} 
\usepackage{cleveref}      
\mciteErrorOnUnknownfalse   

\usepackage{ctable}       
\usepackage[para]{threeparttable} 
\usepackage{colortbl}     

\usepackage{tikz}         
\usetikzlibrary{shapes}

\usepackage{doi}          

\makeatletter
\newcommand{\setlabel}[1]{\edef\@currentlabel{#1}\label} 
\makeatother

\DeclareMathOperator{\csch}{csch} 

\newcommand{\greymidrule}{\arrayrulecolor{black!10}\midrule\arrayrulecolor{black}} 

\newcommand{\markerbluecircle}{\tikz{\node[circle,scale=0.5,draw=blue,fill=none,fill opacity=0](){};}}
\newcommand{\markerfilledbluecircle}{\tikz{\node[circle,scale=0.5,draw=blue,fill=blue,fill opacity=1](){};}}
\newcommand{\markerredcircle}{\tikz{\node[circle,scale=0.5,draw=red,fill=none,fill opacity=0](){};}}
\newcommand{\markerfilledredcircle}{\tikz{\node[circle,scale=0.5,draw=red,fill=red,fill opacity=1](){};}}
\newcommand{\markerredtriangle}{\tikz{\node[regular polygon, regular polygon sides=3,scale=0.5,draw=red,fill=none,fill opacity=0](){};}}
\newcommand{\markerfilledredtriangle}{\tikz{\node[regular polygon, regular polygon sides=3,scale=0.5,draw=red,fill=red,fill opacity=1](){};}}
\newcommand{\markerfilledblackcircle}{\tikz{\node[circle,scale=0.5,draw=black,fill=black,fill opacity=1](){};}}

\author{\'{E}lio Pereira}
\affiliation{Instituto de Plasmas e Fus\~{a}o Nuclear, Instituto Superior T\'{e}cnico, Universidade de Lisboa, Av. Rovisco Pais 1, Lisboa, 1049-001, Portugal}
\author{Jorge Loureiro}
\affiliation{Instituto de Plasmas e Fus\~{a}o Nuclear, Instituto Superior T\'{e}cnico, Universidade de Lisboa, Av. Rovisco Pais 1, Lisboa, 1049-001, Portugal}
\author{M\'{a}rio Lino da Silva}
\email{mlinodasilva@tecnico.ulisboa.pt}
\affiliation{Instituto de Plasmas e Fus\~{a}o Nuclear, Instituto Superior T\'{e}cnico, Universidade de Lisboa, Av. Rovisco Pais 1, Lisboa, 1049-001, Portugal}

\title[Vibronic State-Specific Modelling of High-Speed Nitrogen Shocked Flows. Part I: Kinetic Database]{Vibronic State-Specific Modelling of High-Speed Nitrogen Shocked Flows. Part I: Kinetic Database}


\begin{document}







\begin{abstract}
A database of kinetic processes for nitrogen shocked flows was built using vibronic-specific state-to-state models. 
The Forced-Harmonic-Oscillator model (FHO), which is more physically accurate in the high temperature regime than the popular Schwartz-Slawsky-Herzfeld model (SSH), was implemented in the computation of rate coefficients for vibrational transition and dissociation of \ch{N2} and \ch{N2+} by heavy particle impact. Thermal dissociation rate coefficients of $\ch{N2}(\text{X}{}^1\Sigma_\text{g}^+)$ by collisions with $\ch{N2}(\text{X}{}^1\Sigma_\text{g}^+)$ and  $\ch{N}({}^4\text{S}_\text{u})$ were in their turn obtained, which were shown to agree reasonably well with state-of-the-art experimental values. The possibility of extending the well known Landau-Zener and Rosen-Zener-Demkov models (for heavy particle impact excitation of atomic particles) to heavy particle impact vibronic excitation of diatomic particles was ascertained to be impractical. As an alternative, an exponential gap law was considered. By fitting the curve that represents the law to experimentally obtained values for rate coefficients values of several vibronic transitions of \ch{N2} reported in the literature, discrepancies of as much as one order of magnitude were obtained, evidencing some crudeness of the model. Reactions such as ionisation of \ch{N2} by electron impact, charge exchange and dissociative recombination of \ch{N2+} were modelled using process cross sections or rate coefficients from the literature. 
A companion article describes the application of this model to nitrogen shocked flows.

\end{abstract}

\section{Introduction}
\label{sec:Introduction}
\setlabel{Introduction}{sec:introduction}

We present a devised database of kinetic processes describing the physical phenomena that occur in nitrogen shocked flows developing inside shock tubes. These shock tube tests correspond to partial simulations of the conditions attained downstream of a shock wave in the entry of a body in a nitrogen-dominant atmosphere such as the ones of the Earth and Titan. Note that here the term ``partial simulations'' was considered since the other components of the atmosphere' gas, the interactions of the flow with the body and the non-unidimensional flow dynamics which are important in the atmospheric entry phase do not occur in these shock tube tests.  

\begin{figure}[H]
\centering
\centerline{\includegraphics[scale=1]{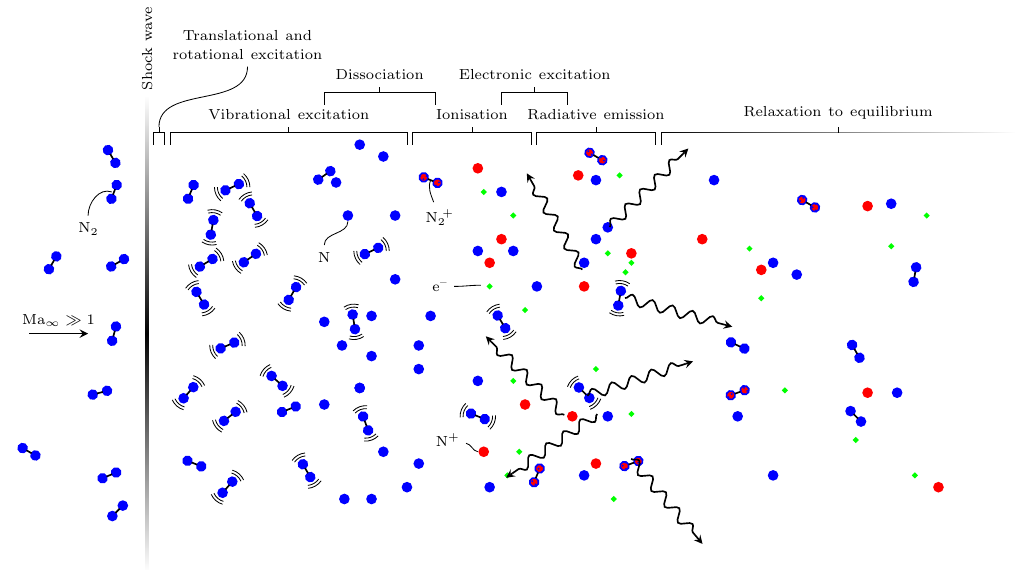}}
\caption{Simplified representation of the post-shock physical phenomena that occur in a pure nitrogen shocked flow.}
\label{fig:Hypersonic_nitrogen_flow}
\end{figure}

The developed database of kinetic processes was in turn implemented in Euler unidimensional simulations of some nitrogen shocked flows generated in  the $62^{\text{th}}$ campaign of Ames Electric Arc Shock Tube (EAST) \cite{brandis2018shock}. The results are reported in a companion paper \cite{epereira2021paperII}. The physical phenomena that occurs in the above-mentioned nitrogen shocked flows are depicted by \Cref{fig:Hypersonic_nitrogen_flow}.
\noindent
The shock wave is a propagating disturbance that causes an abrupt increase in temperature, pressure and density of the gas immediately downstream of the wave. The above-mentioned temperature is associated with the translational energy mode of the particles, the so-called heavy particle translational temperature, $T_\text{tr$_\text{h}$}$. Fast collisions between the gas particles induce a very rapid redistribution of their translational energy - the new distribution can be proved to be a Boltzmann one - and translational equilibrium is said to occur \cite{park1990nonequilibrium}.

Collisions in these circumstances induce an excitation of the rotational and vibrational modes of the molecular particles. The rotational excitation occurs almost simultaneously with the translational equilibration and therefore a rotational temperature $T_\text{rot}$ being equal to the translational temperature $T_\text{tr$_\text{h}$}$ can almost immediately be defined. This equality means that a designation for the combination of both temperatures may be employed: the heavy particle translational-rotational temperature $T_\text{tr$_\text{h}$-rot}=T_\text{tr$_\text{h}$}=T_\text{rot}$. Translational and rotational excitations are much more faster than vibrational excitation, and transrotational equilibrium can be assumed to occur even before the commence of the vibrational excitation \cite{park1990nonequilibrium}.

The excitation of the rotational and vibrational modes,leads to the dissociation of the molecular particles, occurring through overstretching of the vibrational stroke (vibrational dissociation) or centrifugal tearing (rotational dissociation), by collisions with another particles \cite{park1990nonequilibrium}. Such processes produce atomic particles. 

With the presence of atomic particles in the flow, comes the possibility of occurring associative ionisation reactions in which the atomic particles associate, creating excited molecular particles, that spontaneously ionise, producing ionic molecular particles and free electrons (the fluid becomes a plasma). Collisions between the free electrons and atomic particles will excite and ionise them, freeing even more electrons (this is often referred as ionisation avalanche). Similarly, collisions between free electrons and molecular particles will excite and ionise them, and may even cause their dissociation. The electronic excitation of heavy particles by electron impact is much more efficient than the respective excitation by heavy particle impact \cite{park1990nonequilibrium}, and, therefore, the electronic excitation process has its major importance in the region of the flow where free electrons are present.

Excited particles may de-excite through the spontaneous emission of radiation. A radiation field is then created, inducing the emission of even more photons. Some of the emitted photons may be absorbed by the particles in the plasma, exciting them. Also, the emitted photons may even dissociate or ionise the particles \cite{vincenti1965introduction}.

After a considerable amount of collisions, the plasma begins a process of relaxation, i.e. it initiates an evolution to thermodynamic equilibrium. 

All the above-mentioned physical phenomena needs to be accounted for by the devised database of kinetic processes.

\section{State-of-the-art}
\label{sec:stateoftheart}

The accuracy of the numerical simulations is intimately linked to the accuracy of the considered models for the representation of important physical phenomena which occur in the subject of this study. One big difficulty about modelling entry post-shock flows concerns the quantification of non-equilibrium effects. It is here that non-equilibrium thermodynamic and kinetic models come into play. 

\subsection{Thermodynamic models}
\label{subsection:Thermodynamic_models}

There are a lot of different types of non-equilibrium that may occur in the flow, namely chemical, radiative, thermal and what will be termed here as energy mode self-non-equilibrium (in which the energy mode is not in equilibrium with itself). If the  $i$-th energy mode associated with the particles of the $s$-th species is in self-equilibrium, a temperature $T_{s,i}$ can be assigned to it. The population of the particles energy levels in that energy mode would follow a Boltzmann distribution, with the temperature $T_{s,i}$ as parameter. 

Thermodynamic models which assume the energy modes to be in self-equilibrium correspond to the so-called multi-temperature models. Well known examples of multiple-temperature models are the Lee's three-temperature model \cite{lee1984} ($T_\text{tr$_{\text{h}}$-rot}$,  $T_\text{vib}$ and $T_\text{el-tr$_\text{e}$}$) and the Park's two-temperature model \cite{park1989} ($T_\text{tr$_{\text{h}}$-rot}$ and $T_\text{vib-el-tr$_\text{e}$}$). As pointed out by Park \cite{park1992modeling}, numerical results obtained through multi-temperature models are significantly more coherent with the accurate experimental ones for the post-shock conditions than single-temperature models (which assume thermal equilibrium). Several works such as the ones of Candler \cite{candler1989},  Hornung \cite{hornung1972} and Lobb \cite{lobb1964} evidenced that the single-temperature model predicts the post-shock flow to be closer to thermodynamic equilibrium than it actually is. The use of the single-temperature model may then lead for incorrect predictions for the aerodynanamic characteristics of a hypersonic vehicle. And as Park \cite{park1992modeling} said ``[b]ecause of this mistake, most people thought that the flight regime of the most hypersonic vehicles would be in the equilibrium regime, while, in reality, they would be in the nonequilibrium regime. The mistake is caused by using the one-temperature model''.

There is the possibility for the particles vibrational energy mode not being in self-equilibrium in some post-shock flows, as shown by Candler \textit{et al.} \cite{candler1997}, which means that a vibrational temperature $T_\text{vib}$ cannot be assigned, and that the populations of the vibrational energy levels do not follow a Boltzmann distribution. Since the populations distribution of the vibrational energy levels is unknown, there is no possibility to treat all the vibrational levels as a group. The electronic energy mode and even the rotational energy mode may also not be in self-equilibrium\footnote{Although rotational non-equilibrium is rarer owing to the small energy gaps between adjacent rotational levels.}, as referred by Munafó \textit{et al.} \cite{munafo2010}, and a similar procedure with respect to the rotational and electronic energy levels would be needed. Such procedures require models which are specific to the internal energy levels of the involved particles - the so-called state-to-state models (some authors also call them collisional-radiative models) \cite{laux2012}. Dealing with internal levels introduces an extensive set of variables to the problem, requiring much more computational resources than for the case of the multi-temperature models. The higher the number of specified internal energy modes the higher the associated computational costs. Therefore, the simultaneous assumption of self-equilibrium with respect to the fastest equilibrating energy modes, such as the rotational one, may be convenient if not necessary. And in fact, this approach is commonly taken: the most part of the existing state-to-state models are not rotational-specific \cite{laux2012}. In this work a vibronic-specific state-to-state model that assumes the translational and rotational energy modes to be in equilibrium with themselves and with each other will be considered.

\subsection{Kinetic models}
\label{subsection:Kinetic_models}

Models for the plasma kinetics, i.e. the chemical and excitation processes of the particles of the plasma, are required to properly describe the phenomenology of the post-shock flow. These models may be purely theoretical, semi-empirical (with a theoretical form, calibrated by experimental results) or purely empirical (solely described by experimental results). Both valid theoretical and semi-empirical models should agree with the experimental results. However, a significant part of the experimental data is obtained at room temperature (around $300$ K), without specificity of the internal energy levels of the particles. Therefore, the validity of some of the theoretical and semi-empirical models may be only assured for the low temperature regime, and not for each internal energy level but for the overall contribution of the set. When the process is not reasonably well understood, there is no option but to consider some crude assumptions for the dependencies on the temperature and internal levels. Example of such crude assumptions are the expressions associated with the so-called vibrational redistribution procedure, being employed by Annaloro \cite{annaloro2013}, and by Vargas \cite{vargas2021} in their thesis.

\subsubsection{Vibrational transitions by heavy particle impact}

Some important processes that will be further addressed in this work are the vibrational excitation and de-excitation of molecular particles by heavy particle impact.  Adamovich \textit{et al.} \cite{adamovich1995vibrational_a,adamovich1995vibrational_b,adamovich1998vibrational} did a detailed review of the currently available models that can describe these particular processes. One of the simplest ones is the Schwartz-Slawsky-Herzfeld model (SSH) \cite{schwartz1952calculation}. This is a semiclassical model derived under a first-order perturbation theory (FOPT) approach, assuming collinearity of the collision (all nuclei and the collision velocity vector are disposed in a single line), harmonicity of the molecular particles (internuclear forces follow Hooke's law), and an exponential repulsive interaction potential. Due to the first-order perturbation theory approach, only single vibrational energy level jumps are considered to occur in the transition process, which is solely true for cases of small collision speeds. The model should not therefore be employed in numerical simulations of entry post-shock flows, due to the very high heavy particle translational temperatures that occur in those conditions.
The most precise models include the exact quantum mechanical models, like the one suggested by Secrest and Johnson \cite{secrest1966exact}, and semiclassical models like the one developed by Billing \cite{billing1986vibration}. The latter, which corresponds to a Quasi-Classical Trajectory model (QCT) \cite{truhlar1979}, considers three-dimensional collisions, and a more realistic interaction potential. Both the model of Secrest and Johnson, and the one of Billing, require a considerable amount of computational resources, limiting their applicability. There are, though, some recent works which are overcoming this hindrance \cite{esposito1999, jaffe2010, bender2015, macdonald2018}.
An alternative model which is much more accurate than the SSH model, and at the same time more practical than the two above mentioned, is the Forced Harmonic Oscillator model (FHO) \cite{adamovich1998vibrational}. It was originally conceived by Kerner \cite{kerner1958note} and Treanor \cite{treanor1965vibrational}, being only applicable for the case of a molecular particle-atomic particle collision. It consisted of a non-perturbative semiclassical model, assuming collinearity of the collision, harmonicity of the molecular particle and an exponential repulsive interaction potential. It was then generalised by Zelechow \textit{et al.}  \cite{zelechow1968vibrational} for the case of identical molecular particle-molecular particle collisions. At last, Adamovich \textit{et al.} \cite{adamovich1995vibrational_a,adamovich1995vibrational_b,adamovich1998vibrational} applied corrections in the model to account for the anharmonicity of the molecular particles, the attractive long-range part of the interaction potential, the possible non-collinearity of the collision, the case in which the molecular particles are non-identical, and energy conservation. The resultant model was shown to agree considerable well with the results obtained through the state-of-the-art Billing's model \cite{billing1986vibration}. The work of M. Lino da Silva \textit{et al.} \cite{dasilva2007}, which considers the  model of Adamovich \textit{et al.} additionally regarding a more accurate method for the computation of the energies for the vibrational levels of the colliding particles and a more consistent modelling of the dissociation processes, endorses this agreement.

\subsubsection{Vibronic transitions by heavy particle impact}

Other important processes that should be accounted for are the vibronic transitions of molecular particles by heavy particle impact. A search in the literature unveiled that these are particularly difficult to model. All the accurate vibronic-specific rate coefficients for the respective processes which the authors found were experimentally obtained for low vibrational levels at room temperature. Due to the lack of data, simple postulatory models are usually considered. For instance, Annaloro and Bultel \cite{annaloro2014b} considered models for a full-set of vibronic processes which at a fixed heavy particle translational temperature only differ between themselves through the energy defect (the variation in the internal energy of the collision partners) or the optical allowability of the transition. Park \cite{park2008} computed electronic-specific rates using models that are simply determined from the the optical allowability and what will be termed here as effective process cross section, $\sigma_{p,\text{eff}}=\int_{0}^{\infty}\sigma_p(v)\,v\,f(v)\,dv/\int_{0}^{\infty}v\,f(v)\,dv$ (where $v$ is the relative speed of the collision partners, $\sigma_p(v)$ is the process cross section, and $f(v)$ is the distribution of relative speeds), at room temperature. Some researchers, such as Pierrot \textit{et al.} \cite{pierrot1998,pierrot1999}, do not even consider vibronic transtions by collisions with heavy particles, but solely with free electrons (such approach is not completely erroneous in the case of highly ionised flows since free electrons are much more effective in excitation than heavy particles \cite{park1990nonequilibrium}). 

There are two well-known theoretical models for electronic transitions of atomic particles by atomic particle impact: the Landau-Zener model \cite{landau1932,zener1932} and Rosen-Zener-Demkov model \cite{rosen1932,demkov1964}. The possibility of these models being extended to the case of vibronic transitions of molecular particles by heavy particles is questionable. For example, the Rosen-Zener-Demkov model cannot be straightforwardly extended since the coupling described by this model only occurs for internuclear distance values where a molecular particle is already dissociated. Regarding the Landau-Zener model, the classical equations of motion of the nuclei would need to be solved and the model applied at the internuclear distances associated with the crossing or pseudo-crossing of the potential curves - this is the so-called Trajectory Surface Hopping Approach (TSHA) \cite{tully1971}. However, such approach requires the knowledge of the perturbation induced by the collision partner on the potential curves. No data for the parameters that define this perturbation were found in the literature for the case of the heavy particle-impact vibronic transition of the two molecular particles considered in this work: \ch{N2} and \ch{N2+}. An overall lack of such data is indeed acknowledged by the scientific community, as underlined by Capitelli \textit{et al.} \cite{capitelli2000}, who say ``[...] it is practically impossible to carry out reliable theoretical calculations of the corresponding transition probabilities owing to the lack of accurate information on the structure and intersections of the colliding particles' electronic terms''.
There has been a recent attempt, done by Kirillov \cite{kirillov2004a,kirillov2004b}, to account for both Landau-Zener and Rosen-Zener-Demkov models through a general analytic expression using experimental results to calibrate the values of the involved parameters. However, not only the derivation of the analytic expression is somewhat questionable, but the approach per se lacks physical coherency, since Kirillov applied both models assuming that the respective original formulae are directly valid for vibronic transitions of colliding molecular and heavy particles beyond electronic transitions of colliding atomic particles. The relative speed between the nuclei which appears in Landau-Zener and Rosen-Zener-Demkov formulae should be the one associated with the nuclei of the same molecular particle, due to the fact that the considered vibronic transitions are the ones between electronic terms of this molecular particle. Kirillov erroneously considered the relative speed of the nuclei as the relative speed of the collision partners, and used the above-mentioned expressions to obtain rate coefficients considering a distribution of the former relative speeds. Curiously, the expression proposed by Kirillov for the effective process cross section resembles the exponential gap law regarded by Katayama \textit{et al.} \cite{katayama1979,katayama1983,katayama1984}, which is a semi-empirical law that takes into account the so-called Franck-Condon factors of the isolated molecular particles.  Conclusions reached by several researchers rose concern and doubt about the application of such model. Bachmann \textit{et al.} reported that the results of their works \cite{bachmann1992,bachmann1993} had no obvious correlation with the Franck-Condon factors, and that their use would be questionable since factors associated with the overlap of wave functions for the whole collision system should be regarded instead of Franck-Condon factors associated with isolated molecular particles. Bachmann \textit{et al.} also added that in some works, such as the one of Bondybey \textit{et al.} \cite{bondybey1978}, the ones of and Katayama \textit{et al.} \cite{katayama1979,katayama1983,katayama1984} and the one of Dentamaro \textit{et al.} \cite{dentamaro1989}, the dependence on the Franck-Condon factors is in qualitative agreement with the observations, but for others, such as the one of Jihua \textit{et al.} \cite{jihua1986}, no correlation with the Franck-Condon factors was reported. In the latter, the authors referred that their results were ``clearly inconsistent'' with Katayama's exponential gap law. Also, Piper \cite{piper1988} reported that, according to his results, the model fails to predict the energy level distributions of the species involved in the various studied energy transfer processes even qualitatively. The exponential function which appears in Katayama's exponential law has been found to be incorrect. In a work of Katayama \textit{et al.} \cite{katayama1983} a decrease of the effective process cross section with a decrease of the temperature was observed when the inverse was expected. Katayama \textit{et al.} \cite{katayama1987} said that the regarded exponential factor is characteristic of a repulsive interaction and that an attractive interaction would be more appropriate. They then suggested a substitution of the symmetric of the energy defect absolute value in the argument of the exponential function by the well depth of the interaction potential. Such substitution is complacent with the results of the works of Parmenter \textit{et al.} \cite{parmenter1979,lin1979}.
Bachmann \textit{et al.} \cite{bachmann1992,bachmann1993} studied a different type of exponential gap law which disregards the Franck-Condon factors. A good agreement of this law with experimentally obtained values was obtained in one of his works \cite{bachmann1992}. However, in a more recent work \cite{bachmann1993}. it was observed that some effective cross sections for endothermic vibronic transitions departed from the ones associated with the exothermic transitions,  the latter being favoured over the former, contrasting with the law.

\subsubsection{Examples of kinetic databases}

The main objective of this work is to devise a physically consistent database of kinetic processes which would be valid up to the highest temperatures. There has been a significant number of databases of kinetic processes regarding nitrogen vibronic-specific state-to-state models which were developed in the last decades, and the ones built by CORIA (the so-called CoRaM-\ch{N2} database) \cite{bultel2006,annaloro2014a}, EM2C \cite{pierrot1998,pierrot1999,mariotto2019}, CNR Bari \cite{capitelli1998,esposito2006,laricchiuta2006,laporta2014} and STELLAR \cite{STELLAR, lopez2013b} are examples of such databases.

\section{Theory}
\label{sec:theory}
\setlabel{Theory}{sec:theory}

In this section, a detailed physical and mathematical characterisation of the phenomena that occur in the regarded nitrogen shocked flows is presented. Such characterisation addresses the post-shock species, their energy levels and populations, the quantification of general collisional and radiative processes, and a description of particular models for some collisional processes which this work tries to focus on.

\subsection{Species and their energy levels}
\label{ssec:energy_levels}

The species which were considered in the post-shock flow were the molecular nitrogen \ch{N2}, molecular nitrogen ion \ch{N2+}, atomic nitrogen \ch{N}, atomic nitrogen ion \ch{N+} and free electron \ch{e-}. 

\subsubsection{Electronic energy levels}
\label{sssec:ele_energy_levels}

The electronic sensible energies $\epsilon_{s,\text{el},e}$ (also labelled by $T_e$)   of the molecular nitrogen \ch{N2} and molecular nitrogen ion \ch{N2+}, as well as the ionisation energy $\epsilon_{\text{el}}^+$  of \ch{N2} were taken from the literature. \Cref{tab:spec_N2} and \Cref{tab:spec_N2+} in \ref{sec:appendixA} present the electronic sensible energies and the respective references from which they were taken from. Electronic sensible energies for atomic nitrogen \ch{N} and atomic nitrogen ion \ch{N+}, as well as the ionisation energy of \ch{N} can be obtained from the National Institute of Standards and Technology (NIST) database \cite{NIST}. The NIST database takes into account the fine structure of the electronic levels which comes into play when relativistic effects, that depend on the spin of the electrons of the particle, are regarded \cite{landau1977}. When taking into account relativistic effects, electronic levels that ignored them should now be ``split'' into distinct (though very close) levels that differ in the value of the quantum number for the total angular momentum of the electrons in the particle. The NIST database issues this set of split electronic levels. Due to the small difference in the sensible energy of the split electronic levels, the major part of the studies on collisional and radiative processes that are found in the literature disregard spin-orbit splitting. Therefore, the applicability of the compiled data for such processes in this work also requires the authors to make this assumption, and single representative levels for each set of split electronic levels need to be obtained. In this work the sensible energy of the representative electronic level associated with a set of split electronic levels was defined as the mean sensible energy that a state belonging to that set may have, as Julien Annaloro also did in his thesis \cite{annaloro2013modeles}. The procedure of obtaining representative electronic levels from split ones will be termed here as ``lumping procedure''.

\subsubsection{Vibrational energy levels}
\label{sssec:vib_energy_levels}

The vibrational energies $\epsilon'_{s,\text{vib},e,v}$ (also labelled by $G_v$) of \ch{N2} and \ch{N2+} were obtained using the Fourier Grid Hamiltonian method (FGH) \cite{marston1989} which allows one to compute the eigenvalues of the Schr{\"o}dinger equation that governs vibration at each electronic level. This is an alternative to the simpler approach of expressing the vibrational energy through a Dunham expansion \cite{dunham1932} that involves the spectroscopic vibrational constants $Y_{i0}$ and the vibrational quantum number $v$, i.e. $G_v=\sum_{i=0}^{\infty} Y_{i0}\left(v+\frac{1}{2}\right)^i$. Since not much more than the first constants are known, the expansion is not valid for the higher vibrational levels and a better method is required, such as the herein proposed FGH method. To solve the  Schr{\"o}dinger equation, internuclear potential curves are needed. Let these be $V(r)$, where $r$ is the internuclear separation. Such potential curves were obtained in this work by implementation of the Rydberg \cite{rydberg1932,rydberg1933}-Klein \cite{klein1932}-Rees \cite{rees1947} method allied with extrapolation, as previously considered by M. Lino da Silva \textit{et al.} \cite{linodasilva2008} in the computation of a potential curve for the ground electronic level of molecular nitrogen, \ch{N2}(X${}^1\Sigma_\text{g}^+$). The RKR method has as input variables the spectroscopic constants that describe vibration and rotation of the molecule. Since solely a limited set of these constants is known, the valid part of the potential curve obtained through the RKR method corresponds to its middle - let this part be $V_\text{RKR}(r)$. Not one but two extrapolations are actually required for the other parts of the wanted curve: one for the short-range (left) part and another for the long-range (right) part of the potential. Let $V_\text{sr}(r)$ be the former and $V_\text{lr}(r)$ the latter. The short-range part of the potential was assumed to have the form
\begin{equation}
V_\text{sr}(r)=\alpha r^{-\beta}\text{ ,}
\label{eq:V_sr}
\end{equation}
where $\alpha$ and $\beta$ are positive constants. These constants can be determined by fitting the curve $V_{sr}(r)$ to the repulsive part of $V_\text{RKR}(r)$. In this work, the three leftest points of $V_\text{RKR}(r)$ were chosen as the fitting data. The long-range part of the potential was assumed to have one of two possible forms:
\begin{equation}
V_\text{HH}(r)=D_e\left\{\left[1-e^{-\gamma\left(r-r_e\right)}\right]^2+\delta\gamma^3\left(r-r_e\right)^3e^{-2\gamma\left(r-r_e\right)}\left[1+\zeta\gamma\left(r-r_e\right)\right]\right\}\text{ ,}
\label{eq:V_HH}
\end{equation}

\begin{equation}
V_\text{ER}(r)=D_e-D_e\left[1+\gamma\left(r-r_e\right)+\delta\left(r-r_e\right)^2+\zeta\left(r-r_e\right)^3\right]e^{-\gamma\left(r-r_e\right)}\text{ ,}
\label{eq:V_ER}
\end{equation}
where $r_e$ is the equilibrium internuclear distance, and $D_e$ is the depth of the potential well. These are \textit{a priori} known. The constants $\gamma$, $\delta$ and $\zeta$ are three fittable parameters. In this work, they were obtained by fitting the respective curve to the three rightest points of $V_\text{RKR}(r)$. Equation \eqref{eq:V_HH} represents a Hulburt-Hirschfelder potential \cite{hulburt1941} and equation \eqref{eq:V_ER} represents an Extended Rydberg potential \cite{huxley1983}. In contrast with the Extended Rydberg potential, the fit resultant Hulburt-Hirschfelder potential may express an upward ``bump'' at the right of the potential well\footnote{The states of the molecule associated with this ``bump'' are called quasi-bound states.}, which may or may not exist in reality. If it is known that such ``bump'' does not exist in reality\footnote{To verify if the molecule assumes, or not, quasi-bound states, accurate potentials in the literature (namely, the ones of the works of Hochlaf \textit{et al.} \cite{hochlaf2010a,hochlaf2010b}), which correspond to the curves that the authors intend to reconstruct, were reviewed.}, then the Extended Rydberg potential should be considered instead. \Cref{tab:spec_N2} and \Cref{tab:spec_N2+} in in \ref{sec:appendixA} present all the parameters which were used to generate the potential curves. And \Cref{fig:Ve_N2,fig:Ve_N2+}, also in \ref{sec:appendixA}, present the obtained curves.

All considered species and their energy levels are reported in \Cref{tab:lev_synopsis}.

\begingroup
\centerline{\begin{threeparttable}
\setlength\tabcolsep{8.25pt} 
\renewcommand{\arraystretch}{1.5} 
\centering
\caption{Considered species and respective energy levels. The interval that appears between parenthesis immediately after the molecular term symbols of the electronic levels of \ch{N2} and \ch{N2+} correspond to the set of values of vibrational quantum numbers for which bound vibrational levels were computed.}
\begin{scriptsize}
\begin{tabular}{cccc}
\toprule
Type & Species & Energy levels & Reference \\
\midrule
Molecule  & \ch{N2} & \parbox{10cm}{\centering \text{ }\\ X${}^1\Sigma_\text{g}^+\left(\left[0,61\right]\right)$, A${}^3\Sigma_\text{u}^+\left(\left[0,31\right]\right)$, B${}^3\Pi_\text{g}\left(\left[0,32\right]\right)$, W${}^3\Delta_\text{u}\left(\left[0,44\right]\right)$, B$'{}^3\Sigma_\text{u}^-\left(\left[0,47\right]\right)$, a$'{}^1\Sigma_\text{u}^-\left(\left[0,57\right]\right)$, a${}^1\Pi_\text{g}\left(\left[0,52\right]\right)$, w${}^1\Delta_\text{u}\left(\left[0,49\right]\right)$, A$'{}^5\Sigma_\text{g}^+\left(\left[0,5\right]\right)$, C${}^3\Pi_\text{u}\left(\left[0,4\right]\right)$, b${}^1\Pi_\text{u}\left(\left[0,28\right]\right)$, c$_3{}^1\Pi_\text{u}\left(\left[0,4\right]\right)$, c$_4'{}^1\Sigma_\text{u}^+\left(\left[0,8\right]\right)$, b$'{}^1\Sigma_\text{u}^+\left(\left[0,54\right]\right)$ and o$_3{}^1\Pi_\text{u}\left(\left[0,4\right]\right)$ \\ \text{}}  & This work \\
Molecular ion & \ch{N2+} & X${}^2\Sigma_\text{g}^+\left(\left[0,65\right]\right)$, A${}^2\Pi_\text{u}\left(\left[0,66\right]\right)$, B${}^2\Sigma_\text{u}^+\left(\left[0,38\right]\right)$, D${}^2\Pi_\text{g}\left(\left[0,38\right]\right)$ and C${}^2\Sigma_\text{u}^+\left(\left[0,13\right]\right)$ & This work\\
Atom & \ch{N} & ${}^4\text{S}_\text{u}$, ${}^2\text{D}_\text{u}$, ${}^2\text{P}_\text{u}$, ${}^4\text{P}$, ${}^2\text{P}$, ... (131 levels) (\tnote{a} ) & NIST\cite{NIST}\\
Atomic ion & \ch{N+} & ${}^3\text{P}$, ${}^1\text{D}$, ${}^1\text{S}$, ${}^5\text{S}_\text{u}$, ${}^3\text{D}_\text{u}$, ... (81 levels) (\tnote{a} ) & NIST\cite{NIST}\\
Free electron & \ch{e-} & --- & ---\\ 
\bottomrule
\end{tabular}
\end{scriptsize}
\label{tab:lev_synopsis}
\begin{scriptsize}
\begin{tablenotes}
\item[a]{As a reminder to the reader: the electronic levels were obtained through a lumping procedure applied on the ones issued by NIST.}\\
\end{tablenotes}
\end{scriptsize}
\end{threeparttable}}
\endgroup

\subsubsection{Populations of the energy levels}
\label{sssec:pop_energy_levels}

In this work a two-temperature vibronic-specific state-to-state model was considered assuming transrotational equilibrium. Therefore, a temperature $T_\text{tr$_\text{h}$}$ was assigned to the translational and rotational energy modes of the heavy particles, and a temperature $T_\text{tr$_\text{e}$}$ was assigned to the free electron translational energy mode. In this model, assuming that the $s$-th species is molecular, the number of particles of the $s$-th species in the $n$-th translational, $J$-th rotational, $v$-th vibrational and $e$-th electronic levels corresponds to
\begin{equation}
N_{s,n,J,v,e}(N_{s,v,e},T_\text{tr$_\text{h}$})=N_{s,v,e}\frac{g_{s,\text{tr},n}\cdot g_{s,\text{rot},J,v,e}\,e^{-\frac{\epsilon_{s,\text{tr},n}}{k_B T_\text{tr$_\text{h}$}}-\frac{\epsilon_{s,\text{rot},J,v,e}}{k_B T_\text{tr$_\text{h}$}}}}{Q_{s,\text{tr}}(T_\text{tr$_\text{h}$})\cdot Q_{s,\text{rot},v,e}(T_\text{tr$_\text{h}$})}\text{ ,} 
\label{eq:Boltzmann_distribution_state_to_state}
\end{equation}
in which $k_B$ is the Boltzmann constant, $N_{s,v,e}$ is the number of particles of the $s$-th species in the $v$-th vibrational and $e$-th electronic level, $\epsilon_{s,\text{tr},n}$ is the sensible energy of the $n$-th translational level, $\epsilon_{s,\text{rot},e,v,J}$ is the sensible energy of the $J$-th rotational level associated with the $e$-th vibrational and $v$-th vibrational levels, $g_{s,\text{tr},n}$ and $g_{s,\text{rot},J,v,e}$ are the respective degrees of degeneracy, and $Q_{s,\text{tr}}(T_\text{tr$_\text{h}$})$ and $Q_{s,\text{rot},v,e}(T_\text{tr$_\text{h}$})$ are the respective partition functions. By using the model of the free particle in a rigid rectangular box \cite{vincenti1965introduction} for the translation of the particles, and the model of the rigid rotor \cite{vincenti1965introduction} for the rotation of a diatomic particle, one may show that these partition functions are given by
\vspace{-30pt}
\begin{multicols}{2}
\begin{equation}
Q_{s,\text{tr}}(T_\text{tr$_\text{h}$},V)=V\left(\frac{2\pi m_sk_BT_\text{tr$_\text{h}$}}{h^2}\right)^{\frac{3}{2}}\text{ ,}
\label{eq:Partition_function_tr_2}
\end{equation}

\begin{equation}
Q_{s,\text{rot},e,v}(T_\text{tr$_\text{h}$})=\frac{k_BT_\text{tr$_\text{h}$}}{\sigma_s B_{s,e,v}}\text{ ,}
\label{eq:Partition_function_rot_3}
\end{equation}
\end{multicols}
\noindent
respectively, with $h$ being the Planck constant, $V$ the volume of the local element of a fluid, $m_s$ the mass of the $s$-th species particle, $\sigma_s$ its nuclear symmetry factor (which is $1$ if the nuclei are different and $2$ if they are identical\footnote{This is the case of the molecular particles considered in this work, \ch{N2} and \ch{N2+}, which are homonuclear.}), and $B_{s,e,v}$ the first spectroscopic rotational function of the $s$-th species associated with the $e$-th electronic and $v$-th vibrational levels (usually labelled by $B_v$). In this work, due to the current limitations of the employed CFD code, SPARK \cite{lopez2016}, the vibrational dependence of the function $B_{s,e,v}$ was disregarded, and the electronic dependence was approximated by a dependence on the ground electronic level - let this be X - and, therefore, $B_{s,e,v}\approx B_{s,\text{X}}$.

\subsection{Collisional processes} 

In respect of collisional processes, solely binary processes, which result from collisions between two particles, were considered in this work. Such processes are described by a chemical equation of the form 
\begin{equation}
\ch{A} + \ch{B} \ch{<=>} \nu'_\text{C} \ch{C} + \ch{...}\text{ ,} 
\label{eq:chemical_reaction_A_B_C}
\end{equation}
meaning that particles \ch{A} and \ch{B} collide with each other, producing particles C and possibly others. 
Under translational equilibrium, the rate coefficient $k_f$ of the process is related to the process cross section \cite{kuppermann1968chemical} $\sigma_p(v)$ and the distribution of relative speeds $f(v,T_\text{c})$ through
\begin{equation}
k_f(T_\text{c})=\int_{0}^{\infty}\sigma_p(v)\,v\,f(v,T_\text{c})\,dv=\frac{\sqrt{\frac{8k_BT_\text{c}}{\pi\mu}}}{1+\delta_\text{AB}}\cdot\underbrace{\frac{\int_{0}^{\infty}\sigma_p(v)v^3e^{-\frac{\mu v^2}{2k_BT_\text{c}}}\,dv}{2\left(\frac{k_BT_\text{c}}{\mu}\right)^2}}_{=:\sigma_{p,\text{eff}}(T_\text{c})}\text{ ,}
\label{eq:kf_Maxwell}
\end{equation}
with $v$ being the relative speed of the collision partners, $T_c$ the controlling temperature ($T_{\text{tr}_\text{h}}$ if A and B are heavy particles, or $T_{\text{tr}_\text{e}}$ if one of them is a free electron), $\mu=m_\text{A}\cdot m_\text{B}/\left(m_\text{A}+m_\text{B}\right)$ the reduced mass of the particles, $\delta_\text{AB}$ a Kronecker delta (giving 1 if $\ch{A}=\ch{B}$, and $0$ if not), and $\sigma_{p,\text{eff}}(T_\text{c})=\int_{0}^{\infty}\sigma_p(v)\,v\,f(v,T_\text{c})\,dv/\int_{0}^{\infty}v\,f(v,T_\text{c})\,dv$ is the effective process cross section (which was already mentioned in the \ref{sec:introduction} section). This latter quantity may be expressed using the relative kinetic energy of the collision partners, $E=\frac{1}{2}\mu v^2$, or the respective dimensionless variable $u=\frac{E}{k_B T_{\text{tr}_\text{h}}}$, instead of the relative speed by performing a change of variables:
\begin{equation}
\sigma_{p,\text{eff}}(T_\text{c})=\frac{\int_{0}^{\infty}\sigma_p(v)v^3e^{-\frac{\mu v^2}{2k_BT_\text{c}}}\,dv}{2\left(\frac{k_BT_\text{c}}{\mu}\right)^2}=\frac{\int_{0}^{\infty}\sigma_p(E)Ee^{-\frac{E}{k_BT_\text{c}}}\,dE}{\left(k_BT_\text{c}\right)^2}=\int_{0}^{\infty}\sigma_p(u,T_\text{c})\,u\,e^{-u}\,du\text{ .}
\label{eq:ss_av}
\end{equation}
The process cross section is given by $\sigma_p(v)=\sigma(v)\cdot P(v)$, with $\sigma(v)$ being the collisional cross section, and $P(v)$ the process probability. 

The rate coefficient values obtained from \eqref{eq:kf_Maxwell} were conveniently modelled though curve fitting using the modified Arrhenius equation
\begin{equation}
k_f(T_\text{c})=A{T_\text{c}}^n e^{-\frac{E_a}{k_BT_\text{c}}},
\label{eq:modifiedArrhenius}
\end{equation}
or a function which was previously employed by Lopez \textit{et al.} in one of their works \cite{lopez2013},
\begin{multline}
    \frac{k_f(T_\text{c})}{\left[k_f\right]} =  \exp\left[\vphantom{a_1\left(\frac{T_\text{c}}{T_{c,\text{ref}}}\right)^{-3}+a_2\left(\frac{T_\text{c}}{T_{c,\text{ref}}}\right)^{-2}+a_3\left(\frac{T_\text{c}}{T_{c,\text{ref}}}\right)^{-1}+a_4\ln\left(\frac{T_\text{c}}{T_{c,\text{ref}}}\right)+ a_5+a_6\frac{T_\text{c}}{T_{c,\text{ref}}}+a_7\left(\frac{T_\text{c}}{T_{c,\text{ref}}}\right)^{2}+a_8\left(\frac{T_\text{c}}{T_{c,\text{ref}}}\right)^{3}+a_9\left(\frac{T_\text{c}}{T_{c,\text{ref}}}\right)^{4}} a_1\left(\frac{T_\text{c}}{T_{c,\text{ref}}}\right)^{-3}+a_2\left(\frac{T_\text{c}}{T_{c,\text{ref}}}\right)^{-2}+a_3\left(\frac{T_\text{c}}{T_{c,\text{ref}}}\right)^{-1}+a_4\ln\left(\frac{T_\text{c}}{T_{c,\text{ref}}}\right)\right.\\
     \left. + a_5 +a_6\frac{T_\text{c}}{T_{c,\text{ref}}}+a_7\left(\frac{T_\text{c}}{T_{c,\text{ref}}}\right)^{2}+a_8\left(\frac{T_\text{c}}{T_{c,\text{ref}}}\right)^{3}+a_9\left(\frac{T_\text{c}}{T_{c,\text{ref}}}\right)^{4} \vphantom{a_1\left(\frac{T_\text{c}}{T_{c,\text{ref}}}\right)^{-3}+a_2\left(\frac{T_\text{c}}{T_{c,\text{ref}}}\right)^{-2}+a_3\left(\frac{T_\text{c}}{T_{c,\text{ref}}}\right)^{-1}+a_4\ln\left(\frac{T_\text{c}}{T_{c,\text{ref}}}\right)+ a_5+a_6\frac{T_\text{c}}{T_{c,\text{ref}}}+a_7\left(\frac{T_\text{c}}{T_{c,\text{ref}}}\right)^{2}+a_8\left(\frac{T_\text{c}}{T_{c,\text{ref}}}\right)^{3}+a_9\left(\frac{T_\text{c}}{T_{c,\text{ref}}}\right)^{4}}\right]\text{ .}
\label{eq:Poly9thOrder}
\end{multline}
Note that, $A$, $n$ and $E_a$ in \eqref{eq:modifiedArrhenius}, and $\left\{a_i\right\}$, with $i=1, ..., 9$ in \eqref{eq:Poly9thOrder} are the adjustable parameters. Also, in the \eqref{eq:Poly9thOrder}, $T_{c,\text{ref}}=1000\text{K}$ corresponds to a reference temperature, and $\left[k_f\right]$ are the units of the rate coefficient values (since  \eqref{eq:Poly9thOrder} is dimensionless). Function \eqref{eq:Poly9thOrder} was employed when \eqref{eq:modifiedArrhenius} (which has a lesser number of adjustable parameters) could not properly model the rate coefficient values.

All the above-mentioned procedure was followed whenever process cross sections $\sigma_p$ or effective ones $\sigma_{p,\text{eff}}$ were found in the literature, or whenever theoretical or semi-empirical models were used in the computation of these variables. When solely rate coefficients were found in the literature, these were taken, and the curve \eqref{eq:modifiedArrhenius} or \eqref{eq:Poly9thOrder} was fitted to them. The data for some processes involving the vibrational levels of $\ch{N_2}\left(\text{X}{}^1\Sigma_{\text{g}}^+\right)$, which were reported in the literature and were employed in this work, considered a database of vibrational levels that was different from the one regarded here. Therefore, to make these data applicable, it was determined to linearly interpolate the issued values with respect to the vibrational energies if the issued energies were comprised to the in-work database, or to linearly extrapolate if they were not - this procedure will be termed by ``Adaption in respect of the Database for the Vibrational energy levels'' (ADV). Also, whenever rate coefficients for a particular vibrational quantum number or for the contribution of all of them were available but not for a full set of vibrational qunatum numbers, a vibrational redistribution procedure (VRP), based on the one considered by Julien Annaloro in his PhD thesis \cite{annaloro2013} (which will be described further bellow), was used to compute the latter.

Table \ref{tab:kin_h_synopsis} in \ref{sec:appendixB} makes a synopsis on the collisional processes due to heavy particle impact which were regarded in this work. Table \ref{tab:kin_e_synopsis} makes a synopsis on the collisional processes due to electron impact. The two tables present the type of process, chemical equation, additional remarks, and reference from which data were taken to compute the forward rate coefficients $k_f$. 

\subsubsection{Vibrational-translational process}

\begin{figure}[H]
\centering
\centerline{\includegraphics[scale=1]{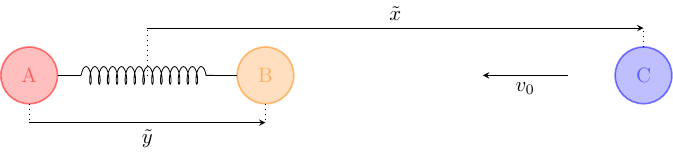}}
\caption{Collinear collision between a diatomic molecular particle AB and an atomic particle C.}
\label{fig:FHO_AB_C}
\end{figure}

In this work the probability of a molecular particle AB to transit from some vibrational level to another by a collision with an atomic particle C - the so-called vibrational-translational process (V-T) - was modeled by the FHO model \cite{rapp1969theory,kerner1958note,treanor1965vibrational}. In this model, the collision is assumed to be collinear (all the nuclei are positioned in the same line) as depicted in Figure \ref{fig:FHO_AB_C}.
\noindent
The molecular particle is assumed to be a harmonic oscillator, and, therefore, the force that the nucleus A imposes on nucleus B follows Hooke's law, being given by $F_{\text{AB}}=-f\left(\tilde{y}-\tilde{y}_0\right)$, where $f$ is the force constant, $\tilde{y}$ is the separation between the nuclei of the molecular particle, and $\tilde{y}_0$ is the respective equilibrium value. Also, the interaction potential of the collision partners is regarded to have a form that resembles the one of the Morse potential:
\begin{equation}
V'(\tilde{x},\tilde{y})=E_\text{M}\left[e^{-\alpha\left(\tilde{x}-\tilde{x}_0\right)}-2e^{-\frac{\alpha}{2}\left(\tilde{x}-\tilde{x}_0\right)}\right]e^{\alpha\gamma\left(\tilde{y}-\tilde{y}_0\right)}\text{ .}
\label{eq:Morse_adapted_VT}
\end{equation}
where $E_\text{M}$ is the potential well depth, $\alpha$ is a reciprocal length parameter, $\gamma=m_\text{A}/(m_\text{A}+m_\text{B})$ is the ratio between the mass of the nucleus A and the mass of the molecular particle, $\tilde{x}$ is the separation between the collision partners' centres of masses, and $\tilde{x}_0$ is the respective equilibrium value. 
The FHO model is semiclassical since it considers the separation between the collision partners $\tilde{x}$, and the molecular particle's elongation $\tilde{y}-\tilde{y}_0$, to be in agreement with Classical and Quantum Mechanics, respectively.
It can be shown that the probability of the molecular particle to transit from the $v$-th vibrational level to the $v'$-th one is
\begin{equation}
P_v^{v'}=v!v'!\eta_0^{v+v'}e^{-\eta_0}\left(\sum_{k=0}^l\frac{(-1)^k\eta_0^{-k}}{(v-k)!(v'-k)!k!}\right)^2\text{ ,}
\label{eq:P_nm}
\end{equation}
where $l=\text{min}\left(v,\,v'\right)$, and $\eta_0$ is a parameter given by
\begin{equation}
\eta_0=\frac{8\pi^2\omega\tilde{m}^2\gamma^2}{\hbar\mu\alpha^2}\csch^2\left(\frac{2\pi \omega}{\alpha v_0}\right)\cosh^2\left[\frac{2\pi\omega}{\alpha v_0}\left(\frac{1}{2}+\frac{\phi}{\pi}\right)\right]\text{ ,}
\label{eq:eta0_VT_Morse}
\end{equation}
In \eqref{eq:eta0_VT_Morse}, $\tilde{m}=(m_\text{A}+m_\text{B})\cdot m_\text{C}/(m_\text{A}+m_\text{B}+m_\text{C})$ is the reduced mass of the collision partners, $v_0$ is their initial relative speed, $\mu=m_\text{A}\cdot m_\text{B}/(m_\text{A}+m_\text{B})$ is the reduced mass of the molecular particle, $\omega=\sqrt{f/\mu}$ is its natural angular frequency of oscillation, and $\phi$ is a parameter given by $\phi=\arctan\left(\sqrt{E_\text{M}/\frac{1}{2}\tilde{m}v_0^2}\right)$. 

The model may be corrected with respect to the anharmonicity of the molecular particle, conservation of energy, and non-collinearity of the collisions, as detailed in the works of Billing \cite{billing1973}, Billing and Fisher \cite{billing1976}, and Adamovich \textit{et al.} \cite{adamovich1995vibrational_a}. 

\subsubsection{Vibrational-dissociative processes}\label{subsection:VD_processes}

In this work, the vibrational-dissociative processes were modelled by taking the assumption of M. Lino da Silva \textit{et al.} \cite{dasilva2007}: dissociation occurs if the final vibrational level is a quasi-bound level, i.e. a level whose energy $G_v$ is equal or higher than the potential well depth $D_e$. Let $v_D$ be the vibrational quantum number associated with the lowest quasi-bound level. The probability of a particle, initially in the $v$-th vibrational level, to dissociate after a collision is then
\begin{equation}
P_{v}^{D}=\sum_{v'\geq v_D}P_{v}^{v'}\text{ ,}
\label{eq:P_diss_VT}
\end{equation}
with $P_{v}^{v'}$ being given by \eqref{eq:P_nm}. The number of quasi-bound levels involved in the sum is completely arbitrary, and may be conveniently chosen such that the resultant probability matches a benchmark one.

\subsubsection{Vibrational-vibrational-translation processes}\label{subsection:V-V-T processes}

\begin{figure}[H]
\centering
\centerline{\includegraphics[scale=1]{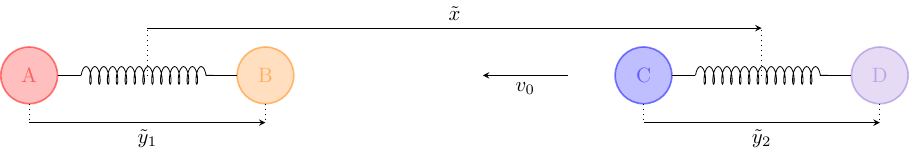}}
\caption{Collinear collision between two diatomic molecular particles AB and CD.}
\label{fig:FHO_AB_CD}
\end{figure}

Transitions between vibrational levels in a collision between two diatomic molecular particles, AB and CD - the so-called vibrational-vibrational-translational process (V-V-T) - were also considered in this work. Such collision is depicted by \Cref{fig:FHO_AB_CD}. 

The probability of the particle AB to transit from the vibrational level $v_1$ to $v'_1$ and of the particle CD to transit from $v_2$ to $v'_2$, $P_{v_1,v_2}^{v'_1,v'_2}$ may be computed through a well-based theoretical model which was originally developed by Zelechow \textit{et al.}  \cite{zelechow1968vibrational} and adapted by Adamovich \textit{et al.} \cite{adamovich1998vibrational}. Its implementation in this work would, however, be impractical due to the amount of computational resources which it requires. Therefore, it was decided to consider the approximation suggested by Adamovich \textit{et al.} \cite{adamovich1998vibrational}: if $T_\text{tr$_\text{h}$}\gg T_\text{vib}$, the transition probability of a V-V-T process corresponds to a multiplication of two uncoupled probabilities with V-T process resemblance:
\begin{equation}
P_{v_1,v_2}^{v_1',v_2'}=P_{v_1}^{v_1'}\cdot P_{v_2}^{v_2'}\text{ ,}
\label{eq:P_red}
\end{equation}
where $P_{v_1}^{v_1'}$ is given by \eqref{eq:P_nm} with the respective parameters substituted by the ones of the particle AB. The analogous follows for $P_{v_2}^{v_2'}$.

Note that the rate of change of the amount concentration of molecular particles AB($v_1'$), i.e. [AB($v_1'$)], due to collisions between AB($v_1$) and CD($v_2$), in which $v_2'$ may represent a quasi-bound level (hence meaning a dissociation of CD), is given by
\begin{equation}
\left(\frac{d\left[\text{AB}(v_1')\right]}{dt}\right)_{v_1,v_2}^{v_1',v_2'}=\left(\nu'_{\text{AB}(v_1')}-\nu_{\text{AB}(v_1')}\right)\,k_{v_1,v_2}^{v_1',v_2'}\,[\text{AB}(v_1)][\text{CD}(v_2)]\text{ ,}
\label{eq:variaton_AB_v_due_to_coll_with_C}
\end{equation}
with $k_{v_1,v_2}^{v_1',v_2'}$ being the rate coefficient, and $\nu_{\text{AB}(v_1')}$ and $\nu'_{\text{AB}(v_1')}$ the stoichiometric coefficients of $\text{AB}(v_1')$ at the reactants and products sides of the respective chemical equation. It can be shown that $\nu'_{\text{AB}(v_1')}-\nu_{\text{AB}(v_1')}=1-\delta_{v_1,v_1'}+\delta_{\text{AB},\text{CD}}\cdot\left(\delta_{v_2',v_1'}-\delta_{v_2,v_1'}\right)$, with $\delta$ being the Kronecker delta. The rate coefficient corresponds to
\begin{equation}
k_{v_1,v_2}^{v_1',v_2'}(T_\text{tr$_\text{h}$})=\int_{0}^{\infty}\sigma_{v_1,v_2}^{v_1',v_2'}(v_0)\,v_0\,f(v_0,T_\text{tr$_\text{h}$})\,dv_0\text{ ,}
\label{eq:kf_2}
\end{equation}
where $\sigma_{v_1,v_2}^{v_1',v_2'}(v_0)=\sigma(v_0)\cdot P_{v_1,v_2}^{v_1',v_2'}(v_0)$ is the process cross section and $\sigma(v_0)$ is the collisional cross section. Due to the collinearity of the collisions assumed by the FHO model, it may not be appropriate to consider as the distribution function of relative speeds $f(v_0,T_\text{tr$_\text{h}$})$ the one previously introduced where $v_0$ is the norm $g$ of a relative velocity $\vec{g}$ that may take any orientation with respect to the line of the centres. It was decided to follow the rationale of Adamovich \textit{et al.} \cite{adamovich1998vibrational} who defined $f(v_0,T_\text{tr$_\text{h}$})$ as the distribution of absolute values of the projection of the relative velocity on the line of the centres, $v_0=g\cos\psi$, with $\psi$ being the angle that $\vec{g}$ makes with the line. Such distribution is given by
\begin{equation}
f(v_0,T_\text{tr$_\text{h}$})=\frac{4}{1+\delta_{\text{AB}(v_1),\text{CD}(v_2)}}\sqrt{\frac{\mu}{2\pi k_BT_\text{tr$_\text{h}$}}}e^{-\frac{\mu v_0^2}{2k_B{T_\text{tr$_\text{h}$}}}}\text{ .}
\label{eq:f_v_axial}
\end{equation}
One comment should be made about the uncoupling approximation of the transition probabilities for the case of V-V-T processes, given by equation \eqref{eq:P_red}. This approximation allows one to compute a rate coefficient for a vibrational transition of a particle, say from $v_1$ to $v_1'$, that accounts for all the possible transitions as well as dissociation of the other particle, from $v_2$ to $v_2'$ (including the $v_2'$ quasi-bound levels, meaning dissociation of CD). It can be shown that the rate of change of the amount concentration of molecular particles AB($v_1'$) due to collisions between AB($v_1$) with CD($v_2$) for all $v_2$ and $v_2'$ is
\begin{equation}
\left(\frac{d\left[\text{AB}(v_1')\right]}{dt}\right)_{v_1}^{v_1'}=\sum_{v_2,v_2'}\left(\frac{d\left[\text{AB}(v_1')\right]}{dt}\right)_{v_1,v_2}^{v_1',v_2'}\approx\left(1-\delta_{v_1,v_1'}\right)\,k_{v_1}^{v_1'}(T_\text{tr$_\text{h}$})\,[\text{AB}(v_1)][\text{CD}]\text{ ,}
\label{eq:variaton_AB_v_due_to_coll_with_CD_sum_2}
\end{equation}
where the involved Kronecker deltas which give $1$ when $\text{AB}(v_1)=\text{CD}(v_2)$, $\text{AB}(v'_1)=\text{CD}(v_2)$, or $\text{AB}(v'_1)=\text{CD}(v'_2)$ were neglected due the fact that such cases occur in a much lesser frequency than the opposite ones. The referred rate coefficient is
\begin{equation}
k_{v_1}^{v_1'}(T_\text{tr$_\text{h}$})=\int_{0}^{\infty}\sigma(v_0)P_{v_1}^{v_1'}(v_0)\,v_0\,f(v_0,T_\text{tr$_\text{h}$})\,dv_0\text{ .}
\label{eq:k_VVT}
\end{equation}
And for the case of dissociation of a particle independently of the fate of the collision partner, the rate coefficient is given by
\begin{equation}
k_{v_1}^{D}(T_\text{tr$_\text{h}$})=\int_{0}^{\infty}\sigma(v_0)P_{v_1}^{D}(v_0)\,v_0\,f(v_0,T_\text{tr$_\text{h}$})\,dv_0\text{ .}
\label{eq:k_VVT_VD}
\end{equation}

\subsubsection{Vibrational-electronic processes}
By taking into account the state-of-the-art of the models for vibrational-electronic processes (V-E), i.e. vibronic transitions of molecular particles by impact with heavy particles, described in the \ref{sec:introduction} section, it was decided to employ in this work an expression for the effective process cross sections with the vibronic dependence suggested by Bachmann \textit{et al.} \cite{bachmann1992,bachmann1993} and the thermal dependence suggested by Katayama \textit{et al.} \cite{katayama1987}:
\begin{equation}
\sigma_{p,\text{eff}}\left(T_{\text{tr}_\text{h}}\right)=\sigma'_0\,e^{-\frac{|\Delta E|}{E_0}+\frac{\varepsilon}{k_BT_{\text{tr}_\text{h}}}}\text{ ,}
\label{eq:ss_av_elio}
\end{equation}
where $\sigma'_0=\sigma_0e^{-\frac{\varepsilon}{k_BT_{\text{ref}}}}$ is described by a characteristic cross section $\sigma_0$, a characteristic energy $E_0$, and a reference temperature $T_{\text{ref}}$. The quantity $\varepsilon$ is the well depth of the potential energy curve associated with interaction between the collision partners. $\Delta E$ is the energy defect, i.e. the difference between the initial and final internal energies of the collision partners.  From \eqref{eq:kf_Maxwell}, the respective rate coefficient is given by
\begin{equation}
k_f(T_{\text{tr}_\text{h}})=\frac{\sigma'_0\,e^{-\frac{|\Delta E|}{E_0}}}{1+\delta_{\text{AB}(e_1,v_1),\text{M}}}\sqrt{\frac{8k_BT_{\text{tr}_\text{h}}}{\pi\mu}}e^{\frac{\varepsilon}{k_BT_{\text{tr}_\text{h}}}}\text{ .}
\label{eq:kf_elio}
\end{equation}
which may be conveniently expressed through a modified Arrhenius function \eqref{eq:modifiedArrhenius}.

\subsubsection{Electronic excitation and ionisation of atomic particles by electron impact}
\label{subsection:Drawin}

The authors considered the theoretical values obtained by Berrington \textit{et al.} \cite{berrington1975} for the process cross sections associated with the excitation of \ch{N$\left({}^4\text{S}_{\text{u}}\right)$} to \ch{N$\left({}^2\text{D}_{\text{u}}\right)$} and \ch{N$\left({}^2\text{P}_{\text{u}}\right)$}, as well as the excitation of \ch{N$\left({}^2\text{D}_{\text{u}}\right)$} to \ch{N$\left({}^2\text{P}_{\text{u}}\right)$}. The experimental values for the process cross sections obtained by Brook \textit{et al.} \cite{brook1978} were considered for the ionisation of \ch{N$\left({}^4\text{S}_{\text{u}}\right)$}, and the numerical values for the process cross sections computed by Wang \textit{et al.} \cite{wang2014} were regarded for the ionisation of \ch{N$\left({}^2\text{D}_{\text{u}}\right)$} and \ch{N$\left({}^2\text{P}_{\text{u}}\right)$}. Since no experimental or accurate theoretical data were found for the other electronic levels of \ch{N}, it was necessary to rely on empirical correlations to complete the database. The same applied for the case electronic excitation of \ch{N+} for all of its electronic levels. 
Rate coefficients for the excitation and ionisation of an atomic particle by electron impact can be computed through the well-known empirical correlations originally obtained by Drawin \cite{drawin1963,drawin1968}. In this work, the Drawin expressions adapted by Panesi \textit{et al.} \cite{panesi2009} were preferred due to their simplicity.

\subsubsection{Electronic excitation and ionisation of atomic particles by heavy particle impact}
\label{subsection:E-a-h}

Due to an overall lack of experimental data for the electronic excitation and ionisation of \ch{N} and \ch{N+} by heavy particle impact, it was decided to rely on semi-empirical formulae to compute the respective rate coefficients. For this we considered the relation obtained by Annaloro \textit{et al.} \cite{annaloro2014a}.
%
The collision partners of \ch{N} and \ch{N+} were considered to be \ch{N} and \ch{N2} in similarity to the work of Annaloro et al. \cite{annaloro2014a}. The relation of Annaloro is an approximation which is based on a suggestion of Park \cite{park1988b}, that tells that the process cross section $\sigma_p$ depends on the relative kinetic energy of the collision partner $E$ through $\sigma_p(E)=\sigma_0\ln(E/\Delta \epsilon)/(E/\Delta \epsilon)$. This suggestion is in turn based on the empirical correlation established by Lotz \cite{lotz1968} for the ionisation of atomic particles by impact with free electrons in the limit $E\gg\Delta \epsilon$. 

\subsection{Radiative processes}

The radiative processes which were considered in this work were solely the spontaneous emission processes. These are quantified by appropriate Einstein coefficients $A_{s,e,v}^{e',v'}$. 
Einstein coefficients for spontaneous emission  of \ch{N2} and \ch{N2+} were directly extracted from the literature or were computed through a theoretical expression involving the so-called sums of the electronic-vibrational transition moments $\left(\sum R_e^2\right)_{s,e,v}^{e',v'}$ \cite{laux1992}, with these being also taken from the literature. Such expression corresponds to \cite{whiting1974}
\begin{equation}
A_{s,e,v}^{e',v'}=\frac{16\pi^3}{3\varepsilon_0c^3h}\left(\nu_{s,e,v}^{e',v'}\right)^3\frac{\left(\sum R_e^2\right)_{s,e,v}^{e',v'}}{\left(2-\delta_{0,\Lambda}\right)\left(2S+1\right)}\text{ ,}
\label{eq:A_whiting}
\end{equation}
where $\nu_{s,e,v}^{e',v'}$ is the frequency of the emitted photon, $\Lambda$ is the initial quantum number for the projection of the total electronic orbital angular momentum vector on the internuclear axis, and $S$ is the initial total spin quantum number. Also, $c$ is the speed of light and $\varepsilon_0$ is the vacuum permittivity.
Table \ref{tab:rad_mol_synopsis} in \ref{sec:appendixB} assembles all the accounted molecular spontaneous emission processes, listing the name of the electronic system, the initial and final electronic levels, the maximum initial and final vibrational quantum numbers, as well as the reference from which the data were taken from.  

For the case of \ch{N} and \ch{N+}, Einstein coefficients for spontaneous emission were extracted from the NIST database \cite{NIST}. These coefficients are in respect of electronic levels that take into account fine structure. Since this work considers representative electronic levels computed from the ones with fine structure, it was necessary to compute Einstein coefficients in respect of these representative electronic levels as well. 
A synopsis on all of the considered atomic spontaneous emission processes is presented in Table \ref{tab:rad_ato_synopsis} in \ref{sec:appendixB}.

\subsection{A vibrational redistribution procedure (VRP)}
\label{section:VRP}

Let one consider some important process involving the vibrational energy mode, which is not reasonably well understood. Yet, its rate coefficient can be measured for some particular vibrational quantum number or for the overall contribution of the full set of vibrational quantum numbers. To use these rate coefficients in vibronic-specific state-to-state simulations it is firstly necessary to obtain vibronic-specific ones from them. Since proper theoretical models are unavailable, it is necessary to rely on some general semi-empirical rules. In this section an approach to compute a full set of vibronic-specific rate coefficients from an overall rate coefficient or from a single vibronic-specific rate coefficient is described. Such approach is based on the vibrational redistribution procedure (VRP) considered by Julien Annaloro in his PhD thesis \cite{annaloro2013}.

\subsubsection{VRP on the final vibrational quantum number, from an overall rate coefficient}
\label{subsection:VRP_vp_overall}

Let one consider a vibronic-specific process described by the chemical equation
\begin{equation}
\ch{...}\ch{ -> X}(e',v')\ch{ + ...}\text{ ,}
\label{eq:VRP_ep_vp}
\end{equation}
where the ellipses represent reactants and the other products. Vibronic-specific rate coefficients of the form $k^{v'}(T_\text{c})$ are to be computed from an overall rate coefficient $k(T_\text{c})$, the latter being with respect to the electronic-specific counterpart process of \eqref{eq:VRP_ep_vp}, i.e.: 
\begin{equation}
\ch{...}\ch{ -> X}(e')\ch{ + ...}\text{ .}
\label{eq:VRP_ep}
\end{equation}
Note that $T_\text{c}$ corresponds to the controlling temperature of the process. One may show that the two types of rate coefficients are related to each other through
\begin{equation}
k(T_\text{c})=\sum_{v'}k^{v'}(T_\text{c})\text{ .}
\label{eq:k_VRP_ep}
\end{equation}
Now, let one take the assumption that both rate coefficients may be described by modified Arrhenius functions \eqref{eq:modifiedArrhenius}:
\begin{subnumcases}{}
k(T_\text{c})=A\,{T_\text{c}}^n e^{-\frac{E_a}{k_BT_\text{c}}}\text{ ,}
\label{eq:k_VRP_ep_model}
\\
k^{v'}(T_\text{c})=A^{v'}(T_\text{c})\,{T_\text{c}}^{n^{v'}} e^{-\frac{E_a^{v'}}{k_BT_\text{c}}}\text{ .}
\label{eq:k_VRP_ep_vp_model}
\end{subnumcases}
with the pre-exponential factor of the vibronic-specific rate $A^{v'}$ being allowed to be dependent on the controlling temperature $T_\text{c}$ (hence the presence of the respective symbol in its argument). The parameters $A$, $n$ and $E_a$ of the overall rate coefficient are all known. Annaloro \cite{annaloro2013} assumed that the power on the temperature of the vibronic-specific rate is the same as the one of the overall rate: 
\begin{equation}
n^{v'}=n\text{ .}
\label{eq:n_vp_VRP}
\end{equation}
The activation energy of vibronic-specific rate was defined as the difference between the internal energy of the products and the internal energy of the reactants, if non-negative, or as zero, if negative. One may express this in terms of the previously introduced energy defect\footnote{The energy defect $\Delta E$ was defined as the difference between the internal energy of the reactants and the internal energy of the products, precisely the symmetric of the difference of energies mentioned in the text.} $\Delta E$ as
\begin{equation}
E_a^{v'}=-\Delta E^{v'} \cdot\operatorname{H}(-\Delta E^{v'})\text{,}
\label{eq:E_a_vp_VRP}
\end{equation}
with $\operatorname{H}(-\Delta E^{v'})$ being the Heaviside function, which gives $0$ if $-\Delta E^{v'}<0$ and $1$ if $-\Delta E^{v'}\ge 0$. Regarding the pre-exponential factor $A^{v'}(T_\text{c})$, Annaloro assumed that it increases linearly with $|\Delta E^{v'}|$ if $\Delta E^{v'}<0$ or with the inverse of $|\Delta E^{v'}|$ if $\Delta E^{v'}>0$. The proportionality coefficients for each cases are defined through a common coefficient, say $B(T_\text{c})$, and the minimum values of $|\Delta E^{v'}|$ associated with the respective signals, $|\Delta E^{-}|_\text{min}$ and $|\Delta E^{+}|_\text{min}$. The case $|\Delta E^{v'}|=0$ is not regarded, being assumed to not occur. One has
\begin{equation}
A^{v'}(T_\text{c})=B(T_\text{c})\left[\frac{|\Delta E^{v'}|}{|\Delta E^{-}|_\text{min}}\operatorname{H}(-\Delta E^{v'})+\frac{|\Delta E^{+}|_\text{min}}{|\Delta E^{v'}|}\operatorname{H}(\Delta E^{v'})\right]\text{.}
\label{eq:A_vp_VRP}
\end{equation}
The common coefficient $B(T_\text{c})$ is determined from condition \eqref{eq:k_VRP_ep}, which gives
\begin{equation}
B(T_\text{c})=\frac{A\,e^{-\frac{E_a}{k_B T_\text{c}}}}{\sum_{v'}\left[\frac{|\Delta E^{v'}|}{|\Delta E^{-}|_\text{min}}\,e^{\frac{\Delta E^{v'}}{k_B T_\text{c}}}\operatorname{H}(-\Delta E^{v'})+\frac{|\Delta E^{+}|_\text{min}}{|\Delta E^{v'}|}\operatorname{H}(\Delta E^{v'})\right]}\text{.}
\label{eq:B_VRP}
\end{equation}

\subsubsection{VRP on the final vibrational quantum number, from a single vibronic-specific rate coefficient}
\label{subsection:VRP_vp_single}

Let one consider the case in which solely a rate coefficient of \eqref{eq:VRP_ep_vp} for a particular vibrational quantum number $v'=v'_\text{ref}$ is known, being given by a modified Arrhenius function:
\begin{equation}
k^{v'_\text{ref}}(T_\text{c})=A\,{T_\text{c}}^{n} e^{-\frac{E_a}{k_BT_\text{c}}}\text{ .}
\label{eq:k_VRP_ref}
\end{equation}
By assuming that the vibronic-specific rate coefficients $k^{v'}(T_\text{c})$ follow the law \eqref{eq:k_VRP_ep_vp_model}, with $n^{v'}=n$, $E_a^{v'}$ given by \eqref{eq:E_a_vp_VRP}, and $A^{v'}(T_\text{c})$ by \eqref{eq:A_vp_VRP} one can easily show from \eqref{eq:k_VRP_ref} that
\begin{equation}
B(T_\text{c})=\frac{A\,e^{-\frac{E_a}{k_B T_\text{c}}}}{\frac{|\Delta E^{v'_\text{ref}}|}{|\Delta E^{-}|_\text{min}}\,e^{\frac{\Delta E^{v'_\text{ref}}}{k_B T_\text{c}}}\operatorname{H}(-\Delta E^{v'_\text{ref}})+\frac{|\Delta E^{+}|_\text{min}}{|\Delta E^{v'_\text{ref}}|}\operatorname{H}(\Delta E^{v'_\text{ref}})}\text{.}
\label{eq:B_VRP_single}
\end{equation}

\subsubsection{VRP on the initial vibrational quantum number, from a single vibronic-specific rate coefficient}
\label{subsection:VRP_v_single}

If one instead deals with a vibronic-specific process of the form
\begin{equation}
\ch{X}(e,v)\ch{ + ...}\ch{ -> ...}\text{ ,}
\label{eq:VRP_e_v}
\end{equation}
and knows the rate coefficient for a particular vibrational quantum number $v=v_\text{ref}$, it is possible the obtain the rate coefficients for all the other if the model \eqref{eq:k_VRP_ep_vp_model} (constrained by \eqref{eq:n_vp_VRP} and \eqref{eq:A_vp_VRP}) with $v'$ simply substituted by $v$ is regarded. The parameters are then the ones obtained in the previous section, also with $v'$ substituted by $v$.

\section{Results}
\label{sec:results}

In this section, rate coefficients were computed for the addressed kinetic processes. Some comparisons with other models mentioned in the literature are also carried out.

\subsection{Electronic excitation and ionisation of \ch{N}}
\label{subsection:Drawin}

It is important to know how the mole fractions of the nitrogen atoms in their electronic states $x_{\ch{N},e}/g_{\ch{N},\text{el},e}$ change with time as well as how much they deviate from equilibrium conditions when under recombinative or ionisative conditions.  Therefore, it was decided to perform a zero-dimensional simulation of an ionisation of a system with the initial equilibrium mole fractions $x_{\ch{N}}=0.98$, $x_{\ch{N+}}=0.01$ and $x_{\ch{e-}}=0.01$, initial internal temperature $T_{\text{int}}=6,000\,\text{K}$, initial static pressure $p=100\,\text{Pa}$ and high constant values for the translational temperatures $T_{\text{tr}_\text{h}}=T_{\text{tr}_\text{e}}=T_{\text{tr}}=30,000\,\text{K}$. It was also decided to do a zero-dimensional simulation for the recombination of a system with the initial equilibrium mole fractions $x_{\ch{N}}=0.384$, $x_{\ch{N+}}=0.308$ and $x_{\ch{e-}}=0.308$, initial internal temperature $T_{\text{int}}=10,000\,\text{K}$, initial static pressure $p=235\,\text{Pa}$ and low constant values for the translational temperatures $T_{\text{tr}_\text{h}}=T_{\text{tr}_\text{e}}=T_{\text{tr}}=5,000\,\text{K}$. Identical simulations were performed by Lopez \textit{et al.}\cite{lopez2016b} in their work. And in similarity to it, solely excitation and ionisation of \ch{N} by electron impact was taken into account and \ch{N+} was considered to be in its ground electronic level.

\begin{figure}[H]
\centering
\centerline{\includegraphics[scale=1]{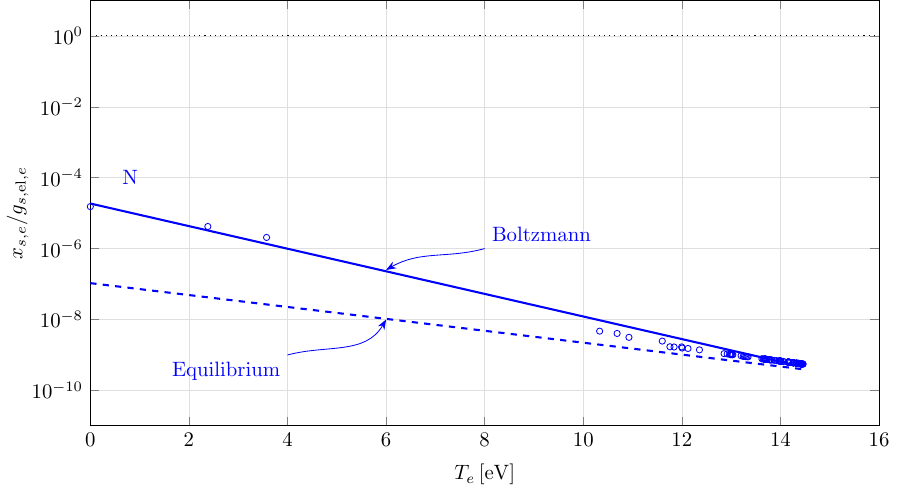}}
\caption{Mole fractions of the nitrogen atoms in their electronic states $x_{\ch{N},e}/g_{\ch{N},\text{el},e}$ (markers), the respective Boltzmann representatives $x_{\ch{N},e}^\text{B}/g_{\ch{N},\text{el},e}$ (solid line) and respective equilibrium ones $x_{\ch{N},e}^*/g_{\ch{N},\text{el},e}$ at $t=2.1\times10^{-5}\,\text{s}$, for the case of the ionisation simulation.}
\label{fig:Saha_io_st}
\end{figure}

\begin{figure}[H]
\centering
\centerline{\includegraphics[scale=1]{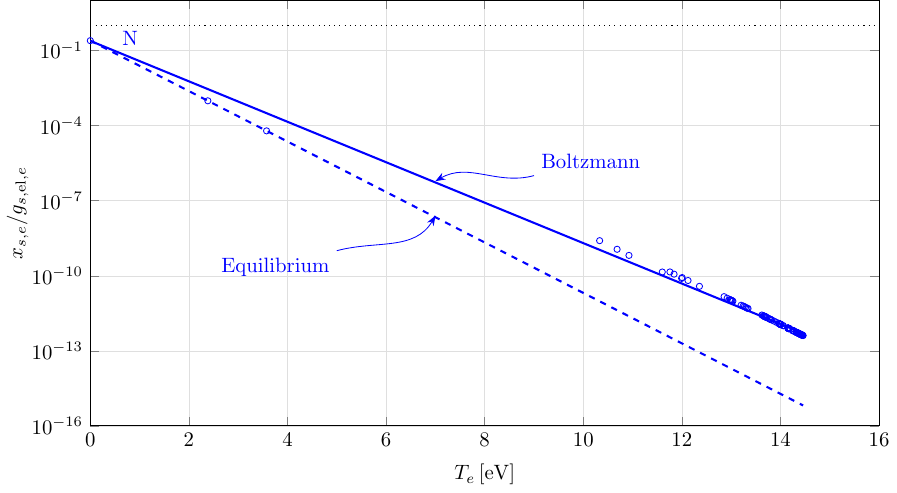}}
\caption{Mole fractions of the nitrogen atoms in their electronic states $x_{\ch{N},e}/g_{\ch{N},\text{el},e}$ (markers), the respective Boltzmann representatives $x_{\ch{N},e}^\text{B}/g_{\ch{N},\text{el},e}$ (solid line) and respective equilibrium ones $x_{\ch{N},e}^*/g_{\ch{N},\text{el},e}$ at $t=0.10\,\text{s}$, for the case of the recombination simulation.}
\label{fig:Saha_re_st}
\end{figure}

The degree of ionisation $\phi=n_{\ch{N+}}/(n_{\ch{N+}}+n_{\ch{N}})$ in equilibrium conditions at a temperature $T_\text{tr}$, may be determined through the so-called Saha ionisation equation \cite{vincenti1965introduction}, 
\begin{equation}
\frac{(\phi^*)^2}{1-(\phi^*)^2}=\frac{2}{n}\left[\frac{2\pi\left(1-\frac{m_\text{e}}{m_{\ch{N}}}\right)m_\text{e}k_BT_\text{tr}}{h^2}\right]^{\frac{3}{2}}\frac{Q_{\ch{N+},\text{el}}(T_\text{tr})}{Q_{\ch{N},\text{el}}(T_\text{tr})}e^{-\frac{\epsilon_{\ch{N}}^+}{k_BT_\text{tr}}}\text{ .}
\label{eq:Saha}
\end{equation}
Note that in \eqref{eq:Saha} $n$ is the global number density of particles, and $[\,]^*$  denotes equilibrium conditions (where $[\,]$ is the operand). The mole fractions of nitrogen atoms in such conditions would then correspond to $x_{\ch{N}}^*=(1-\phi^*)/(1+\phi^*)$. Since conditions are given with respect to an equilibrium temperature $T_{\text{tr}}$, the mole fractions of the nitrogen atoms in their electronic states would be given by a Boltzmann distribution \cite{vincenti1965introduction} with $T_{\text{tr}}$ as parameter,
\begin{equation}
\frac{x_{\ch{N},e}^*}{g_{\ch{N},\text{el},e}}=x_{\ch{N}}^*\frac{e^{-\frac{\epsilon_{\ch{N},\text{el},e}}{k_BT_\text{tr}}}}{Q_{\ch{N},\text{el}}(T_\text{tr})}\text{ .}
\label{eq:Eq_dist}
\end{equation}
It may be appropriate to compute the mole fractions of the nitrogen atoms in their electronic states as if they were at a best fitting electronic temperature $T_\text{el}$, such that deviations from self-equilibrium of the electronic energy mode may be quantified. Let these mole fractions be denoted by $x_{\ch{N},e}^{\text{B}}/g_{\ch{N},\text{el},e}$, and labelled as ``Boltzmann representatives''. Such quantities may be computed by fitting the curve
\begin{equation}
\ln\left(\frac{x_{\ch{N},e}}{g_{\ch{N},\text{el},e}}\right)=-\frac{1}{k_BT_\text{el}}\cdot \epsilon_{\ch{N},\text{el},e}+\ln\left[\frac{x_{\ch{N}}}{Q_{\ch{N},\text{el}}(T_\text{el})}\right]
\label{eq:Eq_dist}
\end{equation}
to the set of points $(\epsilon_{\ch{N},\text{el},e},\,\ln(x_{\ch{N},e}/g_{\ch{N},\text{el},e}))$ obtained in the simulations, with $T_\text{el}$ regarded as the adjustable parameter. The mole fractions of the nitrogen atoms in their electronic states $x_{\ch{N},e}/g_{\ch{N},\text{el},e}$, the respective Boltzmann representatives $x_{\ch{N},e}^\text{B}/g_{\ch{N},\text{el},e}$ and respective equilibrium ones $x_{\ch{N},e}^*/g_{\ch{N},\text{el},e}$ at $t=2.1\times10^{-5}\,\text{s}$ for the case of the ionisation simulation are presented by \Cref{fig:Saha_io_st}. The quantities at $t=0.10\,\text{s}$ for the case of the recombination simulation are presented by by \Cref{fig:Saha_re_st}.
One may conclude from \Cref{fig:Saha_io_st} that for ionisation the populations of the higher energy states tend to be smaller than their Boltzmann representatives, getting closer to the equilibrium values than the populations of the lower energy states. This is physically coherent, since the atoms \ch{N} in the higher energy states require less energy to ionise, and, therefore, their populations tend to decrease more significantly than the populations of the lower energy states, the higher ones being the main contributors to the production of the ions \ch{N+}. For the case of the recombination, the populations of the higher energy states tend to be greater than their Boltzmann representatives and farther from the equilibrium values, since recombination of the ions \ch{N+} to the highest energy states of \ch{N} involves a smaller change of internal energy, being preferred over recombination to the lower energy states.

The same simulations were tried considering collisions with heavy particles instead of free electrons and the evolution of the populations in the energy levels of \ch{N} followed the same pattern as the ones obtained as in \Cref{fig:Saha_io_st,fig:Saha_re_st}.

\subsection{Vibrational transition and dissociation of \ch{N2} and \ch{N2+} by heavy particle impact}
\label{subsection:implementation_FHO}

By recalling the previously introduced theory of the Forced Harmonic Oscillator model (in the \ref{sec:theory} section), one may compute rates for the vibrational transition and dissociation of molecular particles by collision with heavy species. Since in this work the considered heavy species were \ch{N}, \ch{N+}, \ch{N2} and \ch{N2+}, it is necessary to compute rates for seven different interactions: \ch{N2} - \ch{N}, \ch{N2} - \ch{N+}, \ch{N2} - \ch{N2}, \ch{N2} - \ch{N2+}, \ch{N2+} - \ch{N}, \ch{N2+} - \ch{N+} and \ch{N2+} - \ch{N2+}. Note that one needs to consider the electronic and vibrational levels of the collision partners, and therefore, from each one of these species-specific interactions there is a full set of vibronic-specific interactions to be accounted for. To compute rates of vibrational transition and dissociation, the knowledge of some parameters that describe the interactions is necessary. These parameters are the collisional cross section $\sigma$, and the reciprocal characteristic length $\alpha$ and potential well depth $E_\text{M}$ of the adapted Morse interaction potential given by \eqref{eq:Morse_adapted_VT}. Regarding the Morse parameters $\alpha$ and $E_\text{M}$, only values for the interaction \ch{N2}$(\text{X}{}^1\Sigma_\text{g}^+)$ - \ch{N2}$(\text{X}{}^1\Sigma_\text{g}^+)$ were found in the literature. These were defined by Adamovich \textit{et al.} \cite{adamovich1998vibrational}, who also employed the same values for two other cases: \ch{N2} - \ch{O2} and \ch{O2} - \ch{O2}. It was then decided to extrapolate this assumption for all the Morse parameters of the seven interactions, independently of the electronic levels of the collision partners. Regarding the collisional cross section $\sigma$, only a value for the case \ch{N2}$(\text{X}{}^1\Sigma_\text{g}^+)$ - \ch{N2}$(\text{X}{}^1\Sigma_\text{g}^+)$ was obtained, which was estimated from another one found in the literature (Svehla's technical report \cite{svehla1962estimated}). The raw value from which the  collisional cross section was obtained from corresponds to the parameter $d$ of the Lennard-Jones (12-6) potential, called collision diameter, for the interaction \ch{N2}$(\text{X}{}^1\Sigma_\text{g}^+)$ - \ch{N2}$(\text{X}{}^1\Sigma_\text{g}^+)$. The collision diameter parameter of the Lennard-Jones (12-6) potential corresponds to the intermolecular separation value for which the potential is null. For intermolecular distances lower than this value, the potential curve becomes very steep (almost vertical), implying repulsive force. If one neglects the long-range part of the Lennard-Jones (12-6) potential (which is justifiable for high relative speeds of the particles) the collision diameter designation makes sense, corresponding to the distance between the centres of the two hard spheres that represent the collision partners A and B, i.e. $d=d_\text{AB}$. According to the definition of the collisional cross section for hard spheres, one has $\sigma=\pi d_\text{AB}^2=\pi d^2$. And using particular notation for the \ch{N2}$(\text{X}{}^1\Sigma_\text{g}^+)$ - \ch{N2}$(\text{X}{}^1\Sigma_\text{g}^+)$ case, the formula translates itself to $\sigma_{\ch{N2}\text{ - }\ch{N2}}=\pi d_{\ch{N2}\text{ - }\ch{N2}}^2$, with $d_{\ch{N2}\text{ - }\ch{N2}}$ being the respective collision diameter. Svehla's technical report also provides a collision parameter value for the \ch{N}$({}^4\text{S}_\text{u})$ - \ch{N}$({}^4\text{S}_\text{u})$ case. This value in combination with the one associated with the \ch{N2}$(\text{X}{}^1\Sigma_\text{g}^+)$ - \ch{N2}$(\text{X}{}^1\Sigma_\text{g}^+)$ case can be used to estimate the collisional cross section for the \ch{N2}$(\text{X}{}^1\Sigma_\text{g}^+)$ - \ch{N}$({}^4\text{S}_\text{u})$ interaction. By recalling the hard spheres model again and invoking the approximation of the Lennard-Jones potential by a hard-spheres potential, one has
\begin{equation}
\sigma_{\ch{N2}\text{ - }\ch{N}}=\pi d_{\ch{N2}\text{ - }\ch{N}}^2=\pi\left(r_{\ch{N2}}+r_{\ch{N}}\right)^2=\pi\left(\frac{d_{\ch{N2}\text{ - }\ch{N2}}}{2}+\frac{d_{\ch{N}\text{ - }\ch{N}}}{2}\right)^2=\frac{\pi}{4}\left(d_{\ch{N2}\text{ - }\ch{N2}}+d_{\ch{N}\text{ - }\ch{N}}\right)^2\text{ ,}
\label{eq:ss_N2_N}
\end{equation}
with $r_{\ch{N2}}$ and $r_{\ch{N}}$ being the radii of the hard-spheres representing the \ch{N2} and \ch{N} species, respectively. Since no data were found in the literature for the interactions that involve the excited electronic levels of \ch{N2} and \ch{N}, or the ions \ch{N2+} and \ch{N+}, it was assumed that the respective collisional cross sections values were the same as the ones that involve the ground electronic levels of the counterpart neutral species. Note that the approximation regarding the ions is quite rough since a positive ion has one less bounded electron than the respective counterpart neutral species, which influences the electromagnetic force acting in the collision partner, and in turn influences the collisional cross section value. Table \ref{tab:interaction_data} makes a synopsis of collisional cross sections $\sigma$ and Morse parameters $\alpha$ and $E_\text{M}$ considered for the seven different interactions.
\begin{table}[H]
\centering
\caption{Collisional cross sections $\sigma$, and Morse parameters $\alpha$ and $E_\text{M}$ for the collisions m - a and m - m, in which $\text{a}\in\left\{\ch{N},\ch{N+}\right\}$ represents an atomic particle, and $\text{m}\in\left\{\ch{N2},\ch{N2+}\right\}$ represents a molecular particle.}
\begin{scriptsize}
\centerline{\begin{tabular}{cccc}
 \toprule
 Collision & $\sigma[\text{\AA}^2]$ & $\alpha[\text{\AA}^{-1}]$ & $E_\text{M}/k_B[\text{K}]$  \\
 \midrule
 m - a & 39.547 & \multirow{2}{*}{4.0} & \multirow{2}{*}{200.0}   \\
 m - m & 45.317 &  &  \\
 \bottomrule
\end{tabular}}
\end{scriptsize}
\label{tab:interaction_data}
\end{table}

\begin{figure}[H]
\centering
\centerline{\includegraphics[scale=1.25]{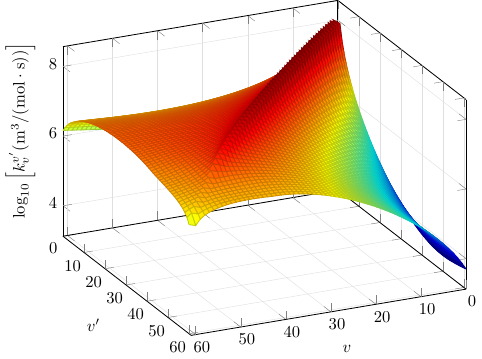}}
\vspace{-15pt}
\caption{Surface plot of rate coefficient values for vibrational transition of \ch{N2}$(\text{X}{}^1\Sigma_\text{g}^+)$ by collision with \ch{N}, at a heavy particle translational temperature $T_{\text{tr}_\text{h}}=20,000$ K.}
\label{fig:k_N2Xv2_N1_VT_T=20000K}
\end{figure}

Rate coefficients for vibrational transition $k_{v}^{v'}(T_{\text{tr}_\text{h}})$ and dissociation $k_{v}^{D}(T_{\text{tr}_\text{h}})$ of \ch{N2} and \ch{N2+} in their electronic levels, due to collisions with \ch{N2}, \ch{N2+}, \ch{N} and \ch{N+}, were computed. To illustrate one of those results, \Cref{fig:k_N2Xv2_N1_VT_T=20000K} depicts the rate coefficients for vibrational transition of \ch{N2}$(\text{X}{}^1\Sigma_\text{g}^+)$ by collision with \ch{N}, at a heavy particle translational temperature $T_{\text{tr}_\text{h}}=20,000$ K. 
\noindent
The vibrational transition process is more effective if the transition is between two adjacent levels, and even more effective if the two adjacent levels are of low energy. For the case of transitions between non-adjacent levels, it is 
observed that rate coefficient values for de-excitation are in general greater than the ones for excitation. Excitation from the lowest energy levels to the highest ones are the less effective. This is in agreement with theoretical trends.

It is of paramount importance to validate the computed rates of vibrational transition $k_{v}^{v'}$ as well as the rates of dissociation $k_{v}^{D}$, for all the considered interactions. Unfortunately, experimental results expressed in the exact same form as these obtained numerically were not found in the literature. What was found instead were experimental results for the dissociation of \ch{N2}$(\text{X}{}^1\Sigma_\text{g}^+)$ by impact with \ch{N} and with \ch{N2}, in thermal equilibrium conditions. Note that the numerical computation of dissociation rates using the FHO model implies the assignment of a value for a free parameter that cannot be \textit{a priori} computed. That parameter is the number of quasi-bound levels for the sum involved in the dissociation rate formula  \eqref{eq:P_diss_VT}, as presented in the \ref{sec:theory} section. Instead of the number of quasi-bound levels, one can define a quantity corresponding to the difference between the upper bound value for the vibrational energy of the quasi-bound levels $G_{v'_\text{u}}$ and the potential well depth $D_e$, i.e. $G_{v'_\text{u}}^{D_e}:=G_{v'_\text{u}}-{D_e}$. Therefore, the considered quasi-bound levels are the ones with a vibrational energy $G_{v'}$ higher than the potential well depth $D_e$ and lower or equal to the upper bound vibrational energy $G_{v'_\text{u}}$, i.e. $D_e< G_{v'} \le G_{v'_\text{u}}$. Such free parameter gives opportunity for calibration of the model by comparison with the aforementioned experimental results\footnote{With careful cross-check that individual probabilities never exceed the unity.}, increasing the reliability on the numerical dissociation rates, and at the same, on the numerical vibrational transition rates, since the former uses the latter for their computation. With this said, vibrational transition and dissociation rates of \ch{N2}$(\text{X}{}^1\Sigma_\text{g}^+)$ due to impact with \ch{N+} and \ch{N2+} cannot be calibrated since there is no experimental data for them. The same happens for the vibrational transition and dissociation rates of \ch{N2+} and the electronically excited \ch{N2}, from impact with \ch{N}, \ch{N2}, \ch{N+} and \ch{N2+}. In the former case, it was decided to set the differences between the upper bound vibrational energy and the potential well for the interactions with \ch{N+} and \ch{N2+} as the ones obtained from calibration of the interactions \ch{N2}$(\text{X}{}^1\Sigma_\text{g}^+)$ - \ch{N} and \ch{N2}$(\text{X}{}^1\Sigma_\text{g}^+)$ - \ch{N2}, respectively. In the latter case, it was decide to solely account for the first quasi-bound level\footnote{Although other numbers of quasi-bound levels were tried, the differences in the dissociation rates were negligible.}.

A numerical rate coefficient for thermal dissociation in respect of the interactions \ch{N2}$(\text{X}{}^1\Sigma_\text{g}^+)$ - \ch{N2} and \ch{N2}$(\text{X}{}^1\Sigma_\text{g}^+)$ - \ch{N} can be computed from the obtained vibronic-specific rate coefficients for the interactions \ch{N2}$(\text{X}{}^1\Sigma_\text{g}^+,v)$ - \ch{N2} and \ch{N2}$(\text{X}{}^1\Sigma_\text{g}^+,v)$ - \ch{N}, respectively. The reasoning presented in the paragraphs below should be followed. 

The rate coefficient for the dissociation of particles of the  X species in the $e_1$-th electronic level and $v_1$-th vibrational level due to collisions with particles of the M species, corresponds to $k^{D}_{\text{X}(e_1,v_1)\text{ - }\text{M}}$, such that the variation of the amount concentration of the first particles due to that process is given by 
\begin{equation}
\left(\frac{d[\text{X}(e_1,v_1)]}{dt}\right)^D_{\text{X}(e_1,v_1)\text{ - }\text{M}}=-k^{D}_{\text{X}(e_1,v_1)\text{ - }\text{M}}[\text{X}(e_1,v_1)][\text{M}]\text{ .}
\label{eq:X_e_v_variation}
\end{equation}
The variation of the amount concentration of the X species in the $e_1$-th electronic level due to the above-mentioned dissociative process corresponds to a sum on all vibrational levels $v_1$ of the vibronic-specific contributions given by \eqref{eq:X_e_v_variation}:
\begin{equation}
\centerline{$\displaystyle \left(\frac{d[\text{X}(e_1)]}{dt}\right)^D_{\text{X}(e_1)\text{ - }\text{M}}=\sum_{v_1}\left(\frac{d[\text{X}(e_1,v_1)]}{dt}\right)^D_{\text{X}(e_1,v_1)\text{ - }\text{M}}=-\left\{\sum_{v_1}k^{D}_{\text{X}(e_1,v_1)\text{ - }\text{M}}[\text{X}(e_1,v_1)]\right\}[\ch{M}]$}
\label{eq:X_e_variation}
\end{equation}
Since we assumed that the rotational constants do not depend on the vibrational level, the rotational and vibrational energy modes are decoupled from each other, and it can be easily proved that the amount concentration of particles of the X species in the $e_1$-th electronic level and $v_1$-th vibrational level at thermal equilibrium corresponds to
\begin{equation}
[\text{X}(e_1,v_1)](T)=[\text{X}(e_1)]\frac{g_{\text{X},\text{vib},e_1,v_1}e^{-\frac{\epsilon_{\text{X},\text{vib},e_1,v_1}}{k_B T}}}{Q_{\text{X},\text{vib},e_1}(T)}\text{ ,}
\label{eq:X_e_v_t}
\end{equation}
where $[\text{X}_{e_1}]$ is the amount concentration of particles of the X species in the $e_1$-th electronic level. The degree of degeneracy of the vibrational level is $g_{\text{X},\text{vib},e_1,v_1}=1$. By substituting \eqref{eq:X_e_v_t} in \eqref{eq:X_e_variation}, one may identify the rate coefficient of thermal dissociation for the interaction $\text{X}(e_1)\text{ - }\text{M}$, i.e. $k^{D}_{\text{X}(e_1)\text{ - }\text{M}}$(T): 
\begin{equation}
\centerline{$\displaystyle \left(\frac{d[\text{X}(e_1)]}{dt}\right)^{D}_{\text{X}(e_1)\text{ - }\text{M}}(T)=-\underbrace{\left\{\sum_{v_1}\frac{e^{-\frac{\epsilon_{\text{X},\text{vib},e_1,v_1}}{k_B T}}}{Q_{\text{X},\text{vib},e_1}}k^{D}_{\text{X}(e_1,v_1)\text{ - }\text{M}}\right\}}_{=:k^{D}_{\text{X}(e_1)\text{ - }\text{M}}(T)}[\text{X}(e_1)][\text{M}]\text{ ,}$}
\label{eq:X_e_variation_t}
\end{equation}
with $k^{D}_{\text{X}(e_1)\text{ - }\text{M}}(T)$ therefore given by
\begin{equation}
k^{D}_{\text{X}(e_1)\text{ - }\text{M}}(T)=\sum_{v_1}\frac{e^{-\frac{\epsilon_{X,\text{vib},e_1,v_1}}{k_B T}}}{Q_{X,\text{vib},e_1}}k^{D}_{\text{X}(e_1,v_1)\text{ - }\text{M}}\text{ .}
\label{eq:k_D_t}
\end{equation}

The experimentally obtained thermal dissociation rates are listed in Table \ref{tab:VDt_experimental}. 

\begingroup
\centerline{\begin{threeparttable}
\centering
\caption{Coefficients $A$, $n$ and $E_a/k_B$ of the modified Arrhenius function \eqref{eq:modifiedArrhenius} for experimentally obtained rates for the thermal dissociation of $\ch{N2}(\text{X}{}^1\Sigma_\text{g}^+)$ by collision with \ch{N} and $\ch{N2}$, as well as the respective interval of temperatures $T\in[T_\text{min},\,T_\text{max}]$ in which they are valid.}
\begin{scriptsize}
\begin{tabular}{cccccc}
\toprule
Experiment & Interaction & $A\,[\text{cm}^3\cdot \text{K}^{-\eta}/(\text{mol}\cdot \text{s})]$ & $n$ & $E_a/k_B\,[\text{K}]$  & $[T_\text{min},\,T_\text{max}]\,[\text{K}]$\\
\midrule
\multirow{2}{*}{Cary (1965) \cite{cary1965}} & \ch{N2}$(\text{X}{}^1\Sigma_\text{g}^+)$ - \ch{N} & $7.1\times10^{19}$ & $-1.0$ & \multirow{2}{*}{$113,310$} & \multirow{2}{*}{$[6,000;\,10,000]$}   \\
& \ch{N2}$(\text{X}{}^1\Sigma_\text{g}^+)$ - \ch{N2} & $5.6\times10^{22}$ & $-1.7$ &  &  \\
\hline
\multirow{2}{*}{Byron (1966) \cite{byron1966}} & \ch{N2}$(\text{X}{}^1\Sigma_\text{g}^+)$ - \ch{N} & $4.3\times10^{22}$ & $-1.5$ & \multirow{2}{*}{$113,200$} & \multirow{2}{*}{$[6,000;\,9,000]$}   \\
& \ch{N2}$(\text{X}{}^1\Sigma_\text{g}^+)$ - \ch{N2} & $4.8\times10^{17}$ & $-0.5$ &  &  \\
\hline
\multirow{2}{*}{Appleton \textit{et al.} (1968) \cite{appleton1968}} & \ch{N2}$(\text{X}{}^1\Sigma_\text{g}^+)$ - \ch{N} & $1.6\times10^{22}$ & $-1.6$ & \multirow{2}{*}{$113,200$} & \multirow{2}{*}{$[8,000;\,15,000]$}   \\
& \ch{N2}$(\text{X}{}^1\Sigma_\text{g}^+)$ - \ch{N2} & $3.7\times10^{21}$ & $-1.6$ &  &  \\
\hline
\multirow{2}{*}{Hanson and Baganoff (1972) \cite{hanson1972}} & \ch{N2}$(\text{X}{}^1\Sigma_\text{g}^+)$ - \ch{N} & $2.2\times10^{26}$ & $-2.5$ & \multirow{2}{*}{$113,000$} & \multirow{2}{*}{$[5,700;\,12,000]$}   \\
& \ch{N2}$(\text{X}{}^1\Sigma_\text{g}^+)$ - \ch{N2} & $3.9\times10^{33}$ & $-4.5$ &  &  \\
\hline
\multirow{2}{*}{Kewley and Hornung (1974) \cite{kewley1974}} & \ch{N2}$(\text{X}{}^1\Sigma_\text{g}^+)$ - \ch{N} & $8.5\times10^{25}$ & $-2.5$ & \multirow{2}{*}{$113,200$} & \multirow{2}{*}{$[6,000;\,14,000]$}   \\
& \ch{N2}$(\text{X}{}^1\Sigma_\text{g}^+)$ - \ch{N2} & $2.3\times10^{29}$ & $-3.5$ &  &  \\
\hline
\multirow{2}{*}{Park (1988) \cite{park1988} (\tnote{a} )} & \ch{N2}$(\text{X}{}^1\Sigma_\text{g}^+)$ - \ch{N} & $3\times10^{22}$ & $-1.6$ & \multirow{2}{*}{$113,200$} & \multirow{2}{*}{$[6,000;\,13,000]$}   \\
& \ch{N2}$(\text{X}{}^1\Sigma_\text{g}^+)$ - \ch{N2} & $7\times10^{21}$ & $-1.6$ &  &  \\
\bottomrule
\end{tabular}
\label{tab:VDt_experimental}
\begin{tablenotes}
\item[a]{Although the work done by Park is labelled here as an experiment, it is actually a theoretical study involving the other experimental works.}\\
\end{tablenotes}
\end{scriptsize}
\end{threeparttable}}
\endgroup

All the experiments previous to the ones done by Hanson and Baganoff \cite{hanson1972}, and by Kewley and Hornung \cite{kewley1974} considered a test gas composed by molecular nitrogen and a diluted inert gas. Such mixture was chosen instead of a pure molecular nitrogen gas, so that dissociation of nitrogen could occur at lower shock speeds. As referred by Park \cite{park1988}: ``this was done because \ch{N2} has a large dissociation energy and hence its dissociation requires a large shock speed, attainable only with a sophisticated facility, unless a large concentration of argon is included''. Therefore, at first glance, one may consider the experiments of Hanson and Baganoff \cite{hanson1972} and of  
Kewley and Hornung \cite{kewley1974} to be more reliable than the ones of Cary \cite{cary1965}, Byron \cite{byron1966} and 
Appleton \textit{et al.} \cite{appleton1968}. The study performed by Park \cite{park1988} was not an experiment but a reinterpretation of the previously obtained experimental results. Such reinterpretation accounted for the possibility of non-thermal equilibrium between the translational and vibrational modes of the particles, i.e. $T_{\text{tr}_\text{h}}\neq T_\text{vib}$, during the dissociation of \ch{N2}. The results presented by the other authors assumed thermal equilibrium, which may not be reasonable. It is important to mention that the temperature in the modified Arrhenius function \eqref{eq:modifiedArrhenius} for the case of thermal dissociation rates of Park is actually a geometrically averaged temperature $T_\text{a}=\sqrt{T_{\text{tr}_\text{h}}T_\text{vib}}$. Anyway, since in thermal equilibrium one has $T=T_{\text{tr}_\text{h}}=T_\text{vib}$, and the averaged temperature matches the single-temperature, i.e. $T_\text{a}=T$. Due to the thoroughness of Park, his results are currently considered to be the ``state-of-the-art'' ones. Jaffe \textit{et al.} \cite{jaffe2018} wrote `` [...] dissociation rate coefficients from the Park [$T_{\text{tr}_\text{h}}$ - $T_\text{vib}$] hypersonic nonequilibrium chemistry model, which is currently the \textit{de facto} standard for aerothermodynamic modeling''. Candler and Olejniczak \cite{candler1997b} said ``[t]he results of Park's two-temperature interpretation of the experimental data are now accepted as the most widely accurate expressions for the equilibrium dissociation rates''. These statements convinced the authors of the present work to use the Park's results in the calibration of the ones which were herein obtained. 

\begin{figure}[H]
\centering
\centerline{\includegraphics[scale=1]{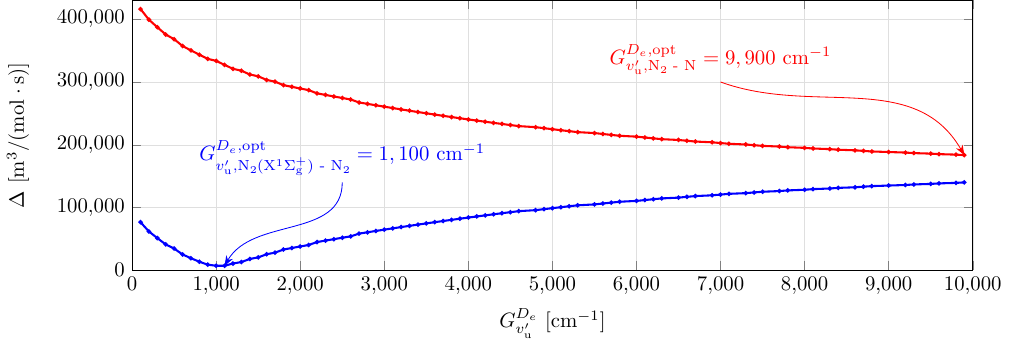}}
\vspace{-15pt}
\caption{Curves of root mean squared deviation $\Delta_{\ch{N2}(\text{X}{}^1\Sigma_\text{g}^+)\text{ - }\ch{N2}}(G^{D_e}_{v_\text{u}'})$ (blue) and $\Delta_{\ch{N2}(\text{X}{}^1\Sigma_\text{g}^+)\text{ - }\ch{N}}(G^{D_e}_{v_\text{u}'})$ (red), obtained by a sweeping procedure with $G^{D_e}_{v_\text{u}'}\in[0;\,10,000]\,\text{cm}^{-1}$, with an increment step of $100\,\text{cm}^{-1}$.}
\label{fig:DD_min}
\end{figure}

The calibration of the numerical rate coefficients for thermal dissociation in respect of the interactions $\ch{N2}(\text{X}{}^1\Sigma_\text{g}^+)$ - $\ch{N2}$ and $\ch{N2}(\text{X}{}^1\Sigma_\text{g}^+)$ - \ch{N}, i.e. $k^D_{\ch{N2}(\text{X}{}^1\Sigma_\text{g}^+)\text{ - }\ch{N2}}(T)$ and $k^D_{\ch{N2}(\text{X}{}^1\Sigma_\text{g}^+)\text{ - }\ch{N}} (T)$, consisted in the minimisation of the root mean square deviation (a cost function) between these and Park's rate coefficients  by sweeping the difference between the upper bound value for the vibrational energy of the quasi-bound levels and the potential well $G^{D_e}_{v_\text{u}'}$. Let $G^{D_e,\text{opt}}_{v_\text{u}',\ch{N2}(\text{X}{}^1\Sigma_\text{g}^+)\text{ - }\ch{N2}}$ and $G^{D_e,\text{opt}}_{v_\text{u}',\ch{N2}(\text{X}{}^1\Sigma_\text{g}^+)\text{ - }\ch{N}}$ be the respective optimum values. The root mean square deviations of $k^D_{\ch{N2}(\text{X}{}^1\Sigma_\text{g}^+)\text{ - }\ch{N2}}(T)$ and $k^D_{\ch{N2}(\text{X}{}^1\Sigma_\text{g}^+)\text{ - }\ch{N}} (T)$ relatively to the Park's rate coefficients are defined as
\begin{equation}
\Delta_{\ch{N2}(\text{X}{}^1\Sigma_\text{g}^+)\text{ - }\ch{M}}(G^{D_e}_{v_\text{u}'})=\sqrt{\frac{\sum_{n}\left[k^D_{\ch{N2}(\text{X}{}^1\Sigma_\text{g}^+)\text{ - }\ch{M}}(T_n)-k^{D,\text{(Park)}}_{\ch{N2}(\text{X}{}^1\Sigma_\text{g}^+)\text{ - }\ch{M}}(T_n)\right]^2}{N}}\text{ ,}
\label{eq:RMSD_N2_M}
\end{equation}
with $\ch{M}\in\{\ch{N},\ch{N2}\}$ being the collision partner, $k^{D,\text{(Park)}}_{\ch{N2}(\text{X}{}^1\Sigma_\text{g}^+)\text{ - }\ch{M}}(T_n)$ the Park's rate coefficient, $T_n$ the $n$-th temperature value at which the rate coefficients are evaluated, and $N$ the number of temperature values for evaluation. Only the experimentally valid range of temperatures was considered for the optimisation process.
The curves $\Delta_{\ch{N2}(\text{X}{}^1\Sigma_\text{g}^+)\text{ - }\ch{N2}}(G^{D_e}_{v_\text{u}'})$ and $\Delta_{\ch{N2}(\text{X}{}^1\Sigma_\text{g}^+)\text{ - }\ch{N}}(G^{D_e}_{v_\text{u}'})$ obtained by the sweeping procedure are represented in Figure \ref{fig:DD_min}.

For the case of the interaction $\ch{N2}(\text{X}{}^1\Sigma_\text{g}^+)\text{ - }\ch{N2}$, a minimum for the root mean squared deviation was found, with $G^{D_e,\text{opt}}_{v_\text{u}',\ch{N2}(\text{X}{}^1\Sigma_\text{g}^+)\text{ - }\ch{N2}}=1,100$ cm$^{-1}$ being the minimiser. For the case of the interaction $\ch{N2}(\text{X}{}^1\Sigma_\text{g}^+)\text{ - }\ch{N}$, no minimum was found in the range $G^{D_e}_{v_\text{u}'}\in[0;\,10,000]\,\text{cm}^{-1}$. Only some of the decreasing part of the full curve is supported by that domain. A sweeping procedure for the higher $(G^{D_e}_{v_\text{u}'})$ values would be required to obtain the minimum point. However, it was found that an increase of the number of considered quasi-bound vibrational levels would not make a meaningful difference in the root mean squared deviation value, due to the even lower probabilities of transition to the higher quasi-bound vibrational levels. On the other hand, the higher the number of the considered quasi-bound vibrational levels, the higher the amount of computations for the evaluation of the rate coefficients, making a new sweeping not worth it. It was then decided to choose the optimum difference between the upper bound value for the vibrational energy of the quasi-bound levels and the potential well for the $\ch{N2}(\text{X}{}^1\Sigma_\text{g}^+)\text{ - }\ch{N}$ case as the one that minimised the root mean squared deviation in the range $G^{D_e}_{v_\text{u}'}\in[0;\,10,000]\,\text{cm}^{-1}$, i.e. $G^{D_e,\text{opt}}_{v_\text{u}',\ch{N2}(\text{X}{}^1\Sigma_\text{g}^+)\text{ - }\ch{N}}=9,900\,\text{cm}^{-1}$. 

\begin{figure}[H]
\centering
\centerline{\includegraphics[scale=1]{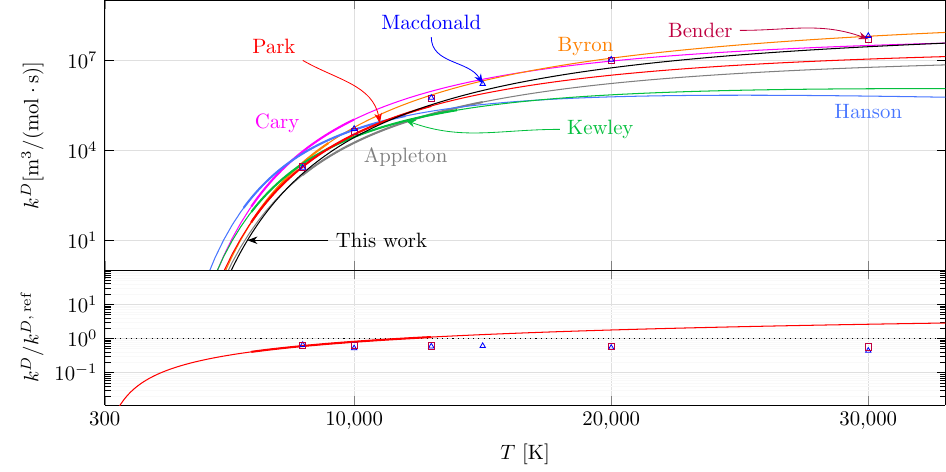}}
\vspace{-15pt}
\caption{Rate coefficient values for thermal dissociation of $\ch{N2}(\text{X}{}^1\Sigma_\text{g}^+)$ due to the interaction $\ch{N2}(\text{X}{}^1\Sigma_\text{g}^+)\text{ - }\ch{N2}$, and ratio between the FHO result and the one obtained by Park \cite{park1988}, as well as the ones obtained by the QCT model (Bender \textit{et al.} \cite{bender2015}, and Macdonald \textit{et al.} \cite{macdonald2018}). The thick part of the lines for each of the experiments listed in Table \ref{tab:VDt_experimental}, is associated with the respective experimentally valid domains.}
\label{fig:k_D_N2_N2_comparison}
\end{figure}

\begin{figure}[H]
\centering
\centerline{\includegraphics[scale=1]{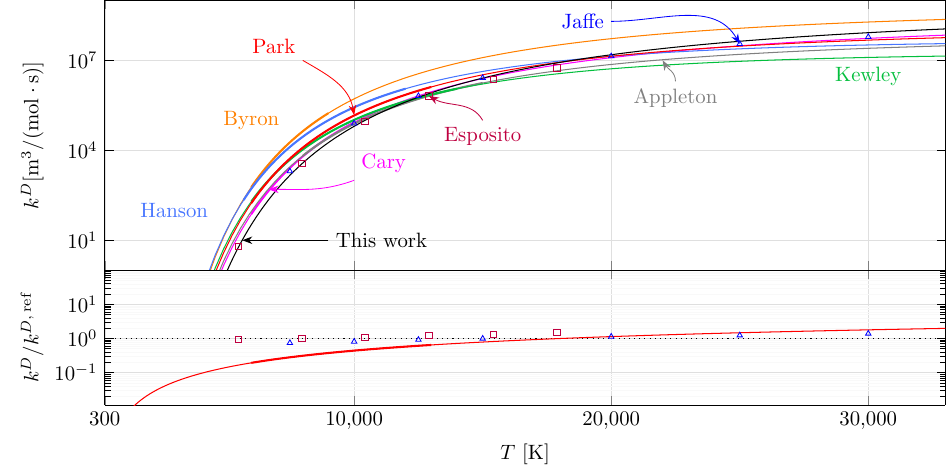}}
\vspace{-15pt}
\caption{Rate coefficient values for thermal dissociation of $\ch{N2}(\text{X}{}^1\Sigma_\text{g}^+)$ due to the interaction $\ch{N2}(\text{X}{}^1\Sigma_\text{g}^+)\text{ - }\ch{N}$, and ratio between the FHO result and the one obtained by Park \cite{park1988}, as well as the ones obtained by the QCT model (Esposito and Capitelli \cite{esposito1999}, and Jaffe \textit{et al.} \cite{jaffe2010}). The thick part of the lines for each of the experiments listed in Table \ref{tab:VDt_experimental}, are associated with the respective experimentally valid domains.}
\label{fig:k_D_N2_N_comparison}
\end{figure}

Figures \ref{fig:k_D_N2_N2_comparison} and \ref{fig:k_D_N2_N_comparison} show the calibrated rate coefficient curves and the experimental ones listed in Table \ref{tab:VDt_experimental}, for the cases $\ch{N2}(\text{X}{}^1\Sigma_\text{g}^+)\text{ - }\ch{N2}$ and $\ch{N2}(\text{X}{}^1\Sigma_\text{g}^+)\text{ - }\ch{N}$, respectively. These figures also show recent numerical rate coefficient values individually obtained by Bender \textit{et al.} \cite{bender2015}, Macdonald \textit{et al.} \cite{macdonald2018}, Esposito and Capitelli \cite{esposito1999}, and Jaffe \textit{et al.} \cite{jaffe2010}, using the Quasi-Classical Trajectory model (QCT), which is considered to be more sophisticated than the FHO model. In the second subfigure of each figure, the ratio between the FHO rate coefficient values and the ones obtained by Park \cite{park1988}, as well as the ones obtained through the QCT model is depicted.

From these two figures, it is possible to observe some overall discrepancies between the experimental results, 
either in offset and in trend. For both interactions $\ch{N2}(\text{X}{}^1\Sigma_\text{g}^+)\text{ - }\ch{N2}$ and $\ch{N2}(\text{X}{}^1\Sigma_\text{g}^+)\text{ - }\ch{N}$, an underestimation of the Park \cite{park1988} results by the FHO model in the low to medium temperatures region, and an overestimation in the high temperatures region occur. There is a better agreement between the two results in the experimentally valid region for the former case (a maximum underestimation of $59.9\%$ and overestimation of $8.9\%$ in contrast with a maximum underestimation of $80.9\%$ and minimum underestimation of $36.1\%$ for the latter case). This was expected since it was only in the former case that the optimisation procedure was successful in the minimisation of the cost function. The FHO results underestimate all the QCT results for the case of the interaction $\ch{N2}(\text{X}{}^1\Sigma_\text{g}^+)\text{ - }\ch{N2}$, with maximum absolute relative deviations of $44.3\%$ and $56.5\%$ from the values of Bender \textit{et al.} \cite{bender2015} and Macdonald \textit{et al.} \cite{macdonald2018}, respectively. For the case of the interaction $\ch{N2}(\text{X}{}^1\Sigma_\text{g}^+)\text{ - }\ch{N}$, one finds that the FHO results underestimate the QCT results at the lower temperatures and overestimate them at the higher ones, with maximum absolute relative deviations of $46.2\%$ and $37.9\%$ from the values of Esposito and Capitelli \cite{esposito1999} and Jaffe \textit{et al.} \cite{jaffe2010}, respectively.

\subsection{Vibronic transition of \ch{N2}$(\text{A}{}^3\Sigma_\text{u}^+,v_1)$ to \ch{N2}$(\text{B}{}^3\Pi_\text{g},v'_1)$ by collision with \ch{N2}$(\text{X}{}^1\Sigma_\text{g}^+,v_2)$ and \ch{N}$({}^4\text{S}_\text{u})$}
\label{subsection:VE_h_implementation_bachmann1993}

Herein the proposed V-E model \eqref{eq:kf_elio} is implemented for the cases of vibronic transitions of \ch{N2}$(\text{A}{}^3\Sigma_\text{u}^+,v_1)$ to \ch{N2}$(\text{B}{}^3\Pi_\text{g},v'_1)$ by collisions of \ch{N2}$(\text{A}{}^3\Sigma_\text{u}^+,v_1)$ with \ch{N2}$(\text{X}{}^1\Sigma_\text{g}^+,v_2)$ and \ch{N}$({}^4\text{S}_\text{u})$. For that purpose, the effective process cross sections experimentally obtained by Bachmann \textit{et al.} \cite{bachmann1993} were used. Bachmann \textit{et al.} studied intramolecular V-E processes of the form:
\begin{equation}
\ch{N2}(\text{A}{}^3\Sigma_\text{u}^+,v_1)\ch{ + M}\ch{ -> N2}(\text{B}{}^3\Pi_\text{g},v'_1)\ch{ + M}\text{ ,}
\label{eq:VEh_intra_bachmann1993}
\end{equation}
\sloppy with \ch{M} being the collision partner which may correspond to an atomic particle, $\ch{M}\in\{\ch{He}, \ch{Ne}, \ch{Ar}, \ch{Kr}, \ch{Xe}\}=:\{\ch{M}_\text{a}\}$, or to a molecular particle, $\sloppy\ch{M}\in\{\ch{H2}, \ch{N2}, \ch{NO}\}=:\{\ch{M}_\text{m}\}$ in its ground energy level. Bachmann \textit{et al.}, however, strongly believed that two intermolecular V-E processes for the case $\ch{M}=\ch{N2}$ were found due to their quasi-resonance \cite{bachmann1993}:
\begin{equation}
\ch{N2}(\text{A}{}^3\Sigma_\text{u}^+,15)\ch{ + N2}(\text{X}{}^1\Sigma_\text{g}^+,0)\ch{ -> N2}(\text{B}{}^3\Pi_\text{g},4)\ch{ + N2}(\text{X}{}^1\Sigma_\text{g}^+,1)\text{ ,}
\label{eq:VEh_N2_inter_1_bachmann1993}
\end{equation}
\begin{equation}
\ch{N2}(\text{A}{}^3\Sigma_\text{u}^+,17)\ch{ + N2}(\text{X}{}^1\Sigma_\text{g}^+,0)\ch{ -> N2}(\text{B}{}^3\Pi_\text{g},5)\ch{ + N2}(\text{X}{}^1\Sigma_\text{g}^+,1)\text{ .}
\label{eq:VEh_N2_inter_2_bachmann1993}
\end{equation}
Such consideration was also accounted for in this work. It is important to point out that the experiment was carried at room temperature $T_{\text{tr}_\text{h}}=300\,\text{K}=:T_\text{ref}$. Although the work of Bachmann \textit{et al.} does not provide data for the case $\ch{M}=\ch{N}$, a model applied on the available results would allow one to infer values for such a case. The final objective would be to obtain rate coefficients through the law \eqref{eq:kf_elio} for the V-E processes represented by the chemical equations
\begin{equation}
\ch{N2}(\text{A}{}^3\Sigma_\text{u}^+,v_1)\ch{ + N2}(\text{X}{}^1\Sigma_\text{g}^+,v_2)\ch{ -> N2}(\text{B}{}^3\Pi_\text{g},v'_1)\ch{ + N2}(\text{X}{}^1\Sigma_\text{g}^+,v'_2),\,\,\,\,\,\,\,\,\,\forall\,v_1,\,v'_1,\,v_2,\text{ and }v'_2\text{ ,}
\label{eq:VEh_N2_bachmann1993}
\end{equation}
\begin{equation}
\ch{N2}(\text{A}{}^3\Sigma_\text{u}^+,v_1)\ch{ + N}({}^4\text{S}_\text{u})\ch{ -> N2}(\text{B}{}^3\Pi_\text{g},v'_1)\ch{ + N}({}^4\text{S}_\text{u}),\,\,\,\,\,\,\,\,\,\forall\,v_1\text{ and }v'_1\text{ .}
\label{eq:VEh_N_bachmann1993}
\end{equation}

\begin{figure}[H]
\centering
\centerline{\includegraphics[scale=1]{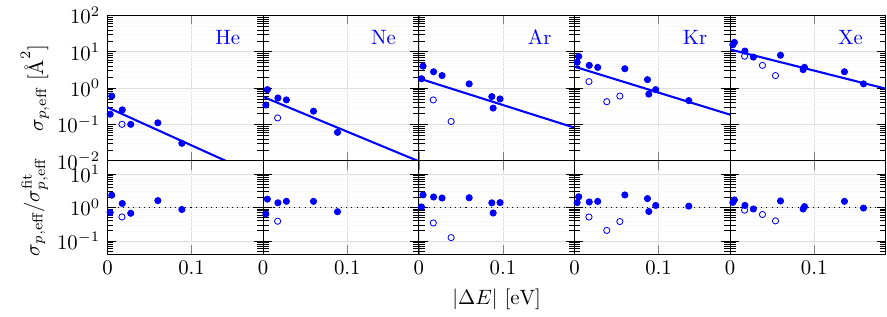}}
\vspace{-15pt}
\caption{Upper plots: data of Bachmann \textit{et al.} \cite{bachmann1993} and fitted curves \eqref{eq:ss_av_elio} for the dependence of the effective process cross sections $\sigma_{p,\text{eff}}$ on the absolute value of the energy defects $|\Delta E|$, regarding the atomic collision partners \ch{He}, \ch{Ne}, \ch{Ar}, \ch{Kr} and \ch{Xe}. Lower plots: values for the ratios between the data and the fitted curves. Intramolecular exothermic processes: \protect\markerfilledbluecircle; intramolecular endothermic processes: \protect\markerbluecircle.}
\label{fig:gap_law_bachmann1993_a}
\end{figure}

\begin{figure}[H]
\centering
\centerline{\includegraphics[scale=1]{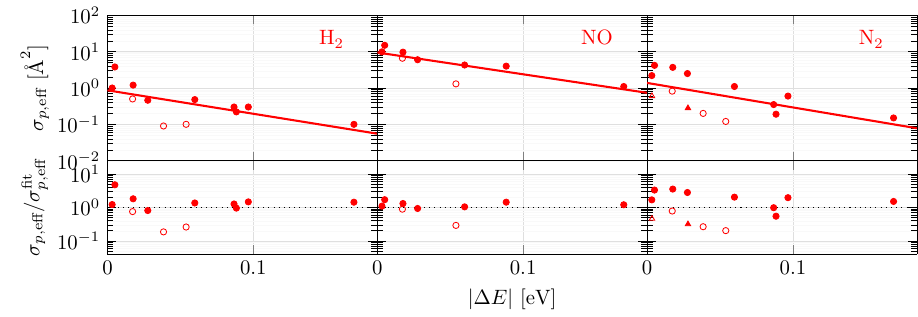}}
\vspace{-15pt}
\caption{Upper plots: data of Bachmann \textit{et al.} \cite{bachmann1993} and fitted curves \eqref{eq:ss_av_elio} for the dependence of the effective process cross sections $\sigma_{p,\text{eff}}$ on the absolute value of the energy defects $|\Delta E|$, regarding the molecular collision partners \ch{H2}, \ch{NO} and \ch{N2}. Lower plots: values for the ratios between the data and the fitted curves. Intramolecular exothermic  processes: \protect\markerfilledredcircle; Intramolecular endothermic processes: \protect\markerredcircle; intermolecular exothermic processes: \protect\markerfilledredtriangle; intermolecular endothermic processes: \protect\markerredtriangle.}
\label{fig:gap_law_bachmann1993_m}
\end{figure}

By fitting the exponential gap law curve \eqref{eq:ss_av_elio} (which at $T_{\text{tr}_\text{h}}=T_\text{ref}$ reduces to the exponential gap law $\sigma=\sigma_0e^{-\frac{|\Delta E|}{E_0}}$) to the set of points constituted by energy defect absolute values $|\Delta E|$ as abscissas, and effective process cross section values $\sigma_{p,\text{eff}}$ as ordinates, it is possible to obtain characteristic cross sections $\sigma_0$ and energies $E_0$ for each collisional partner. The resultant exponential gap law curves are depicted by Figures \ref{fig:gap_law_bachmann1993_a} and \ref{fig:gap_law_bachmann1993_m} for the case of atomic and molecular collision partners, respectively. 
Bachmann \textit{et al.} refer that the data associated with the endothermic processes appear to follow a law which is distinct from the one associated with the exothermic processes. However, the number of points associated with the endothermic processes is not reasonable enough to properly build a different model for them. It was then decided to fit the data altogether, since obtaining rate coefficients for both exothermic and endothermic processes is of capital importance. In general, the fitted curves underestimate the rate coefficients for the exothermic processes and overestimate the rate coefficients for the endothermic ones. The ratio between the data effective process cross sections and the fit ones is as low as $0.1$ and as high as $4.7$. Values of $\sigma_0=1.380$ $\text{\AA}^2$ and $E_0=0.064$ eV were obtained for the case of the collision partner $\text{M}=\ch{N2}$.

\begin{figure}[H]
\centering
\centerline{\includegraphics[scale=1]{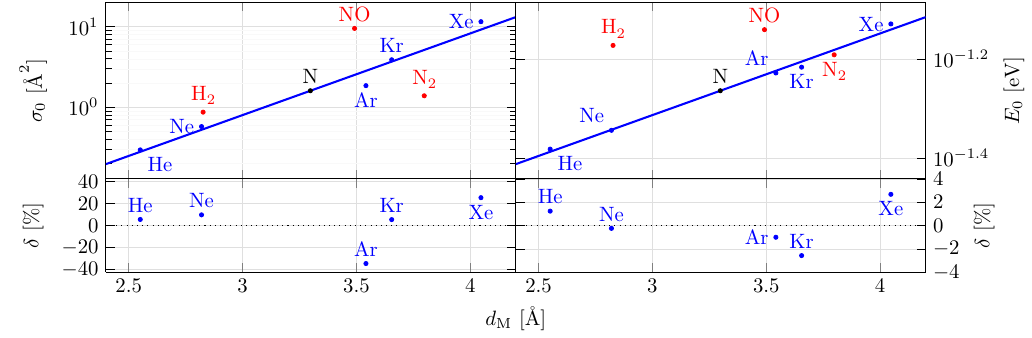}}
\vspace{-15pt}
\caption{Upper plots: exponential curves fitted to the data $(d_\text{M},\sigma_0)$ (at left) and $(d_\text{M},E_0)$ (at right) for the processes involving the atomic collision partners. Lower plots: relative deviations of the data, $\delta=\left(\sigma_0-\sigma_{0,\text{fit}}\right)/\sigma_{0,\text{fit}}$ (at left) and $\delta=\left(E_0-E_{0,\text{fit}}\right)/E_{0,\text{fit}}$ (at right). Atomic collision partners: \protect\markerfilledbluecircle; molecular collision partners: \protect\markerfilledredcircle; nitrogen atom \ch{N}: \protect\markerfilledblackcircle.}
\label{fig:ss0_and_E0_d_bachmann1993}
\end{figure}

By analysing the obtained values for the characteristic cross sections $\sigma_0$ and energies $E_0$ associated with processes involving the atomic collision partners, it was found that these quantities increase exponentially with the hard-sphere diameter $d_\text{M}$ of the latter. It was then decided to obtain the values of $\sigma_0$ and $E_0$ for the processes involving the collision partner $\ch{M}=\ch{N}$ by fitting exponential curves to the data $(d_\text{M},\sigma_0)$ and $(d_\text{M},E_0)$, respectively. The values for the hard-sphere diameters $d_\text{M}$ were taken from the work of Svehla \cite{svehla1962estimated}. The results are depicted by Figure \ref{fig:ss0_and_E0_d_bachmann1993}, which additionally reveal that the data $(d_\text{M},\sigma_0)$ and $(d_\text{M},E_0)$ for the processes involving molecular particles seem to not follow any evident law (though the number of only three data points for each plot is too small to make a proper judgement). The data $\sigma_0$ and $E_0$ values deviate from the fit ones in an interval between $-34.6\%$ and $9.7\%$, and between $-2.6\%$ and $2.7\%$, respectively, thus showing a much greater agreement on the latter quantity than on the former. The fitting procedure resulted in the values $\sigma_0=1.594$ $\text{\AA}^2$ and $E_0=0.054$ eV for the case $\text{M}=\ch{N}$. 
The computation of the corresponding rate coefficients through expression \eqref{eq:kf_elio} requires the knowledge of the potential well depths for the interactions between \ch{N2} and the collision partners \ch{M}, i.e. $\varepsilon:=\varepsilon_{\text{\ch{N2} - M}}$. For this purpose, values of potential well depths associated with Lennard-Jones (12-6) interactions between collision partners of the same species $\varepsilon_{\text{M - M}}:=E_{\text{LJ},\text{M - M}}$ were extracted from the work of Svehla \cite{svehla1962estimated}. The quantities $\varepsilon$ were then assumed to be equal to the geometric mean of $\varepsilon_{\text{\ch{N2} - \ch{N2}}}$ and $\varepsilon_{\text{\ch{M} - \ch{M}}}$, i.e. $\varepsilon=\sqrt{\varepsilon_{\text{\ch{N2} - \ch{N2}}}\cdot\varepsilon_{\text{\ch{M} - \ch{M}}}}$, as suggested by Parmenter \textit{et al.} \cite{parmenter1979,lin1979}. 

The chemical equation \eqref{eq:VEh_N2_bachmann1993} represents a large number of different kinetic processes, each one associated with particular set of values of vibrational quantum numbers $v_1$, $v'_1$, $v_2$ and $v'_2$. Since accounting for all of these kinetic processes would require too much computational resources for the incoming CFD simulations, it was decided to regard only the most significant ones. The accounted kinetic processes of \eqref{eq:VEh_N2_bachmann1993} were the ones for which the ratio between the rate coefficient $k_{v_1,v_2}^{v'_1,v'_2}$ given by \eqref{eq:kf_elio} and the specific collisional frequency\footnote{Which corresponds to the frequency per unit of volume of collisions that occur between the two collision partners being divided by the their number densities.} $Z$  in the limit of the high temperatures (since this will correspond to the simulated regime) was higher or equal to an arbitrated factor of $2\times10^{-2}$, i.e
\begin{equation*}
\lim_{T_{\text{tr}_\text{h}}\to +\infty} \frac{k_{v_1,v_2}^{v'_1,v'_2}}{Z} = \frac{\sigma_0}{\sigma}e^{-\left(\frac{\varepsilon}{k_B T_{\text{ref}}}+\frac{|\Delta E|}{E_0}\right)}\ge 2\times10^{-2}\text{ ,}
\end{equation*}
with $\sigma=\pi d_{\ch{N2}}^2$ being the respective collisional cross section. This reduces the number of regarded processes from the unbearable $4,059,264$ to the reasonably manageable $7,436$.

\subsection{Vibronic transition of \ch{N2}$(\text{W}{}^3\Delta_\text{u},v_1)$ to \ch{N2}$(\text{B}{}^3\Pi_\text{g},v'_1)$ by collision with \ch{N2}$(\text{X}{}^1\Sigma_\text{g}^+,v_2)$ and \ch{N}$({}^4\text{S}_\text{u})$}
\label{subsection:VE_h_W_B}
 
In similarity to the previous section, the proposed V-E model \eqref{eq:kf_elio} will be implemented for the cases of vibronic transitions of \ch{N2}$(\text{W}{}^3\Delta_\text{u},v_1)$ to \ch{N2}$(\text{B}{}^3\Pi_\text{g},v'_1)$ by collisions of \ch{N2}$(\text{W}{}^3\Delta_\text{u},v_1)$ with \ch{N2}$(\text{X}{}^1\Sigma_\text{g}^+,v_2)$ and \ch{N}$({}^4\text{S}_\text{u})$. These are described by the chemical equations
\begin{equation}
\ch{N2}(\text{W}{}^3\Delta_\text{u},v_1)\ch{ + N2}(\text{X}{}^1\Sigma_\text{g}^+,v_2)\ch{ -> N2}(\text{B}{}^3\Pi_\text{g},v'_1)\ch{ + N2}(\text{X}{}^1\Sigma_\text{g}^+,v'_2),\,\,\,\,\,\,\,\,\,\forall\,v_1,\,v'_1,\,v_2,\text{ and }v'_2\text{ ,}
\label{eq:VEh_N2_W_B_N2}
\end{equation}
and
\begin{equation}
\ch{N2}(\text{W}{}^3\Delta_\text{u},v_1)\ch{ + N}({}^4\text{S}_\text{u})\ch{ -> N2}(\text{B}{}^3\Pi_\text{g},v'_1)\ch{ + N}({}^4\text{S}_\text{u}),\,\,\,\,\,\,\,\,\,\forall\,v_1\text{ and }v'_1\text{ ,}
\label{eq:VEh_N_W_B_N}
\end{equation}
respectively. Bachmann \textit{et al.} \cite{bachmann1992,bachmann1993} obtained characteristic cross sections $\sigma_0$ and characteristic energies $E_0$ for the intramolecular endothermic and exothermic vibronic processes regarding atomic collisional partners $\ch{M}\in\{\ch{He},\ch{Ne},\ch{Ar},\ch{Kr},\ch{Xe}\}=:\{\ch{M}_\text{a}\}$ and molecular collisional partners $\ch{M}\in\{\ch{H2},\ch{N2},\ch{NO}\}=:\{\ch{M}_\text{m}\}$ in their ground energy level, by fitting the curve \eqref{eq:ss_av_elio} to the set of points constituted by energy defect absolute values $|\Delta E|$ as abscissas, and experimentally obtained effective process cross sections $\sigma_{p,\text{eff}}$ as ordinates.

\begin{figure}[H]
\centering
\centerline{\includegraphics[scale=1]{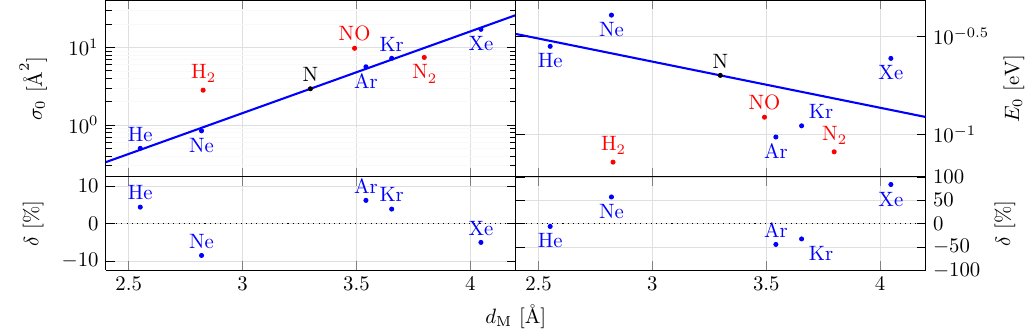}}
\vspace{-15pt}
\caption{Upper plots: exponential curves fitted to the data $(d_\text{M},\sigma_0)$ (at left) and $(d_\text{M},E_0)$ (at right) for the processes involving the atomic collision partners. Lower plots: relative deviations of the data, $\delta=\left(\sigma_0-\sigma_{0,\text{fit}}\right)/\sigma_{0,\text{fit}}$ (at left) and $\delta=\left(E_0-E_{0,\text{fit}}\right)/E_{0,\text{fit}}$ (at right). Atomic collision partners: \protect\markerfilledbluecircle; molecular collision partners: \protect\markerfilledredcircle; nitrogen atom \ch{N}: \protect\markerfilledblackcircle.}
\label{fig:ss0_and_E0_d_bachmann1993_W_B}
\end{figure}

Since no data are available for the case $\ch{M}=\ch{N}$, it was decided to make a study about the dependence of the effective process cross section on the atomic collision partner - in a similar way to what  was done in the previous section - and from it obtain the corresponding data. The characteristic cross section $\sigma_0$ varies exponentially with the hard-sphere diameter $d_\text{M}$ of the atomic collision partners. The same cannot be said about the characteristic energy $E_0$, conversely to the result which was obtained in previous section. Figure \ref{fig:ss0_and_E0_d_bachmann1993_W_B} presents exponential curves fitted to the data $(d_\text{M},\sigma_0)$ and $(d_\text{M},E_0)$ of the atomic collision partners, showing deviations between $-8.5\%$ and $6.2\%$ with respect to the former, and between $-44.8\%$ and $83.6\%$ with respect to the latter.

Values of $\sigma_0=2.921$ $\text{\AA}^2$ and $E_0=0.200$ eV were obtained for the case $\text{M}=\ch{N}$. Values of $\sigma_0=7.400$ $\text{\AA}^2$ and $E_0=0.081$ eV were obtained for the case $\text{M}=\ch{N2}$.

The accounted kinetic processes of \eqref{eq:VEh_N2_W_B_N2} were the ones for which the ratio between the rate coefficient $k_{v_1,v_2}^{v'_1,v'_2}$ given by \eqref{eq:kf_elio} and the specific collisional frequency $Z$ in the limit of the high temperatures was higher or equal to an arbitrated factor of $1.2\times10^{-1}$. This allowed a reduction on the number of regarded processes from $1,682,816$ to $5,720$.

\subsection{Vibronic transition of \ch{N2}$(\text{A}{}^3\Sigma_\text{u}^+,v_1)$ to \ch{N2}$(\text{B}{}^3\Pi_\text{g},v'_1)$ and \ch{N2}$(\text{C}{}^3\Pi_\text{u},v'_1)$, by collision with \ch{N2}$(\text{A}{}^3\Sigma_\text{u}^+,v_2)$, which in turn transits to \ch{N2}$(\text{X}{}^1\Sigma_\text{g}^+,v'_2)$}
\label{subsection:VE_h_implementation_piper1988}

\begin{figure}[H]
\centering
\centerline{\includegraphics[scale=1]{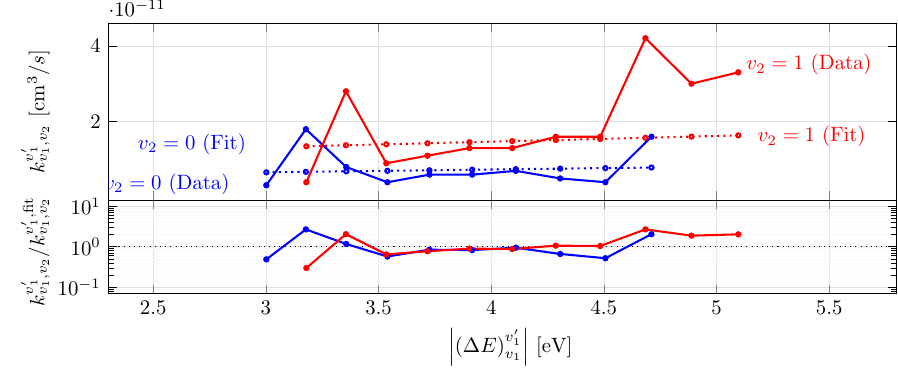}}
\vspace{-15pt}
\caption{Upper plot: points of the fitted curve \eqref{eq:k_v1_v2_vp_1__elio} to the data $(\{\left(\Delta E\right)_{v_1,v_2}^{v'_1,v'_2}\},k_{v_1,v_2}^{v'_1})$ with the rate coefficients $k_{v_1,v_2}^{v'_1}$ issued by Piper \cite{piper1988b} for process \eqref{eq:VE_m_h_N2A_B_N2A_X_general}. Lower plot: values of ratios between the data rate coefficients and the fit rate coefficients, $r=k_{v_1,v_2}^{v'_1}/k_{v_1,v_2}^{v'_1,\text{fit}}$. The obtained data are with respect to $v_1=0$, $v_2\in\{0,1\}$, and $v'_1\in[1,11]$.}
\label{fig:k_v1_v2_vp_1__N2A_B_N2A_X_v_DD_E_v_1_vp_1}
\end{figure}

\begin{figure}[H]
\centering
\centerline{\includegraphics[scale=1]{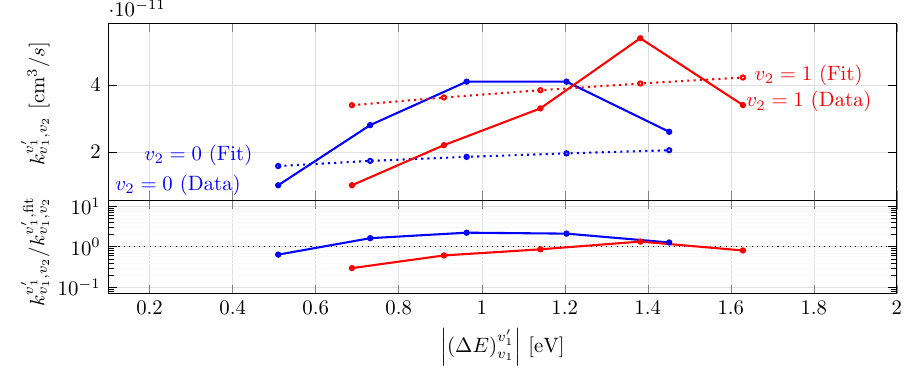}}
\vspace{-15pt}
\caption{Upper plot: points of the fitted curve \eqref{eq:k_v1_v2_vp_1__elio} to the data $(\{\left(\Delta E\right)_{v_1,v_2}^{v'_1,v'_2}\},k_{v_1,v_2}^{v'_1})$ with the rate coefficients $k_{v_1,v_2}^{v'_1}$ issued by Piper \cite{piper1988a} for process \eqref{eq:VE_m_h_N2A_C_N2A_X_general}. Lower plot: values of ratios between the data rate coefficients and the fit rate coefficients, $r=k_{v_1,v_2}^{v'_1}/k_{v_1,v_2}^{v'_1,\text{fit}}$. The obtained data are with respect to $v_1=0$, $v_2\in\{0,1\}$, and $v'_1\in[0,4]$.}
\label{fig:k_v1_v2_vp_1__N2A_C_N2A_X_v_DD_E_v_1_vp_1}
\end{figure}

A nitrogen molecule in the $\text{A}{}^3\Sigma_\text{u}^+$  electronic level may transit to $\text{B}{}^3\Pi_\text{g}$ or to $\text{C}{}^3\Pi_\text{u}$ by colliding with another nitrogen molecule also in the $\text{A}{}^3\Sigma_\text{u}^+$ electronic level. Piper \cite{piper1988b,piper1988a} studied such processes considering the simultaneous transition of the collision partner to the ground electronic level, i.e
\begin{equation}
\ch{N2}(\text{A}{}^3\Sigma_\text{u}^+,v_1)\ch{ + }\ch{N2}(\text{A}{}^3\Sigma_\text{u}^+,v_2)\ch{ -> N2}(\text{B}{}^3\Pi_\text{g},v'_1)\ch{ + }\ch{N2}(\text{X}{}^1\Sigma_\text{g}^+,v'_2)\text{ ,}
\label{eq:VE_m_h_N2A_B_N2A_X_general}
\end{equation}
and
\begin{equation}
\ch{N2}(\text{A}{}^3\Sigma_\text{u}^+,v_1)\ch{ + }\ch{N2}(\text{A}{}^3\Sigma_\text{u}^+,v_2)\ch{ -> N2}(\text{C}{}^3\Pi_\text{u},v'_1)\ch{ + }\ch{N2}(\text{X}{}^1\Sigma_\text{g}^+,v'_2)\text{ .}
\label{eq:VE_m_h_N2A_C_N2A_X_general}
\end{equation}
Piper issues rate coefficients values for \eqref{eq:VE_m_h_N2A_B_N2A_X_general} and \eqref{eq:VE_m_h_N2A_C_N2A_X_general} at room temperature $T_{\text{tr}_\text{h}}=300\text{ K}=:T\text{ref}$, which are specific to the vibrational levels $v_1$, $v_2$ and $v'_1$ but not to $v'_2$, i.e. $k_{v_1,v_2}^{v'_1}(T_\text{ref})=:k_{v_1,v_2}^{v'_1,\text{ref}}$. Since rate coefficients for the full set of vibrational levels $v_1$, $v_2$, $v'_1$ and $v'_2$ and temperatures $T_{\text{tr}_\text{h}}$ are needed, model \eqref{eq:kf_elio} was considered, allowing one to express a relationship between the non $v'_2$-specific rate coefficient $k_{v_1,v_2}^{v'_1,\text{ref}}=\sum_{v'_2}k_{v_1,v_2}^{v'_1,v'_2}(T_\text{ref})$ and the $v'_2$-specific rate coefficients $k_{v_1,v_2}^{v'_1,v'_2}(T_\text{ref})$ through
\begin{equation}
k_{v_1,v_2}^{v'_1,\text{ref}}=\frac{\sigma_0}{1+\delta_{v_1,v_2}}\sqrt{\frac{8k_BT_\text{ref}}{\pi\mu}}\left(\sum_{v_2'}e^{-\frac{\left|\left(\Delta E\right)_{v_1,v_2}^{v'_1,v'_2}\right|}{E_0}}\right)\text{ .}
\label{eq:k_v1_v2_vp_1__elio}
\end{equation}
By fitting curve \eqref{eq:k_v1_v2_vp_1__elio} to the data issued by Piper, values for the  characteristic cross section $\sigma_0$ and characteristic energy $E_0$ may be obtained. Figures \ref{fig:k_v1_v2_vp_1__N2A_B_N2A_X_v_DD_E_v_1_vp_1} and \ref{fig:k_v1_v2_vp_1__N2A_C_N2A_X_v_DD_E_v_1_vp_1} depict such fitted curves, with the absolute value of the partial energy defect $\left|\left(\Delta E\right)_{v_1}^{v'_1}\right|$ in the abscissae axes\footnote{Since the issued rate coefficients $k_{v_1,v_2}^{v'_1,\text{ref}}$ are not $v'_2$-specific the energy defect $\left(\Delta E\right)_{v_1,v_2}^{v'_1,v'_2}$ cannot be used to label the process.}. This partial energy defect corresponds to the difference between the initial and final internal energies of only the first collision partner, i.e. $\left(\Delta E\right)_{v_1}^{v'_1}=T_{e_1v_1}-T_{e'_1v'_1}$. Quantities $\sigma_0=0.132\text{ }\text{\AA}^2$ and $E_0=1.960\text{ eV}$, and $\sigma_0=0.873\text{ }\text{\AA}^2$ and $E_0=1.071\text{ eV}$, were obtained for processes \eqref{eq:VE_m_h_N2A_B_N2A_X_general} and \eqref{eq:VE_m_h_N2A_C_N2A_X_general}, respectively. The fit points did however depart significantly from the data points, in both value and behaviour. The ratio between the data rate values and the fit rate values was as low as $0.3$ and as high as $2.7$. 

The rate coefficients $k_{v_1,v_2}^{v'_1,v'_2}(T_{\text{tr}_\text{h}})$ may be then computed through law \eqref{eq:kf_elio} constrained to the obtained $\sigma_0$ and $E_0$ values. Since the number of processes described by the chemical equations \eqref{eq:VE_m_h_N2A_B_N2A_X_general} and \eqref{eq:VE_m_h_N2A_C_N2A_X_general} is too large, only the ones for which the ratio between the rate coefficient and the specific collisional frequency in the limit of the high temperatures was higher or equal to the arbitrated factors $2.28\times10^{-3}$ and $1.3\times10^{-2}$, respectively, were considered. Again, this procedure allowed a reduction of the number of accounted processes from $2,095,104$ to $6,067$ for the case of \eqref{eq:VE_m_h_N2A_B_N2A_X_general}, and from $317,440$ to $8,135$ for the case of \eqref{eq:VE_m_h_N2A_C_N2A_X_general}.

\subsection{Vibronic transition of \ch{N2}$(\text{A}{}^3\Sigma_\text{u}^+,v_1)$ to \ch{N2}$(\text{X}{}^1\Sigma_\text{g}^+,v'_1)$ by collision with \ch{N2}$(\text{X}{}^1\Sigma_\text{g}^+,v_2)$ and \ch{N}$({}^4\text{S}_\text{u})$, which in turn transits to \ch{N}$({}^2\text{P}_\text{u})$}
\label{subsection:VE_m_h_N2A_X}

Let one consider the vibronic transition of a nitrogen molecule in the $\text{A}{}^3\Sigma_\text{u}^+$ electronic level to the ground electronic level, by collision with a heavy particle:
\begin{equation}
\ch{N2}(\text{A}{}^3\Sigma_\text{u}^+,v)\ch{ + M}\ch{ -> N2}(\text{X}{}^1\Sigma_\text{g}^+,v')\ch{ + M}\text{ .}
\label{eq:VE_m_h_N2A_X_general}
\end{equation}

There is some data available in the literature that may be useful in this work. These correspond to experimentally obtained values for the rate coefficient of the process \eqref{eq:VE_m_h_N2A_X_general} at room temperature $T_{\text{tr}_\text{h}}=300\text{ K}=:T_\text{ref}$ concerning the particular cases
\begin{equation}
\ch{N2}(\text{A}{}^3\Sigma_\text{u}^+,v_1)\ch{ + N2}(\text{X}{}^1\Sigma_\text{g}^+,v_2)\ch{ -> N2}(\text{X}{}^1\Sigma_\text{g}^+,v'_1)\ch{ + N2}(\text{X}{}^1\Sigma_\text{g}^+,v'_2)\text{ ,}
\label{eq:VE_m_h_N2A_X_N2X}
\end{equation}
and
\begin{equation}
\ch{N2}(\text{A}{}^3\Sigma_\text{u}^+,v_1)\ch{ + N}({}^4\text{S}_\text{u})\ch{ -> N2}(\text{X}{}^1\Sigma_\text{g}^+,v'_1)\ch{ + N}({}^2\text{P}_\text{u})\text{ ,}
\label{eq:VE_m_h_N2A_X_NS}
\end{equation}
with $v_1\in\{0,1\}$ and $v_2=0$. 

Table \ref{tab:VE_m_h_N2A_X_N2X} presents the values obtained in the works of Drewer and Pener \cite{dreyer1973}, Vidaud \textit{et al.} \cite{vidaud1976} and Levron and Phelps \cite{levron1978}, for the rate coefficient associated with process \eqref{eq:VE_m_h_N2A_X_N2X}. Vidaud \textit{et al.} did not treat the vibrational levels $\ch{N2}(\text{A}{}^3\Sigma_\text{u}^+,0)$ and $\ch{N2}(\text{A}{}^3\Sigma_\text{u}^+,1)$ separately. Their issued rate coefficient value is with respect to these two levels lumped together. The values issued by Levron and Phelps do agree with the one of Vidaud \textit{et al.} although not as much with the ones of Drewer and Pener. Also, since Vidaud \textit{et al.} supply a lumped rate coefficient instead of the vibrationally-specific ones of Levron and Phelps, it was decided to consider the results of Levron and Phelps in this work. 

\begingroup
\centerline{\begin{threeparttable}
\setlength\tabcolsep{25pt} 
\renewcommand{\arraystretch}{1} 
\centering
\caption{Experimentally obtained values for the rate coefficient of process \eqref{eq:VE_m_h_N2A_X_N2X} with $v_1\in\{0,1\}$ and $v_2=0$, at $T_{\text{tr}_\text{h}}=300\text{ K}$, i.e. $k_{v_1,0}(T_\text{ref})=:k_{v_1,0}^{\text{ref}}$.}
\begin{scriptsize}
\begin{tabular}{ccc}
\toprule
$k_{0,0}^{\text{ref}}[\text{cm}^3/\text{s}]$ & $k_{1,0}^{\text{ref}}[\text{cm}^3/\text{s}]$ & Reference\\
\midrule
$3.7\times10^{-16}$ & $3.4\times 10^{-16}$ & Drewer and Pener \cite{dreyer1973}\\
\multicolumn{2}{c}{$4.5\times 10^{-17}$ (\tnote{a} )} & Vidaud \textit{et al.} \cite{vidaud1976}\\
$2.6\times10^{-18}$ & $3.8\times 10^{-17}$ & Levron and Phelps \cite{levron1978}\\
\bottomrule
\end{tabular}
\label{tab:VE_m_h_N2A_X_N2X}
\begin{tablenotes}
\item[a]{This value is with respect to $\ch{N2}(\text{A}{}^3\Sigma_\text{u}^+,0)$ and $\ch{N2}(\text{A}{}^3\Sigma_\text{u}^+,1)$ lumped together.}
\end{tablenotes}
\end{scriptsize}
\end{threeparttable}}
\endgroup
\vspace{30pt} 
Process \eqref{eq:VE_m_h_N2A_X_NS} not only describes a vibronic transition of $\ch{N2}(\text{A}{}^3\Sigma_\text{u}^+,v_1)$ to $\ch{N2}(\text{X}{}^1\Sigma_\text{g}^+,v'_1)$ but also an excitation of the collision partner, $\ch{N}({}^4\text{S}_\text{u})$ to $\ch{N}({}^2\text{P}_\text{u})$. This latter excitation was assumed in accordance with the discussions done by Meyer \textit{et al.} \cite{meyer1970} and Young and Dunn \cite{young1975}. There was an attempt, performed by Piper \cite{piper1989}, to quantify the fraction of ground state nitrogen atoms which are actually excited in the process. Such attempt was not however successful, as Piper \cite{piper1998} later showed that electronically excited nitrogen molecules different from $\ch{N2}(\text{A}{}^3\Sigma_\text{u}^+)$ were present in his experiment, and by being undetected and unaccounted, these leaded to wrong results. With only the evidence of Meyer \textit{et al.} and Young and Dunn available, it was decided to assume that excitation of $\ch{N}({}^4\text{S}_\text{u})$ to $\ch{N}({}^2\text{P}_\text{u})$ always occurred. 
Table \ref{tab:VE_m_h_N2A_X_NS} presents experimental values for the rate coefficient of process \eqref{eq:VE_m_h_N2A_X_NS} obtained in the works of Wray \cite{wray1966}, Young and St. John \cite{young1968}, Meyer \textit{et al.} \cite{meyer1970}, Dunn and Young \cite{dunn1976}, Vidaud \textit{et al.} \cite{vidaud1976}, and Piper \cite{piper1989}. The values agree reasonably well with each other, being of the same order of magnitude. It was decided to regard the most recent ones, the results of Piper, in this work. 

\begingroup
\centerline{\begin{threeparttable}
\setlength\tabcolsep{25pt} 
\renewcommand{\arraystretch}{1} 
\centering
\caption{Experimentally obtained values for the rate coefficient of process \eqref{eq:VE_m_h_N2A_X_NS} with $v_1\in\{0,1\}$, at $T_{\text{tr}_\text{h}}=300\text{ K}$, i.e. $k_{v_1}(T_\text{ref})=:k_{v_1}^{\text{ref}}$.}
\begin{scriptsize}
\begin{tabular}{ccc}
\toprule
$k_{0}^{\text{ref}}[\text{cm}^3/\text{s}]$ & $k_{1}^{\text{ref}}[\text{cm}^3/\text{s}]$ & Reference\\
\midrule
$5.4\times10^{-11}$ & --- & Wray \cite{wray1966}\\
$5\times10^{-11}$ & --- & Young and St. John \cite{young1968}\\
$4.3\times10^{-11}$ & --- & Meyer \textit{et al.} \cite{meyer1970}\\
$4.8\times10^{-11}$ & $4.8\times10^{-11}$ & Dunn and Young \cite{dunn1976}\\
\multicolumn{2}{c}{$3.5\times10^{-11}$ (\tnote{a} )} & Vidaud \textit{et al.} \cite{vidaud1976}\\
$4.0\times10^{-11}$ & $4.0\times10^{-11}$ & Piper \cite{piper1989}\\
\bottomrule
\end{tabular}
\label{tab:VE_m_h_N2A_X_NS}
\begin{tablenotes}
\item[a]{This value is with respect to $\ch{N2}(\text{A}{}^3\Sigma_\text{u}^+,0)$ and $\ch{N2}(\text{A}{}^3\Sigma_\text{u}^+,1)$ lumped together.}
\end{tablenotes}
\end{scriptsize}
\end{threeparttable}}
\endgroup
\vspace{30pt}

Values for the rate coefficient for process \eqref{eq:VE_m_h_N2A_X_N2X} for all $v_1$, $v_2$, $v'_1$ and $v'_2$, and values for the rate coefficient for process \eqref{eq:VE_m_h_N2A_X_NS} for all $v_1$, and $v'_1$, in the full set of heavy particle translational temperatures $T_{\text{tr}_\text{h}}$, are required. It is then necessary to make an assumption regarding the dependence of the rate coefficient on the vibrational levels and on the temperature, since the available data only concern some few levels and a room temperature. The VE-m-h model \eqref{eq:kf_elio} may be regarded for that purpose. Let one start by analysing process \eqref{eq:VE_m_h_N2A_X_N2X}. From \eqref{eq:kf_elio}, the respective rate coefficient is given by
\begin{equation}
k_{v_1,v_2}^{v'_1,v'_2}(T_{\text{tr}_\text{h}})=\sigma_0\,e^{-\frac{\left|\left(\Delta E\right)_{v_1,v_2}^{v'_1,v'_2}\right|}{E_0}}\sqrt{\frac{8k_BT_{\text{tr}_\text{h}}}{\pi\mu}}e^{\frac{\varepsilon}{k_B}\left(\frac{1}{T_{\text{tr}_\text{h}}}-\frac{1}{T_\text{ref}}\right)}\text{ .}
\label{eq:kf_v1v2vp1vp2_VE_m_h_N2A_X_N2X}
\end{equation}
The vibrational dependence of the rate coefficient is dictated by the characteristic energy $E_0$. It was decided to make the value of this variable to coincide with one of the obtained set for the analogous V-E processes studied in the previous sections. Since the vibronic transition of \ch{N2}$(\text{W}{}^3\Delta_\text{u},v_1)$ to \ch{N2}$(\text{B}{}^3\Pi_\text{g},v'_1)$ by impact with $\ch{N2}(\text{X}{}^1\Sigma_\text{g}^+,v_2)$ involves a transition of solely one of the collision partners from a higher electronic level to a lower electronic level in similarity to the present case and in dissimilarity to the other ones, the respective obtained $E_0$ value was taken. The variable $\sigma_0$ can in turn be computed using either $k_{0,0}^{\text{ref}}$ or $k_{1,0}^{\text{ref}}$. 
Arbitrarily, it was decided to use $k_{0,0}^{\text{ref}}$ and then to quantify the discrepancy between $k_{1,0}^{\text{ref}}$ and the respective value obtained from the model. The rate coefficient for \eqref{eq:VE_m_h_N2A_X_N2X} with $v_1=0$ and $v_2=0$ regardless of the vibrational levels $v'_1$ and $v'_2$, at $T_{\text{tr}_\text{h}}=T_\text{ref}$, is given by
\begin{equation}
k_{0,0}^{\text{ref}}=\sum_{v'_1,v'_2}k_{0,0}^{v'_1,v'_2}(T_\text{ref})=\sigma_0\sqrt{\frac{8k_BT_{\text{ref}}}{\pi\mu}}\left(\sum_{v'_1,v'_2}e^{-\frac{\left|\left(\Delta E\right)_{0,0}^{v'_1,v'_2}\right|}{E_0}}\right)\text{ .}
\label{eq:kf_v10_VE_m_h_N2A_X_N2X}
\end{equation}
By solving \eqref{eq:kf_v10_VE_m_h_N2A_X_N2X} with respect to $\sigma_0$, one may get
\begin{equation}
\sigma_0=\frac{k_{0,0}^{\text{ref}}}{\sqrt{\frac{8k_BT_{\text{ref}}}{\pi\mu}}\left(\sum_{v'_1,v'_2}e^{-\frac{\left|\left(\Delta E\right)_{0,0}^{v'_1,v'_2}\right|}{E_0}}\right)}\text{ .}
\label{eq:ssp0_VE_m_h_N2A_X_N2X}
\end{equation}
And by inserting the result \eqref{eq:ssp0_VE_m_h_N2A_X_N2X} into \eqref{eq:kf_v1v2vp1vp2_VE_m_h_N2A_X_N2X} a general expression for the rate coefficient of process \eqref{eq:VE_m_h_N2A_X_N2X} may be obtained:
\begin{equation}
k_{v_1,v_2}^{v'_1,v'_2}(T_{\text{tr}_\text{h}})=k_{0,0}^{\text{ref}}\frac{e^{-\frac{\left|\left(\Delta E\right)_{v_1,v_2}^{v'_1,v'_2}\right|}{E_0}}}{\sum_{v'_1,v'_2}e^{-\frac{\left|\left(\Delta E\right)_{0,0}^{v'_1,v'_2}\right|}{E_0}}}\sqrt{\frac{T_{\text{tr}_\text{h}}}{T_{\text{ref}}}}e^{\frac{\varepsilon}{k_B}\left(\frac{1}{T_{\text{tr}_\text{h}}}-\frac{1}{T_{\text{ref}}}\right)}\text{ .}
\label{eq:kf_v1v2vp1vp2_VE_m_h_N2A_X_N2X_final}
\end{equation}

In similarity to what was done in the modelling of the other V-E processes, it was decided to only account for the kinetic processes of \eqref{eq:VE_m_h_N2A_X_N2X} for which the ratio between the rate coefficient $k_{v_1,v_2}^{v'_1,v'_2}$ and the specific collisional frequency $Z$ in the limit of the high temperatures was higher or equal to an arbitrated factor of $3.3\times10^{-10}$, i.e
\begin{equation*}
\lim_{T_{\text{tr}_\text{h}}\to +\infty} \frac{k_{v_1,v_2}^{v'_1,v'_2}}{Z} = \frac{k_{0,0}^{\text{ref}}}{\sigma\sqrt{\frac{8k_BT_\text{ref}}{\pi\mu}}}\frac{e^{-\frac{\left|\left(\Delta E\right)_{v_1,v_2}^{v'_1,v'_2}\right|}{E_0}}}{\sum_{v'_1,v'_2}e^{-\frac{\left|\left(\Delta E\right)_{0,0}^{v'_1,v'_2}\right|}{E_0}}}e^{-\frac{\varepsilon}{k_B T_{\text{ref}}}}\ge 3.3\times10^{-10}\text{ .}
\end{equation*}
This again allows one to reduce the number of regarded processes from 7,626,496 to 7,355. A characteristic cross section  $\sigma_0=2.04\times10^{-8}\,\text{\AA}^2$ was obtained. The ratio between the numerical rate $k_{1,0}(T_{\text{ref}})$ and the experimental one $k_{1,0}^{\text{ref}}$ corresponds to \sloppy $k_{1,0}(T_{\text{ref}})/k_{1,0}^{\text{ref}}=0.06$, evidencing a significant underestimation of this quantity by the model. This shows how crude the assumption on the vibrational dependency of the rate coefficient may be.
For the case of process \eqref{eq:VE_m_h_N2A_X_NS}, it can be shown that if we regard model \eqref{eq:kf_elio} as well as the result $k_{0}^{\text{ref}}$ and the obtained value for $E_0$ in the modelling of vibronic transitions of \ch{N2}$(\text{W}{}^3\Delta_\text{u},v_1)$ to \ch{N2}$(\text{B}{}^3\Pi_\text{g},v'_1)$ by \ch{N}$({}^4\text{S}_\text{u})$, the rate coefficient $k_{v_1}^{v'_1}$ of the process is given by
\begin{equation}
k_{v_1}^{v'_1}(T_{\text{tr}_\text{h}})=k_{0}^{\text{ref}}\frac{e^{-\frac{\left|\left(\Delta E\right)_{v_1}^{v'_1}\right|}{E_0}}}{\sum_{v'_1}e^{-\frac{\left|\left(\Delta E\right)_{0}^{v'_1}\right|}{E_0}}}\sqrt{\frac{T_{\text{tr}_\text{h}}}{T_{\text{ref}}}}e^{\frac{\varepsilon}{k_B}\left(\frac{1}{T_{\text{tr}_\text{h}}}-\frac{1}{T_{\text{ref}}}\right)}\text{ .}
\label{eq:kf_v1vp1_VE_m_h_N2A_X_NS_final}
\end{equation}
A characteristic cross section $\sigma_0=3.33\,\text{\AA}^2$ was accordingly obtained. The relative deviation between the model obtained result $k_{1}(T_{\text{ref}})$ and the experimental one $k_{1}^{\text{ref}}$ corresponds to $10.2\%$, evidencing an overestimation of this quantity by the model.

\subsection{Vibronic transition of \ch{N2}$(\text{A}'{}^5\Sigma_\text{g}^+,v)$ to \ch{N2}$(\text{B}{}^3\Pi_\text{g},v')$ by collision with \ch{N2}$(\text{X}{}^1\Sigma_\text{g}^+,0)$ and \ch{N}$({}^4\text{S}_\text{u})$}
\label{subsection:VE_h_implementation_ottinger1994a}

\begin{figure}[H]
\centering
\centerline{\includegraphics[scale=1]{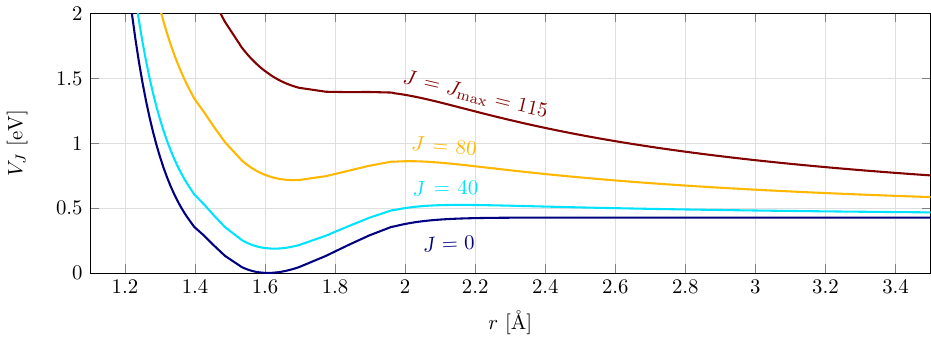}}
\vspace{-15pt}
\caption{Centrifugally corrected potential curves $V_{J}(r)$ for $J=0,$ $40$, $80$ and $J_\text{max}=115$ obtained for $\ch{N2}(\text{A}'{}^5\Sigma_\text{g}^+)$.}
\label{fig:V_J_N2_Ap}
\end{figure}

According to the work of Ottinger \textit{et al.} \cite{ottinger1994a}, there is the possibility of occuring intramolecular vibronic transitions from \ch{N2}$(\text{A}'{}^5\Sigma_\text{g}^+,v,J)$ to \ch{N2}$(\text{B}{}^3\Pi_\text{g},v',J')$ by collisions with heavy particles through the so-called gateway mechanism \cite{gelbart1973,freed1981,tramer1981}:
\begin{equation}
\ch{N2}(\text{A}'{}^5\Sigma_\text{g}^+,v,J)\ch{ + M}\ch{ -> N2}(\text{B}{}^3\Pi_\text{g},v',J')\ch{ + M}\text{ .}
\label{eq:VEh_intra_vJ_vpJp_ottinger1994a}
\end{equation}
The main route is the one associated with $v=0$, $J=12$, $v'=10$ and $J'=12$. The authors issue experimental values for effective process cross sections $\sigma_{p,\text{eff},v,J}^{v',J'}$ associated with this main route considering atomic collision partners, $\ch{M}\in\{\ch{He},\ch{Ne},\ch{Ar},\ch{Kr},\ch{Xe}\}=:\{\ch{M}_\text{a}\}$, as well as molecular collision partners, $\ch{M}\in\{\ch{H2},\ch{N2},\ch{NO},\ch{O2}\}=:\{\ch{M}_\text{m}\}$ in their ground energy levels. The experiment was done at room temperature $T_{\text{tr}_\text{h}}=300\text{ K}=:T_\text{ref}$. In the present work a vibronic-specific rate coefficient for the process 
\begin{equation}
\ch{N2}(\text{A}'{}^5\Sigma_\text{g}^+,v)\ch{ + M}\ch{ -> N2}(\text{B}{}^3\Pi_\text{g},v')\ch{ + M}\text{ ,}
\label{eq:VEh_intra_v_vp_ottinger1994a}
\end{equation}
with $\ch{M}\in\{\ch{N},\,\ch{N2}\}$, $v=0$ and $v'=10$ instead of a rovibronic one for process \eqref{eq:VEh_intra_vJ_vpJp_ottinger1994a} is wanted. The rate coefficient for process \eqref{eq:VEh_intra_vJ_vpJp_ottinger1994a} corresponds to $k_{v,J}^{v',J'}$ such that the variation on time of the concentration of $\ch{N2}(\text{A}'{}^5\Sigma_\text{g}^+,v,J)$ due to this same process is given by
\begin{equation}
\left(\frac{d[\ch{N2}(\text{A}'{}^5\Sigma_\text{g}^+,v,J)]}{dt}\right)_{v,J}^{v',J'}=-k_{v,J}^{v',J'}[\ch{N2}(\text{A}'{}^5\Sigma_\text{g}^+,v,J)][\text{M}]\text{ ,}
\label{eq:dN2ApvJdt_ottinger1994a}
\end{equation}
and, similarly, the rate coefficient for process \eqref{eq:VEh_intra_v_vp_ottinger1994a} corresponds to $k_{v}^{v'}$ such that the variation on time of the concentration of $\ch{N2}(\text{A}'{}^5\Sigma_\text{g}^+,v)$ due to this same process is given by
\begin{equation}
\left(\frac{d[\ch{N2}(\text{A}'{}^5\Sigma_\text{g}^+,v)]}{dt}\right)_{v}^{v'}=-k_{v}^{v'}[\ch{N2}(\text{A}'{}^5\Sigma_\text{g}^+,v)][\text{M}]\text{ .}
\label{eq:dN2Apvdt_ottinger1994a}
\end{equation}
Rate coefficients $k_{v}^{v'}$ and $k_{v,J}^{v',J'}$ are related to each other since from the definition of $\left(d[\ch{N2}(\text{A}'{}^5\Sigma_\text{g}^+,v)]/dt\right)_{v}^{v'}$ one has
\begin{equation}
\centerline{$
\left(\frac{d[\ch{N2}(\text{A}'{}^5\Sigma_\text{g}^+,v)]}{dt}\right)_{v}^{v'}=\sum_{J,J'}\left(\frac{d[\ch{N2}(\text{A}'{}^5\Sigma_\text{g}^+,v,J)]}{dt}\right)_{v,J}^{v',J'}=-\underbrace{\left(\sum_{J,J'}k_{v,J}^{v',J'}\frac{[\ch{N2}(\text{A}'{}^5\Sigma_\text{g}^+,v,J)]}{[\ch{N2}(\text{A}'{}^5\Sigma_\text{g}^+,v)]}\right)}_{=:k_v^{v'}}[\ch{N2}(\text{A}'{}^5\Sigma_\text{g}^+,v)][\text{M}]\text{ ,}
$}
\label{eq:dN2Apvdt_def_ottinger1994a}
\end{equation}
and the rate coefficient  $k_{v}^{v'}$ is then given by
\begin{equation}
k_{v}^{v'}=\sum_{J,J'}k_{v,J}^{v',J'}\frac{[\ch{N2}(\text{A}'{}^5\Sigma_\text{g}^+,v,J)]}{[\ch{N2}(\text{A}'{}^5\Sigma_\text{g}^+,v)]}\text{.}
\label{eq:k_v_vp_ottinger1994a}
\end{equation}
Due to the fact that the population of the rotational levels $\ch{N2}(\text{A}'{}^5\Sigma_\text{g}^+,v,J)$ follows the Boltzmann distribution \eqref{eq:Boltzmann_distribution_state_to_state}, and only the rotational quantum numbers $J=12$ and $J'=12$ contribute to the gateway process \eqref{eq:VEh_intra_vJ_vpJp_ottinger1994a} for $v=0$ and $v'=10$ \cite{ottinger1994a}, the rate coefficient $k_{0}^{10}$, given by relation \eqref{eq:k_v_vp_ottinger1994a}, is reduced to
\begin{equation}
k_{0}^{10}=k_{0,12}^{10,12}\,\frac{g_{\ch{N2},\text{rot},\text{A}',0,12}\cdot e^{-\frac{\epsilon_{\ch{N2},\text{rot},\text{A}',0,12}}{k_B T_\text{ref}}}}{Q_{\ch{N2},\text{rot},\text{A}',0}}\text{ .}
\label{eq:k_0_10_ottinger1994a}
\end{equation}
From \eqref{eq:kf_Maxwell} and \eqref{eq:k_0_10_ottinger1994a}, the respective effective process cross section $\sigma_{\text{p},\text{av},0}^{10}$ is simply given by
\begin{equation}
\sigma_{\text{p},\text{av},0}^{10}=\sigma_{\text{p},\text{av},0,12}^{10,12}\,\frac{g_{\ch{N2},\text{rot},\text{A}',0,12}\cdot e^{-\frac{\epsilon_{\ch{N2},\text{rot},\text{A}',0,12}}{k_B T_\text{ref}}}}{Q_{\ch{N2},\text{rot},\text{A}',0}}\text{ .}
\label{eq:ss_av_p_0_10_ottinger1994a}
\end{equation}
The quantity $Q_{\ch{N2},\text{rot},\text{A}',0}$ corresponds to the rotational partition function for $\ch{N2}(\text{A}'{}^5\Sigma_\text{g}^+,0)$, which according to its definition, is given by
\begin{equation}
Q_{\ch{N2},\text{rot},\text{A}',0}=\sum_{J=0}^{J_\text{max}} g_{\ch{N2},\text{rot},\text{A}',0,J}\cdot e^{-\frac{\epsilon_{\ch{N2},\text{rot},\text{A}',0,J}}{k_B T_\text{ref}}}\text{ ,}
\label{eq:Q_N2_rot_Ap_0_ottinger1994a}
\end{equation}
with $J_\text{max}$ being the maximum rotational quantum number that $\ch{N2}(\text{A}'{}^5\Sigma_\text{g}^+)$ may assume. For $J>J_\text{max}$ the nuclei are subjected to a purely repulsive potential, which makes them depart from each other resulting in the dissociation of the diatomic particle. Therefore, $J_\text{max}$ corresponds to the maximum rotational quantum number $J$ for which the effective internuclear potential, the so-called centrifugally corrected potential $V_J(r)=V(r)+\hbar^2J(J+1)/(2\mu r^2)$ remains with a well in its curve. For higher $J$ values the well disappears, and the curve is transformed into another with a slope which is non-positive throughout all of its extension. The internuclear force, which corresponds to the symmetric value of the slope and has a positive conventional signal in the direction of increasing internuclear distances, is in turn transformed into a non-negative quantity, imposing repulsiveness. One gets $J_\text{max}=115$, with the respective centrifugally corrected potential curve $V_{J_\text{max}}(r)$ being represented in Figure \ref{fig:V_J_N2_Ap}. 
The sensible rotational energy $\epsilon_{\ch{N2},\text{rot},\text{A}',0,J}$ corresponds to $\epsilon_{\ch{N2},\text{rot},\text{A}',0,J}=B_{\ch{N2},\text{rot},\text{A}',0} J(J+1)$, with $B_{\ch{N2},\text{rot},\text{A}',0}$ being the respective spectroscopic vibronic-specific rotational function, usually labelled by $B_v$.
Since \ch{N2} is a homonuclear diatomic particle, the rotational degree of degeneracy $g_{\ch{N2},\text{rot},\text{A}',0,J}$ corresponds to a product between two contributions \cite{herzberg1950}: one due to the nuclear spin, $g_{\ch{N2},\text{rot},\text{A}',0,J}^n$, and another due to the rotational quantum number, $2J+1$, i.e. $g_{\ch{N2},\text{rot},\text{A}',0,J}=g_{\ch{N2},\text{rot},\text{A}',0,J}^n\cdot\left(2J+1\right)$.
The molecular term symbol ${}^5\Sigma_\text{g}^+$ associated with the $A'$ electronic level states that the rotational levels are symmetric for even $J$, and antisymmetric for odd $J$ \cite{herzberg1950}. Additionally, it is known that the nitrogen nuclei follow Bose-Einstein statistics. The contribution of the nuclear spin to the rotational degree of degeneracy of $\ch{N2}(\text{A}'{}^5\Sigma_\text{g}^+,v,J)$ corresponds to \cite{herzberg1950}
\begin{equation}
g_{\ch{N2},\text{rot},\text{A}',0,J}^n=
\begin{cases}
(2I+1)(I+1), & \text{if $J$ is even}\text{ ,}\\
(2I+1)I, & \text{if $J$ is odd}\text{ ,}
\end{cases}
\label{eq:g_N2_rot_Ap_0_ottinger1994a_2}
\end{equation}
with $I$ being the so-called nuclei spin quantum number. For the case of the nitrogen nuclei one has $I=1$ \cite{herzberg1950}, meaning that $g_{\ch{N2},\text{rot},\text{A}',0,J}^n=6$ for even $J$ and $g_{\ch{N2},\text{rot},\text{A}',0,J}^n=3$ for odd $J$.

\begin{figure}[H]
\centering
\centerline{\includegraphics[scale=1]{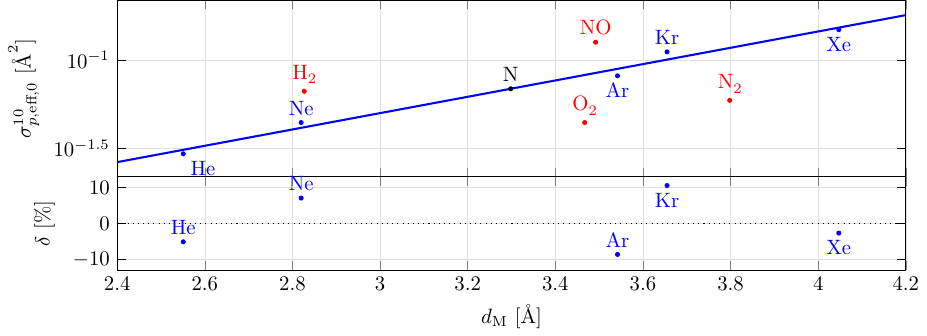}}
\vspace{-15pt}
\caption{Upper plot: exponential curve fitted to the data $(d_\text{M},\sigma_{\text{p},\text{av},0}^{10})$ for process \eqref{eq:VEh_intra_v_vp_ottinger1994a} involving solely the atomic collision partners. Lower plot: relative deviations of the data, $\delta=\left(\sigma_{\text{p},\text{av},0}^{10}-\sigma_{\text{p},\text{av},0}^{10,\text{fit}}\right)/\sigma_{\text{p},\text{av},0}^{10,\text{fit}}$. Atomic collision partners: \protect\markerfilledbluecircle; molecular collision partners: \protect\markerfilledredcircle; nitrogen atom \ch{N}: \protect\markerfilledblackcircle.}
\label{fig:ss_av_d_ottinger1994a}
\end{figure}

The values of vibronic-specific effective process cross sections $\sigma_{\text{p},\text{av},0}^{10}$ obtained through expression \eqref{eq:ss_av_p_0_10_ottinger1994a} are presented in Table \ref{tab:ss_p_av_0_10_ottinger1994a}.
\begin{table}[H]
\centering
\caption{Vibronic-specific effective process cross sections $\sigma_{\text{p},\text{av},0}^{10}$ computed through expression \eqref{eq:ss_av_p_0_10_ottinger1994a} using the rovibronic-specific effective process cross sections $\sigma_{\text{p},\text{av},0,12}^{10,12}$ issued by Ottinger \textit{et al.} \cite{ottinger1994a}.}
\begin{scriptsize}
\centerline{\begin{tabular}{c|ccccccccc}
\toprule
M & \ch{He} & \ch{Ne} & \ch{Ar} & \ch{Kr} & \ch{Xe} & \ch{H2} & \ch{N2} & \ch{NO} & \ch{O2}\\
\midrule
$\sigma_{\text{p},\text{av},0}^{10}[10^{-2}\text{\AA}^2]$ & $2.953$ & $4.430$ & $8.122$ & $11.075$ & $14.767$ & $6.645$ & $5.907$ & $12.552$ & $4.430$\\
\bottomrule
\end{tabular}}
\end{scriptsize}
\label{tab:ss_p_av_0_10_ottinger1994a}
\end{table}
Since no data are available for the case $\ch{M}=\ch{N}$, it was decided to make a study about the dependence of the effective process cross section on the atomic collision partner - in a similar way to what was done in the two first studied V-E processes - and from it obtain the respective data. It was indeed found that the effective process cross section $\sigma_{\text{p},\text{av},0}^{10}$ increased exponentially with the hard-sphere diameter $d_\text{M}$ of the atomic collision partners, although the same cannot be said about the molecular collision partners. By fitting an exponential curve to the data points $(d_\text{M},\sigma_{\text{p},\text{av},0}^{10})$ for the atomic collision partners, the value of the effective process cross section for $\ch{M}=\ch{N}$ was obtained. The respective results are depicted by Figure \ref{fig:ss_av_d_ottinger1994a}. The data values for the effective process cross sections deviate from the fit ones from $-8.6\%$ to $10.6\%$. A value of $\sigma_{\text{p},\text{av},0}^{10}=6.863\times 10^{-2} \text{\AA}^2$ was obtained for the case $\ch{M}=\ch{N}$.
The dependence of the effective process cross section $\sigma_{\text{p},\text{av},0}^{10}$ on the temperature was assumed to be the one referred by Parmenter \textit{et al.} \cite{parmenter1979,lin1979}, and therefore, the respective forward rate coefficient may be mathematically expressed by
\begin{equation}
k_{0}^{10}(T_{\text{tr}_\text{h}})={\sigma'}_{p,\text{av},0}^{\,10,\text{ref}}\,\sqrt{\frac{8k_BT_{\text{tr}_\text{h}}}{\pi\mu}}\,e^{\frac{\varepsilon}{k_BT_{\text{tr}_\text{h}}}}\text{ ,}
\label{eq:kf_ottinger1994a}
\end{equation}
with ${\sigma'}_{p,\text{av},0}^{\,10,\text{ref}}=\sigma_{p,\text{eff},0}^{10,\text{ref}}\,e^{-\frac{\varepsilon}{k_BT_{\text{ref}}}}$, where $\sigma_{p,\text{eff},0}^{10,\text{ref}}$ is the effective process cross section evaluated at the reference temperature $T_\text{ref}$ of Ottinger \textit{et al.} \cite{ottinger1994a} - the object of study considered above, and ambiguously labelled as $\sigma_{\text{p},\text{av},0}^{10}$. Note that relation \eqref{eq:kf_ottinger1994a} may be conveniently expressed through a modified Arrhenius function.

\section{Conclusions}
\label{sec:conclusions}

In this work a extensive set of vibronic energy levels for \ch{N2}, \ch{N2+}, \ch{N} and \ch{N+} was built using the most up-to-date (to the limits of knowledge of the present authors) data available in the literature. A near complete database of vibronic-specific kinetic processes involving these species was congregated in which special caution was taken to ensure physical consistency up to the highest temperature values. The database comprises chemical processes such as dissociation, ionisation, dissociative recombination and charge exchange, as well as non-chemical processes, i.e. excitation and de-excitation of the vibronic energy levels of the particles either due to collisions or spontaneous emission of radiation. Note, however, that the term ``near complete'' was herein used to describe the current state of the database, since it misses collision-induced transitions between the high electronic energy levels of the molecular particles, the bound-bound radiative processes of absorption and induced emission, spontaneous emission processes from some very high vibronic levels of the molecular particles (due to a lack of numerically and experimentally obtained Einstein coefficients), bound-free radiative processes (photodissociation and photoionisation), and free-free radiative processes (bremsstrahlung). Efforts in modelling these should be taken in the future.

This work focused almost entirely on two types of processes: the vibrational and the vibronic transitions of the molecular particles by collisions with heavy particles.
Regarding vibrational transition processes, the Forced Harmonic Oscillator model was used in place of the most commonly considered one, the Schwartz-Slawsky-Herzfeld model, to compute the respective rate coefficients since in contrast with the latter, it is physically consistent at the high temperature values attained in atmospheric entries. The obtained vibrational transition rate coefficients were, in some part, indirectly validated, as thermal dissociation rate coefficients of $\ch{N2}(\text{X}{}^1\Sigma_\text{g}^+)$ by collisions with \ch{N2} and \ch{N} which depend on these transition rates were calibrated using state-of-the-art experimental results. An agreement between $-59.9$ and $8.9\,\%$, and between $-80.9$ and $-36.1\,\%$ were obtained for the former and latter interactions, respectively. Furthermore, the values were compared with the most recent Quasi-Classical Trajectory model calculations deviating by a maximum of $56.5\,\%$. Note that although these deviations seem to be significant, they should be regarded as reasonable, since the deviations are evaluated within an extensive set of temperatures (varying in many thousands of kelvins) for which the rate coefficients suffer changes of several orders of magnitude.

The authors tried to increase the degree of fidelity of the current models for vibronic transitions by collisions with heavy particles by employing the so-called exponential gap law for the effective process cross section. This takes into account two parameters that are specific to the set of electronic levels - the characteristic cross section $\sigma_0$ and characteristic energy $E_0$ - and a dependence on the energy defect $\Delta E$. However, the computed effective process cross sections differed from the experimental ones at room temperature by as much as one order of magnitude, which evidences that other models with a even greater complexity should be tried. Furthermore, since the available experimental data that was used for calibration was only with respect to the low electronic levels of \ch{N2}, it was not possible to model transitions between the highest ones, as well as transitions between the vibronic levels of \ch{N2+}. 

The herein developed database of kinetic processes was employed in Euler unidimensional simulations of several shots of the $62^{\text{nd}}$ campaign of the Electric Arc Shock Tube. The respective results are reported in a companion paper {\color{red} CITE COMPANION PAPER}.

\begin{acknowledgement}

We would like to express gratitude to Vasco Guerra from the N-PRiME group for his literature recommendations which were important for the conception of the \ref{sec:theory} section. 
We also want to thank Dr. Annarita Laricchiuta from CNR Bari for her clarifications and providing us her values for the branching ratios associated with the process of non-dissociative ionisation of nitrogen molecules by electron impact, as well as the used database for the vibrational energies of the nitrogen molecule in its ground electronic level. 
We thank Dr. Steven Guberman from the Institute for Scientific Research for his noteworthy discussion about the valid temperature range for which his numerical data may be used in the modelling of rates of dissociative recombination of molecular nitrogen ions. 
We also acknowledge the clarifications provided by Dr. Julien Annaloro from the National Centre for Space Studies in France about his lumping procedure with respect to the electronic levels and Einstein coefficients for spontaneous emission issued by NIST.

This work has been partially supported by the Portuguese Science Foundation FCT, under Projects UIDB/50010/2020 and UIDP/50010/2020.

\end{acknowledgement}


\bibliography{bibliography}

\newpage

\section*{TOC Graphic}

\begin{figure}[H]
\centering
\centerline{\includegraphics[scale=0.5]{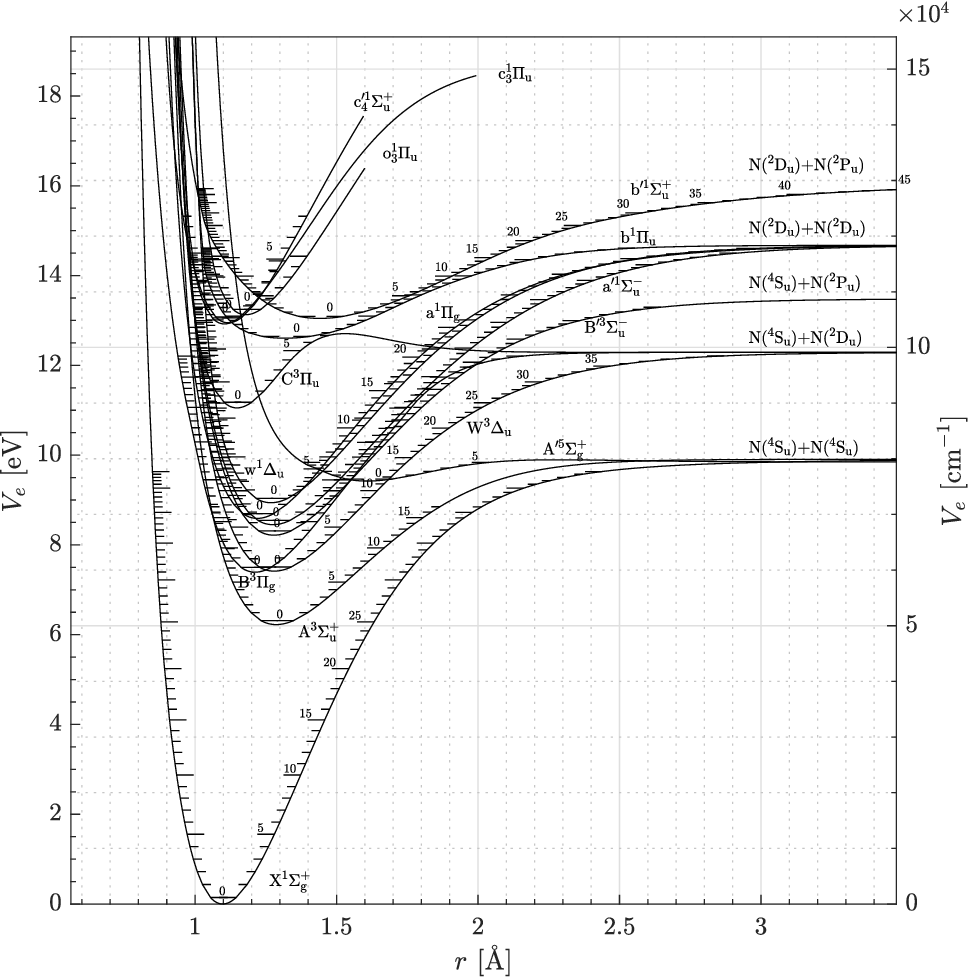}}
\end{figure}

\newpage

\appendix

\section{Appendix A: Potential curves of \ch{N2} and \ch{N2+} in their different electronic levels}
\setlabel{Appendix A}{sec:appendixA}

\begin{figure}[H]
\centering
\centerline{\includegraphics[scale=1]{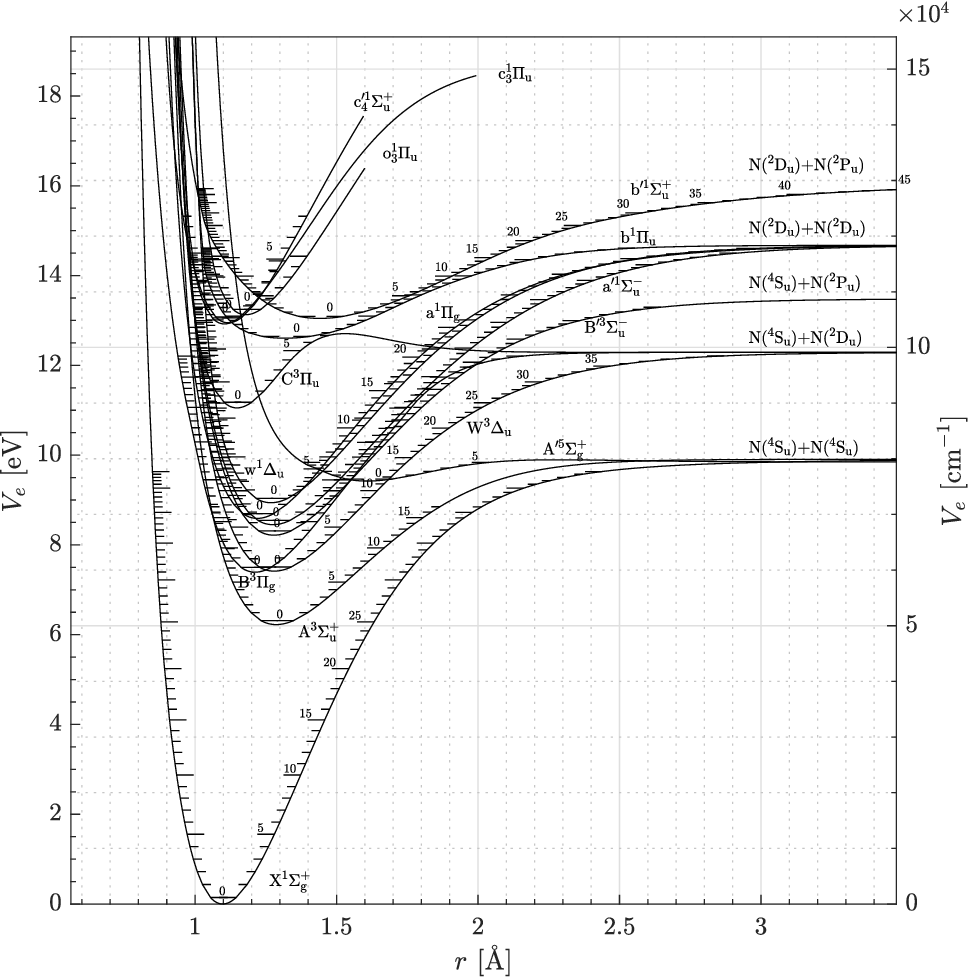}}
\caption{Electronically corrected internuclear potential curves $V_e(r)=V(r)+T_e$ for the different electronic levels of the nitrogen molecule \ch{N_2}. The terms \ch{N}$\left({}^4\text{S}_\text{u}\right)$\ch{ + N}$\left({}^4\text{S}_\text{u}\right)$, \ch{N}$\left({}^4\text{S}_\text{u}\right)$\ch{ + N}$\left({}^2\text{D}_\text{u}\right)$, \ch{N}$\left({}^4\text{S}_\text{u}\right)$\ch{ + N}$\left({}^2\text{P}_\text{u}\right)$, \ch{N}$\left({}^2\text{D}_\text{u}\right)$\ch{ + N}$\left({}^2\text{D}_\text{u}\right)$ and \ch{N}$\left({}^2\text{D}_\text{u}\right)$\ch{ + N}$\left({}^2\text{P}_\text{u}\right)$ represent the dissociation products of the nitrogen molecule in the electronic levels associated with the immediately below potential curves.}
\label{fig:Ve_N2}
\end{figure}

Internuclear potential curves $V(r)$ for the electronic levels X${}^1\Sigma_\text{g}^+$, A${}^3\Sigma_\text{u}^+$, B${}^3\Pi_\text{g}$, W${}^3\Delta_\text{u}$, B$'{}^3\Sigma_\text{u}^-$, a$'{}^1\Sigma_\text{u}^-$, a${}^1\Pi_\text{g}$, w${}^1\Delta_\text{u}$, A$'{}^5\Sigma_\text{g}^+$, C${}^3\Pi_\text{u}$, b${}^1\Pi_\text{u}$, c$_3{}^1\Pi_\text{u}$, c$_4'{}^1\Sigma_\text{u}^+$, b$'{}^1\Sigma_\text{u}^+$ and o$_3{}^1\Pi_\text{u}$ of the nitrogen molecule \ch{N2} were obtained. All curves were constructed by application of the Rydberg \cite{rydberg1932,rydberg1933}-Klein \cite{klein1932}-Rees \cite{rees1947} method and posterior extrapolation, as described in the \ref{sec:theory} section. Values of the parameters $Y_{ij}$ of the Dunham expansion that describes the rovibrational energy of the molecular particles, 
\begin{equation}
T_{vJ}=\sum_{i,j=0}^\infty Y_{ij}\left(v+\frac{1}{2}\right)^i\left[J\left(J+1\right)\right]^j\text{ ,}
\label{eq:Dunham_expansion}
\end{equation}
and are input variables of the RKR method, were required. Table \ref{tab:spec_N2} shows such values, being taken from the literature. The electronically corrected (i.e. with the sensible electronic energy summed) internuclear potential curves $V_e(r)=V(r)+T_e$ obtained for each of the electronic levels of molecular nitrogen \ch{N2} are depicted in Figure \ref{fig:Ve_N2}. All of the dissociation products shown in Figure \ref{fig:Ve_N2} were taken from the work of Lofthus and Krupenie \cite{lofthus1977} with the exception of the one associated with the electronic level A$'{}^5\Sigma_\text{g}^+$, which was taken from the work of Partridge \textit{et al.} \cite{partridge1988}.

\begin{figure}[H]
\centering
\centerline{\includegraphics[scale=1]{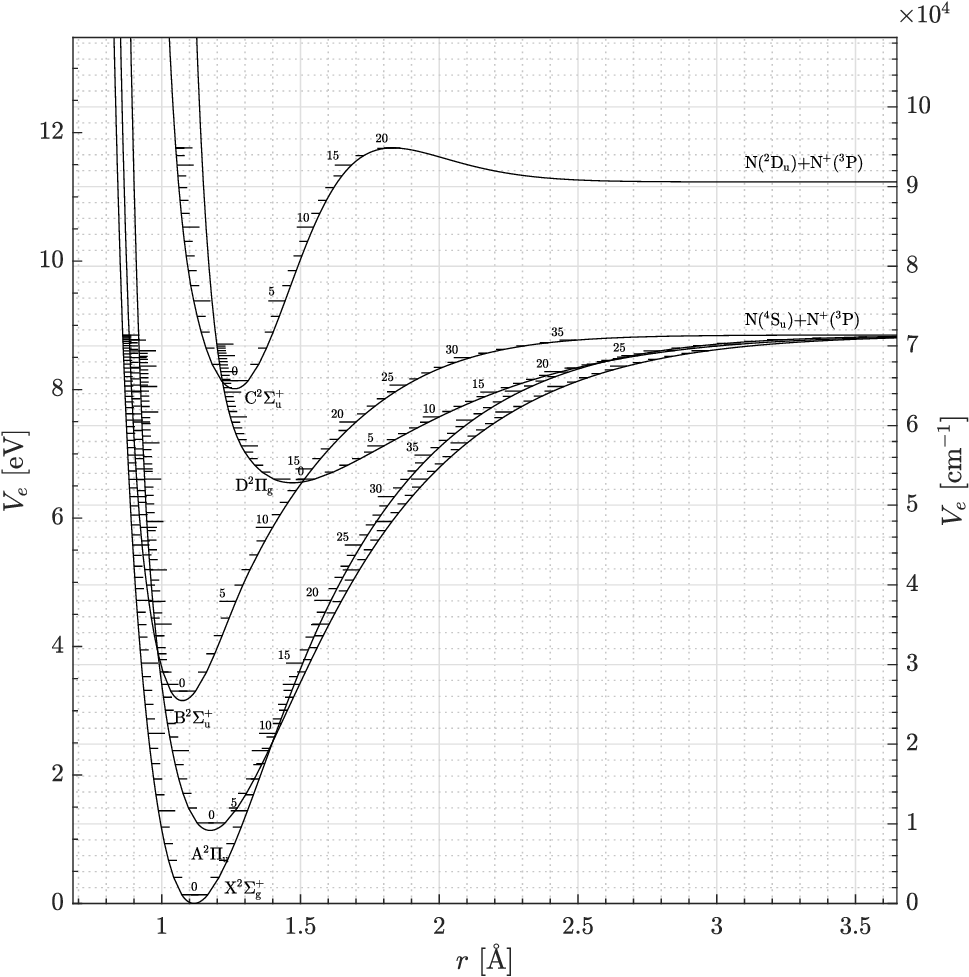}}
\caption{Electronically corrected internuclear potential curves $V_e(r)=V(r)+T_e$ for the different electronic levels of the nitrogen molecular ion \ch{N_2+}. The terms \ch{N}$\left({}^4\text{S}_\text{u}\right)$\ch{ + N+}$\left({}^3\text{P}\right)$ and \ch{N}$\left({}^2\text{D}_\text{u}\right)$\ch{ + N+}$\left({}^3\text{P}\right)$ represent the dissociation products of the nitrogen molecular ion in the electronic levels associated with the immediately below potential curves.}
\label{fig:Ve_N2+}
\end{figure}

All the data required to obtain the internuclear potential curves $V(r)$ for the different electronic levels of the nitrogen molecular ion \ch{N2+} are found in Table \ref{tab:spec_N2+}. The considered electronic levels for this species were X${}^2\Sigma_\text{g}^+$, A${}^2\Pi_\text{u}$, B${}^2\Sigma_\text{u}^+$, D${}^2\Pi_\text{g}$ and C${}^2\Sigma_\text{u}^+$. The resultant electronically corrected internuclear potential curves $V_e(r)$ are depicted in Figure \ref{fig:Ve_N2+}. All of the dissociation products shown in Figure \ref{fig:Ve_N2+} were taken from the work of Lofthus and Krupenie \cite{lofthus1977}.

\newpage

\begin{sidewaystable}
\centering
\setlength\tabcolsep{1pt} 
\begin{threeparttable}
\caption{ Spectroscopic constants $T_e$, $D_e$ and $r_e$, maximum vibrational quantum number $v_\text{max}$ for which the respective Dunham expansion is valid, Dunham parameters $Y_{i0}$ with $i=0,\,1,\,...,\,6$ and $Y_{i1}$ with $i=0,\,1,\,...,\,4$ and the shape of the extrapolated long-range part of the potential $V_\text{lr}$ of the different electronic levels of molecular nitrogen \ch{N2}.}
\begin{scriptsize}
\begin{tabular}{c|ccccccccccccccc}
\toprule
$e$ & X${}^1\Sigma_\text{g}^+$ & A${}^3\Sigma_\text{u}^+$ & B${}^3\Pi_\text{g}$ & W${}^3\Delta_\text{u}$ &  B$'{}^3\Sigma_\text{u}^-$ & a$'{}^1\Sigma_\text{u}^-$ & a${}^1\Pi_\text{g}$ & w${}^1\Delta_\text{u}$ & A$'{}^5\Sigma_\text{g}^+$ & C${}^3\Pi_\text{u}$ & b${}^1\Pi_\text{u}$ & c$_3{}^1\Pi_\text{u}$ & c$_4'{}^1\Sigma_\text{u}^+$ & b$'{}^1\Sigma_\text{u}^+$ & o$_3{}^1\Pi_\text{u}$ \\
\midrule
Ref. for $T_e$  & \cite{chauveau2001} (\tnote{a} ) & \cite{chauveau2001} (\tnote{a} ) & \cite{chauveau2001} (\tnote{a} ) & \cite{laher1991} (\tnote{a} ) & \cite{laher1991} (\tnote{a} ) & \cite{laher1991} (\tnote{a} ) & \cite{laher1991} (\tnote{a} ) & \cite{laher1991} (\tnote{a} ) & \cite{ottinger1994b} & \cite{chauveau2001} (\tnote{a} ) & \cite{chauveau2001} (\tnote{a} ) & \cite{chauveau2001} (\tnote{a} ) & \cite{chauveau2001} (\tnote{a} ) & \cite{chauveau2001} (\tnote{a} ) & \cite{chauveau2001} (\tnote{a} )\\
$T_e$ [cm$^{-1}$] & $0.00$ & $50203.66$ & $59618.77$ & $59805.85$ & $66271.53$ & $68152.79$ & $69283.34$ & $72097.31$ & $75990.03$ & $89137.92$ & $101660.02$ & $104217.77$ & $104418.37$ & $105215.38$ & $105873.81$\\
\hline
Ref. for $D_e$  & \cite{chauveau2001} (\tnote{b} )  & \cite{chauveau2001} (\tnote{b} ) & \cite{chauveau2001} (\tnote{b} ) & \cite{lofthus1977} (\tnote{b} ) & \cite{lofthus1977} (\tnote{b} ) & \cite{lofthus1977} (\tnote{b} ) & \cite{lofthus1977} (\tnote{b} ) & \cite{lofthus1977} (\tnote{b} ) &  \cite{partridge1988} & \cite{chauveau2001} (\tnote{b} ) & \cite{chauveau2001} (\tnote{b} ) & \cite{chauveau2001} (\tnote{b} ) & \cite{chauveau2001} (\tnote{b} ) & \cite{chauveau2001} (\tnote{b} ) & \cite{chauveau2001} (\tnote{b} ) \\
$D_e$ [cm$^{-1}$] & $79886.67$ & $29686.89$ & $39499.33$ & $39308.13$ & $42456.54$ & $50186.28$ & $49055.63$ & $46241.86$ & $3450.00$ & $9980.02$ & $16677.65$ & $47196.90$ & $87946.60$ & $24185.10$ & $88858.57$\\
\hline
Ref. for $r_e$  & \cite{huber1979}  & \cite{huber1979} & \cite{huber1979} & \cite{mcgowan1981} & \cite{lofthus1977} & \cite{lofthus1977} & \cite{lofthus1977} & \cite{lofthus1977} & \cite{hochlaf2010a} & \cite{huber1979} & \cite{huber1979} & \cite{huber1979} & \cite{huber1979} & \cite{huber1979} & \cite{huber1979} \\
$r_e$ [\AA] & $1.09768$ & $1.28660$ & $1.21260$ & $1.28$ & $1.27838$ & $1.27542$ & $1.22025$ & $1.26883$ & $1.60840$ & $1.14860$ & $1.28400$ & $1.11630$ & $1.10800$ & $1.44390$ & $1.17840$\\
\hline
Ref. for $v_\text{max}$  & \cite{chauveau2001}  & \cite{chauveau2001} & \cite{chauveau2001} & \cite{laher1991} & \cite{laher1991} & \cite{laher1991} & \cite{laher1991} & \cite{laher1991} & This work (\tnote{c} )  & \cite{chauveau2001} & \cite{chauveau2001} & \cite{chauveau2001} & \cite{chauveau2001} & \cite{chauveau2001} & \cite{chauveau2001} \\
$v_\text{max}$ & $15$ & $16$ & $21$ & $11$ & $17$ & $18$ & $14$ & $5$ & $5$ & $4$ & $19$ & $4$ & $8$ & $28$ & $4$\\
\hline
Ref. for $Y_\text{i0}$  & \cite{chauveau2001}  & \cite{chauveau2001} & \cite{chauveau2001} & \cite{laher1991} & \cite{laher1991} & \cite{laher1991} & \cite{laher1991} & \cite{laher1991} & (\tnote{c} ) & \cite{chauveau2001} & \cite{chauveau2001} & \cite{chauveau2001} & \cite{chauveau2001} & \cite{chauveau2001} & \cite{chauveau2001} \\
$Y_{00}$ [cm$^{-1}$] & $7.30(-2)$ & $-1.89(-1)$ & $-8.30(-2)$ & --- & $1.38(-1)$ & $1.67(-1)$ & $1.20(-2)$ & $1.05(-1)$ & $-2.50(-2)$ & $-1.79(0)$ & $6.53(0)$ & $3.90(0)$ & $7.27(-1)$ & $2.89(-1)$ & $4.24(0)$\\
$Y_{10}$ [cm$^{-1}$] & $2.36(3)$ & $1.46(3)$ & $1.73(3)$ & $1.51(3)$ & $1.52(3)$ & $1.53(3)$ & $1.69(3)$ & $1.56(3)$ & $7.31(2)$ & $2.05(3)$ & $6.42(2)$ & $2.20(3)$ & $2.17(3)$ & $7.59(2)$ & $1.97(3)$\\
$Y_{20}$ [cm$^{-1}$] & $-1.43(1)$ & $-1.40(1)$ & $-1.44(1)$ & $-1.26(1)$ & $-1.22(1)$ & $-1.21(1)$ & $-1.39(1)$ & $-1.20(1)$ & $-1.09(1)$ & $-2.89(1)$ & $2.17(1)$ & $-2.56(1)$ & $-1.33(1)$ & $-3.40(0)$ & $-1.09(1)$\\
$Y_{30}$ [cm$^{-1}$] & $-3.31(-3)$ & $2.40(-2)$ & $-3.30(-3)$ & $3.09(-2)$ & $4.19(-2)$ & $4.13(-2)$ & $7.94(-3)$ & $4.54(-2)$ & $-2.25(0)$ & $2.25(0)$ & $-1.41(0)$ & --- & $-2.94(-1)$ & $1.78(-2)$ & ---\\
$Y_{40}$ [cm$^{-1}$] & $-1.95(-4)$ & $2.56(-3)$ & $-7.90(-4)$ & $-7.10(-4)$ & $-7.30(-4)$ & $-2.9(-4)$ & $2.9(-4)$ & --- & --- & $-5.51(-1)$ & $2.29(-2)$ & --- & --- & $-1.78(-3)$ & ---\\
$Y_{50}$ [cm$^{-1}$] & --- & --- & $4.20(-5)$ & --- & --- & --- & --- & --- & --- & --- & --- & --- & --- & --- & ---\\
$Y_{60}$ [cm$^{-1}$] & --- & --- & $1.68(-6)$ & --- & --- & --- & --- & --- & --- & --- & --- & --- & --- & --- & ---\\
\hline
Ref. for $Y_\text{i1}$  & \cite{chauveau2001}  & \cite{chauveau2001} & \cite{chauveau2001} & \cite{laher1991} & \cite{laher1991} & \cite{laher1991} & \cite{laher1991} & \cite{laher1991} & \cite{hochlaf2010a} & \cite{chauveau2001} & \cite{chauveau2001} & \cite{chauveau2001} & \cite{chauveau2001} & \cite{chauveau2001} & \cite{chauveau2001} \\
$Y_{01}$ [cm$^{-1}$] & $2.00(0)$ & $1.45(0)$ & $1.64(0)$ & $1.47(0)$ & $1.47(0)$ & $1.48(0)$ & $1.62(0)$ & $1.50(0)$ & $9.31(-1)$ & $1.83(0)$ & $1.39(0)$ & $1.98(0)$ & $1.93(0)$ & $1.16(0)$ & $1.73(0)$\\
$Y_{11}$ [cm$^{-1}$] & $-1.73(-2)$ & $-1.75(-2)$ & $-1.79(-2)$ & $-1.70(-2)$ & $-1.67(-2)$ & $-1.66(-2)$ & $-1.79(-2)$ & $-1.63(-2)$ & $-1.71(-2)$ & $-2.40(-2)$ & $-1.42(-2)$ & $-3.80(-2)$ & $-1.96(-2)$& $-1.04(-2)$ & $-2.75(-2)$\\
$Y_{21}$ [cm$^{-1}$] & $-3.01(-5)$ & $-1.40(-4)$ & $-1.00(-4)$ & $-1.01(-5)$ & $1.84(-5)$ & $2.41(-5)$ & $-2.93(-5)$ & --- & --- & $1.90(-3)$ & $-5.21(-4)$ & --- & --- & $3.90(-4)$ & ---\\
$Y_{31}$ [cm$^{-1}$] & $-6.93(-8)$ & --- & $5.00(-6)$ & $3.30(-7)$ & $-4.50(-7)$ & --- & --- & --- & --- & $-6.00(-4)$ & --- & --- & --- & $-1.73(-5)$ & ---\\
$Y_{41}$ [cm$^{-1}$] & --- & --- & $2.10(-7)$ & --- & --- & --- & --- & --- & --- & --- & --- & --- & --- & --- & ---\\
\hline
$V_\text{lr}$ & $V_\text{HH}$ & $V_\text{ER}$ & $V_\text{ER}$ & $V_\text{ER}$ & $V_\text{ER}$ & $V_\text{ER}$ & $V_\text{ER}$ & $V_\text{ER}$ & $V_\text{HH}$ & $V_\text{HH}$ & $V_\text{ER}$ & (\tnote{d} ) & (\tnote{d} ) & $V_\text{ER}$ & (\tnote{d} )\\
\bottomrule
\end{tabular}
\end{scriptsize}
\label{tab:spec_N2}
\begin{scriptsize}
\begin{tablenotes}
\item{The number between parenthesis in the $Y_{i0}$ and $Y_{i1}$ cells correspond to the orders of magnitude of these quantities.}\\
\item[a]{This reference only reports the value for $\Delta T_{e0}:=T_{e0}-T_{\text{X}0}$, in which $T_{e0}$ is the quantity $T_{ev}=T_e+G_v$ with $v=0$ for the electronic level $e$, and $T_{\text{X}0}$ is the homologous quantity for the ground electronic level X. Therefore, and sticking with the introduced notation, one can obtain $T_e$ by using the relation $T_e=\Delta T_{e0}+G_0^{X}-G_0^{e}$.}\\
\item[b]{This reference only reports the value for the dissociation energy $D_0:=D_e-G_0$, and therefore the potential well depth needs to be computed using the relation $D_e=D_0+G_0$.}\\
\item[c]{The values for $Y_{i0}$, with $i=0,\,1,\,...,\,3$ for the case $e=\text{A}'{}^5\Sigma_\text{g}^+$ were obtained by fitting the vibrational energies values $G_v$, with $v=0,\,1,\,...,\,5$, taken from the literature, considering as fitting function a Dunham expansion of third order, i.e. $G_v=\sum_{i=0}^{3}Y_{i0}\left(v+\frac{1}{2}\right)^i$, with $Y_{00}$ approximated by \cite{child2014semiclassical} $\frac{B_e}{4}+\frac{\alpha_e\omega_e}{12B_e}+\frac{\left(\alpha_e\omega_e\right)^2}{144B_e^3}-\frac{\omega_ex_e}{4}$. Note that in this relation, $B_e=Y_{01}$, $\alpha_e=-Y_{11}$, $\omega_e=Y_{10}$, and $\omega_e x_e=-Y_{20}$. The $G_0$ value was taken from the work of Huber and Vervloet \cite{huber1992}, and the other $G_v$ values were taken from the work of Hochlaf \textit{et al.} \cite{hochlaf2010a} (note that in this article the values appear subtracted by $G_0$).}\\
\item[d]{Since it is known that the potential curve for this electronic level does not have a conventional shape like the $V_\text{HH}(r)$ or $V_\text{ER}(r)$ curves, and its dissociation products were not already determined, no attempt to extrapolate the RKR potential was performed. Therefore, it was assumed that the dissociation of the molecule in this electronic level does not occur, and only the vibrational levels restrained to RKR determined part may be accessed.}\\
\end{tablenotes}
\end{scriptsize}
\end{threeparttable}
\end{sidewaystable}

\newpage

\begin{sidewaystable}
\setlength\tabcolsep{20pt} 
\centering
\begin{threeparttable}
\caption{ Spectroscopic constants $T_e$, $D_e$ and $r_e$, maximum vibrational quantum number $v_\text{max}$ for which the respective Dunham expansion is valid, Dunham parameters $Y_{i0}$ with $i=0,\,1,\,...,\,7$ and $Y_{i1}$ with $i=0,\,1,\,...,\,3$ and the shape of the extrapolated long-range part of the potential $V_\text{lr}$ of the different electronic levels of nitrogen molecular ion \ch{N2+}.}
\begin{scriptsize}
\begin{tabular}{c|ccccc}
\toprule
$e$ & X${}^2\Sigma_\text{g}^+$ & A${}^2\Pi_\text{u}$ & B${}^2\Sigma_\text{u}^+$ & D${}^2\Pi_\text{g}$ & C${}^2\Sigma_\text{u}^+$\\
\midrule
Ref. for $T_e$  & \cite{chauveau2001} (\tnote{a} ) & \cite{chauveau2001} (\tnote{a} ) & \cite{chauveau2001} (\tnote{a} ) & \cite{lofthus1977} (\tnote{b} ) & \cite{chauveau2001} (\tnote{a} )\\
$T_e$ [cm$^{-1}$] & $0.00$ & $9167.34$ & $25462.43$ & $52814.06$ & $64610.79$\\
\hline
Ref. for $D_e$  & \cite{chauveau2001} (\tnote{c} )  & \cite{chauveau2001} (\tnote{c} ) & \cite{chauveau2001} (\tnote{c} ) &  \cite{lofthus1977} (\tnote{c} ) & \cite{chauveau2001} (\tnote{c} ) \\
$D_e$ [cm$^{-1}$] & $71368.43$ & $62201.49$ & $45903.08$ & $18557.87$ & $25988.34$\\
\hline
Ref. for $r_e$  & \cite{huber1979}  & \cite{huber1979} & \cite{huber1979} &  \cite{lofthus1977} & \cite{huber1979}\\
$r_e$ [\AA] & $1.11642$ & $1.17490$ & $1.07400$ & $1.47082$ & $1.26200$\\
\hline
Ref. for $v_\text{max}$  & \cite{chauveau2001}  & \cite{chauveau2001} & \cite{chauveau2001} &  --- & \cite{chauveau2001}\\
$v_\text{max}$ & $21$ & $27$ & $8$ & $5$ & $6$\\
\hline
Ref. for $Y_\text{i0}$  & \cite{chauveau2001}  & \cite{chauveau2001} & \cite{chauveau2001} & \cite{lofthus1977} & \cite{chauveau2001}\\
$Y_{00}$ [cm$^{-1}$] & $-1.20(-1)$ & $1.70(-2)$ & $-1.43(0)$ & $1.67(-1)$ & $-1.80(0)$\\
$Y_{10}$ [cm$^{-1}$] & $2.21(3)$ & $1.90(3)$ & $2.42(3)$ & $9.12(2)$ & $2.07(3)$\\
$Y_{20}$ [cm$^{-1}$] & $-1.62(1)$ & $-1.51(1)$ & $-2.41(1)$ &$-1.26(1)$ & $-8.30(0)$\\
$Y_{30}$ [cm$^{-1}$] & $4.00(-3)$ & $1.12(-2)$ & $-3.00(-1)$ &$5.55(-2)$ & $-6.30(-1)$\\
$Y_{40}$ [cm$^{-1}$] & $-6.10(-3)$ & $-2.70(-4)$ & $-6.67(-2)$ & --- & $1.30(-2)$\\
$Y_{50}$ [cm$^{-1}$] & $3.90(-4)$ & --- & --- & --- & ---\\
$Y_{60}$ [cm$^{-1}$] & $-1.40(-5)$ & --- & --- & --- & ---\\
$Y_{70}$ [cm$^{-1}$] & $2.00(-7)$ & --- & --- & --- & ---\\
\hline
Ref. for $Y_\text{i1}$  & \cite{chauveau2001}  & \cite{chauveau2001} & \cite{chauveau2001} & \cite{lofthus1977} & \cite{chauveau2001}\\
$Y_{01}$ [cm$^{-1}$] & $1.93(0)$ & $1.74(0)$ & $2.09(0)$ & $1.11(0)$ & $1.51(0)$\\
$Y_{11}$ [cm$^{-1}$] & $-1.88(-2)$ & $-1.87(-2)$ & $-2.12(-2)$ & $-2.00(-2)$ & $1.00(-3)$\\
$Y_{21}$ [cm$^{-1}$] & $-6.77(-5)$ & $-6.00(-5)$ & $-5.00(-4)$ & --- & $-1.50(-3)$\\
$Y_{31}$ [cm$^{-1}$] & $-2.32(-6)$ & $-1.10(-6)$ & $-8.80(-5)$ & --- & $6.00(-5)$\\
\hline
$V_\text{lr}$ & $V_\text{ER}$ & $V_\text{ER}$ & $V_\text{ER}$ & $V_\text{ER}$ & $V_\text{HH}$\\
\bottomrule
\end{tabular}
\end{scriptsize}
\label{tab:spec_N2+}
\begin{scriptsize}
\begin{tablenotes}
\item{The number between parenthesis in the $Y_{i0}$ and $Y_{i1}$ cells correspond to the orders of magnitude of these quantities.}\\
\item[a]{This reference only reports the value for $\Delta T_{e0}:=T_{e0}-T_{\text{X}0}$, in which $T_{e0}$ is the quantity $T_{ev}=T_e+G_v$ with $v=0$ for the electronic level $e$, and $T_{\text{X}0}$ is the homologous quantity for the ground electronic level X. Therefore, and sticking with the introduced notation, one can obtain $T_e$ by using the relation $T_e=\Delta T_{e0}+G_0^{X}-G_0^{e}$.}\\
\item[b]{This reference only reports the value for $\Delta_{\text{n}} T_{e0}:=T_{e0}-T_{\text{n},\text{X}0}$, in which $T_{e0}$ is the quantity $T_{ev}=T_e+G_v$ with $v=0$ for the electronic level $e$, and $T_{\text{n},\text{X}0}$ is the homologous quantity for the ground electronic level X of the counterpart neutral species. Therefore, and sticking with the introduced notation, one can obtain $T_e$ by using the relation $T_e=\Delta_{\text{n}} T_{e0}-\Delta_{\text{n}} T_{\text{X}0}+G_0^{X}-G_0^{e}$.}\\
\item[c]{This reference only reports the value for the dissociation energy $D_0:=D_e-G_0$, and therefore the potential well depth needs to be computed using the relation $D_e=D_0+G_0$.}\\
\end{tablenotes}
\end{scriptsize}
\end{threeparttable}
\end{sidewaystable}

\pagebreak

\section{Appendix B: The kinetic processes database}
\setlabel{Appendix B}{sec:appendixB}

\Cref{tab:kin_h_synopsis,tab:kin_e_synopsis,tab:rad_mol_synopsis,tab:rad_ato_synopsis} present the database of kinetic processes developed in this work. In the case of \Cref{tab:kin_h_synopsis,tab:kin_e_synopsis} which are with respect to the regarded collisional processes, a codename for the type of process was defined to simplify the description of the process. It consists of three terms separated by hyphens. If the process solely corresponds to a transition in the internal energy modes of the collision partners, the first term of the codename is defined by a set of capital letters each one representing the involved internal energy modes (``V'' from ``vibrational'' or/and ``E'' from ``electronic''). If the process corresponds to a bond breaking or/and forming of the internal structure, the first term is defined by a set of capital letters, each one representing a type of bond breaking or forming (``I'' from ``ionisation'', ``R'' from ``recombination'', ``D'' from ``dissociation'' or ``A'' from ``association''). The second and third terms correspond to labels of minuscule letters representing the type of the collision partners (``h'' from ``heavy species'', ``m'' from ``molecular particle'', ``a'' from ``atomic particle'', or ``e'' from ``electron'').

\begingroup
\centerline{\begin{threeparttable}
\setlength\tabcolsep{4pt} 
\renewcommand{\arraystretch}{1} 
\centering
\caption{Collisional processes due to heavy particle impact for which forward rate constants were obtained. The symbol after the reference in the column ``Reference'' represents the physical quantity which was extracted from it: process cross section (if $\sigma_p$), average process cross section (if $\sigma_{p,\text{av}}$) or forward rate constant (if $k_f$).}
\begin{scriptsize}
\begin{tabular}{cccc}
\toprule
Type & Chemical equation & Remarks & Reference \\
\midrule
V-m-h & $\ch{N_2}\left(e,v\right)\ch{ + M}\ch{ <=> N2}\left(e,v'\right)\ch{ + M}$ & \parbox{7cm}{\centering \text{ }\\ $e\in\{\text{X},\text{A},\text{B},\text{W},\text{B}',\text{a}',\text{a},\text{w},\text{A}',\text{C},\text{b},\text{c}_3,\text{c}'_4,\text{b}',\text{o}_3\}$, \\ $\forall\,v$, $\forall\,v'>v$ and $\ch{M} \in \{\ch{N2}, \ch{N2+}, \ch{N}, \ch{N+}\}$\\ \text{ }} & This work (\tnote{a} )\\
\greymidrule
V-m-h & $\ch{N_2+}\left(e,v\right)\ch{ + M}\ch{ <=> N2+}\left(e,v'\right)\ch{ + M}$ & \parbox{7cm}{\centering \text{ }\\ $e\in\{\text{X},\text{A},\text{B},\text{D},\text{C}\}$, $\forall\,v$, $\forall\,v'>v$\\ and $\ch{M} \in \{\ch{N2}, \ch{N2+}, \ch{N}, \ch{N+}\}$\\ \text{ }} & This work (\tnote{a} )\\
\midrule
VE-m-a & $\ch{N2}\left(\text{A},v\right)\ch{ + N}({}^4\text{S}_\text{u})\ch{ <=> N2}\left(\text{X},v'\right)\ch{ + N}({}^2\text{P}_\text{u})$ & $\forall\,v,$ and $\forall\,v'$ & \cite{piper1989}-$k_f$\\
\greymidrule
VE-m-a & $\ch{N2}\left(\text{A},v\right)\ch{ + N}({}^4\text{S}_\text{u})\ch{ <=> N2}\left(\text{B},v'\right)\ch{ + N}({}^4\text{S}_\text{u})$ & $\forall\,v,$ and $\forall\,v'$ & \cite{bachmann1993}-$\sigma_{p,\text{av}}$\\
\greymidrule
VE-m-a & $\ch{N2}\left(\text{W},v\right)\ch{ + N}({}^4\text{S}_\text{u})\ch{ <=> N2}\left(\text{B},v'\right)\ch{ + N}({}^4\text{S}_\text{u})$ & $\forall\,v,$ and $\forall\,v'$ & \cite{bachmann1993}-$\sigma_{p,\text{av}}$\\
\greymidrule
VE-m-a & $\ch{N2}\left(\text{A}',0\right)\ch{ + N}({}^4\text{S}_\text{u})\ch{ <=> N2}\left(\text{B},10\right)\ch{ + N}({}^4\text{S}_\text{u})$ & --- & \cite{ottinger1994a}-$\sigma_{p,\text{av}}$\\
\greymidrule
VE-m-m & $\ch{N2}\left(\text{A},v_1\right)\ch{ + N2}\left(\text{X},v_2\right)\ch{ <=> N2}\left(\text{X},v'_1\right)\ch{ + N2}\left(\text{X},v'_2\right)$ & $\forall\,v_1$, $\forall\,v_2$, $\forall\,v'_1$ and $\forall\,\,v_2'$ & \cite{levron1978}-$k_f$\\
\greymidrule
VE-m-m & $\ch{N2}\left(\text{A},v_1\right)\ch{ + N2}\left(\text{X},v_2\right)\ch{ <=> N2}\left(\text{B},v'_1\right)\ch{ + N2}\left(\text{X},v'_2\right)$ & $\forall\,v_1$, $\forall\,v_2$, $\forall\,v'_1$ and $\forall\,\,v_2'$ & \cite{bachmann1993}-$\sigma_{p,\text{av}}$\\
\greymidrule
VE-m-m & $\ch{N2}\left(\text{A},v_1\right)\ch{ + N2}\left(\text{A},v_2\right)\ch{ <=> N2}\left(\text{B},v'_1\right)\ch{ + N2}\left(\text{X},v'_2\right)$ & $\forall\,v_1$, $\forall\,v_2$, $\forall\,v'_1$ and $\forall\,\,v_2'$ & \cite{piper1988b}-$k_f$\\
\greymidrule
VE-m-m & $\ch{N2}\left(\text{A},v_1\right)\ch{ + N2}\left(\text{A},v_2\right)\ch{ <=> N2}\left(\text{C},v'_1\right)\ch{ + N2}\left(\text{X},v'_2\right)$ & $\forall\,v_1$, $\forall\,v_2$, $\forall\,v'_1$ and $\forall\,\,v_2'$ & \cite{piper1988a}-$k_f$\\
\greymidrule
VE-m-m & $\ch{N2}\left(\text{W},v_1\right)\ch{ + N2}\left(\text{X},v_2\right)\ch{ <=> N2}\left(\text{B},v'_1\right)\ch{ + N2}\left(\text{X},v'_2\right)$ & $\forall\,v_1$, $\forall\,v_2$, $\forall\,v'_1$ and $\forall\,\,v_2'$ & \cite{bachmann1993}-$\sigma_{p,\text{av}}$\\
\greymidrule
VE-m-m & $\ch{N2}\left(\text{A}',0\right)\ch{ + N2}\left(\text{X},0\right)\ch{ <=> N2}\left(\text{B},10\right)\ch{ + N2}\left(\text{X},0\right)$ & --- & \cite{ottinger1994a}-$\sigma_{p,\text{av}}$\\
\midrule
E-a-h & $\ch{N}\left(e\right)\ch{ + M <=> N}\left(e'\right)\ch{ + M}$ & $\forall e$, $\forall e'>e$ and $\text{M}\in\{\ch{N},\ch{N2}\}$ & \cite{annaloro2014b}-$k_f$ \\
\greymidrule
E-a-h & $\ch{N+}\left(e\right)\ch{ + M <=> N+}\left(e'\right)\ch{ + M}$ & $\forall e$, $\forall e'>e$ and $\text{M}\in\{\ch{N},\ch{N2}\}$ & \cite{annaloro2014b}-$k_f$ \\
\midrule
D-m-h & $\ch{N2}\left(e,v\right)\ch{ + M}\ch{ <=> N}\left(e'_1\right)\ch{ + N}\left(e'_2\right)\ch{ + M}$ & \parbox{7cm}{\centering \text{ }\\$e\in\{\text{X},\text{A},\text{B},\text{W},\text{B}',\text{a}',\text{a},\text{w},\text{A}',\text{C},\text{b},\text{b}'\}$,\\ $\forall\,v$ and  $\ch{M} \in \{\ch{N2}, \ch{N2+}, \ch{N}, \ch{N+}\}$\\ \text{ }} & This work (\tnote{a} )\\
\greymidrule
D-m-h & $\ch{N2+}\left(e,v\right)\ch{ + M}\ch{ <=> N}\left(e'_1\right)\ch{ + N+}\left(e'_2\right)\ch{ + M}$ & \parbox{7cm}{\centering \text{ }\\$e\in\{\text{X},\text{A},\text{B},\text{D},\text{C}\}$,\\ $\forall\,v$ and  $\ch{M} \in \{\ch{N2}, \ch{N2+}, \ch{N}, \ch{N+}\}$\\ \text{ }} & This work (\tnote{a} )\\
\midrule
I-a-h & $\ch{N}\left(e\right)\ch{ + M <=> N+}\left({}^3\text{P}\right)\ch{ + M + e-}$ & $\forall e$ and $\text{M}\in\{\ch{N},\ch{N2}\}$ & \cite{annaloro2014b}-$k_f$ \\
\midrule
IR-m-a & $\ch{N2}\left(\text{X},v\right)\ch{ + N+}({}^3\text{P})\ch{ <=> N2+}\left(\text{X},v'\right)\ch{ + N}({}^4\text{S}_{\text{u}})$ & VRP on $v$ and $v'$ from case $v=0$ and $\sum_{v'}$ & \cite{freysinger1994}-$\sigma_p$ \\
\bottomrule
\end{tabular}
\label{tab:kin_h_synopsis}
\end{scriptsize}
\begin{scriptsize}
\begin{tablenotes}
\item[a]{Note that although the respective chemical equation does not show any possible transition in the vibrational level (or even dissociation) of the second collision partner (if it is a molecular particle), such possibility is implicit.
}\\
\end{tablenotes}
\end{scriptsize}
\end{threeparttable}}
\endgroup
\begingroup
\centerline{\begin{threeparttable}
\setlength\tabcolsep{0pt} 
\renewcommand{\arraystretch}{0.8} 
\centering
\caption{Collisional processes due to electron impact for which forward rate constants were obtained. The symbol after the reference in the column ``Reference'' represents the physical quantity which was extracted from it: process cross section (if $\sigma_p$), average process cross section (if $\sigma_{p,\text{av}}$) or forward rate constant (if $k_f$).}
\begin{scriptsize}
\begin{tabular}{cccc}
\toprule
Type & Chemical equation & Remarks & Reference \\
\midrule
V-m-e & $\ch{N_2}\left(\text{X}{}^1\Sigma_{\text{g}}^+,\,v\right)\ch{ + e- <=> N_2}\left(\text{X}{}^1\Sigma_{\text{g}}^+,\,v'\right)\ch{ + e-}$ & $\forall v$ and $\forall\,v'>v$, ADV & \cite{laporta2014}-$\sigma_p$ (from \cite{Phys4Entry})\\
\midrule
\multirow{4}{*}{\parbox{1.2cm}{\centering \text{ }\\ \text{ }\\ \text{ }\\ \text{ }\\VE-m-e}} & \multirow{4}{*}{\parbox{5cm}{\centering \text{ }\\ \text{ }\\ \text{ }\\ \text{ }\\$\ch{N2}\left(\text{X},v\right)\ch{ + e-}\ch{ <=> N2}\left(e',v'\right)\ch{ + e-}$}}  & \parbox{7cm}{\centering \text{ }\\$e'\in\{\text{A},\text{B},\text{W},\text{B}',\text{a}',\text{a},\text{w},\text{C}\}$,\\VRP on $v$ and $v'$ from case $v=0$ and $\sum_{v'}$\\ \text{ }} & \cite{brunger2003}-$\sigma_p$\\
 & & \parbox{9cm}{\centering \text{ } \\$e'\in\{\text{c}_3,\text{o}_3\}$, VRP on $v$ and $v'$ from case $v=0$ and $\sum_{v'}$ \\ \text{ }} & \cite{malone2012}-$\sigma_p$\\
 & & \parbox{9cm}{\centering \text{ } \\$e'\in\{\text{b},\text{c}'_4,\text{b}'\}$, VRP on $v$ and $v'$ from case $v=0$ and $\sum_{v'}$\\ \text{ }} & \cite{itikawa2006}-$\sigma_p$\\
 & & \parbox{9cm}{\centering \text{ }\\$e'=\text{A}'$, Assumption of same reference values as for $e'=\text{A}$,\\VRP on $v$ and $v'$ from case $v=0$ and $\sum_{v'}$\\ \text{ }} & ---\\
\greymidrule
\multirow{2}{*}{\parbox{1.2cm}{\centering \text{ }\\ \text{ }\\ \text{ }\\VE-m-e}} & \multirow{2}{*}{\parbox{5cm}{\centering \text{ }\\ \text{ }\\ \text{ }\\$\ch{N2+}\left(\text{X},v\right)\ch{ + e-}\ch{ <=> N2+}\left(e',v'\right)\ch{ + e-}$}} & \parbox{9cm}{\centering \text{ }\\$e'=\text{B}$, $v=0$ and $v'=0$,\\Remainder of $v$ and $v'$: VRP from case $v=0$ and $v'=0$\\ \text{ }}  & \cite{crandall1974}-$\sigma_p$\\
 & & \parbox{9cm}{\centering \text{ }\\$e'\in\{\text{A},\text{D},\text{C}\}$,\\ Assumption of same reference values as for $e'=\text{B}$, $v=0$ and $v'=0$,\\Remainder of $v$ and $v'$: VRP from case $v=0$ and $v'=0$\\ \text{ }}  & ---\\
\midrule
\multirow{2}{*}{E-a-e} & \multirow{2}{*}{$\ch{N}\left(e\right)\ch{ + e- <=> N}\left(e'\right)\ch{ + e-}$ } & $(e,e')\in\{({}^4\text{S}_\text{u},{}^2\text{D}_\text{u}),({}^4\text{S}_\text{u},{}^2\text{P}_\text{u}),({}^2\text{D}_\text{u},{}^2\text{P}_\text{u})\}$ & \cite{berrington1975}-$\sigma_p$\\
 & &  Remainder of $(e,e')$, with $e'>e$ & \cite{panesi2009}-$k_f$ \\
\greymidrule
E-a-e & $\ch{N+}\left(e\right)\ch{ + e- <=> N+}\left(e'\right)\ch{ + e-}$ & $\forall e$ and $\forall e'>e$ & \cite{panesi2009}-$k_f$ \\
\midrule
\multirow{2}{*}{D-m-e} & \multirow{2}{*}{$\ch{N_2}\left(\text{X}{}^1\Sigma_{\text{g}}^+,\,v\right)\ch{ + e- <=> N}\left(e_1'\right)\ch{ + N}\left(e_2'\right)\ch{ + e-}$} & $\forall\,v$, $(e_1',e_2')=({}^4\text{S}_\text{u},{}^4\text{S}_\text{u})$, ADV & \cite{laporta2014}-$\sigma_p$ (from \cite{Phys4Entry}) \\
& & $\forall\,v$, $(e_1',e_2')=({}^4\text{S}_\text{u},{}^2\text{D}_\text{u})$, ADV & \cite{capitelli1998}-$\sigma_p$ (from \cite{Phys4Entry}) \\
\midrule
\multirow{3}{*}{\parbox{1.2cm}{\centering \text{ }\\ \text{ }\\ \text{ }\\ DR-m-e}} & \multirow{2}{*}{\parbox{5cm}{\centering \text{ }\\ \text{ }\\ \text{ }\\ $\ch{N2+}\left(\text{X}{}^2\Sigma_{\text{g}}^+,\,v\right)\ch{ + e^- <=> N}\left(e_1'\right)\ch{ + N}\left(e_2'\right)$}} & \parbox{9cm}{\centering  $v\in\{0,2\}$ and $(e'_1,e'_2)\in\{({}^4\text{S}_\text{u},{}^2\text{D}_\text{u}),({}^4\text{S}_\text{u},{}^2\text{P}_\text{u}),({}^2\text{D}_\text{u},{}^2\text{D}_\text{u})\}$,\\ \text{ }} & \multirow{2}{*}{\parbox{1cm}{\centering \text{ }\\ \text{ }\\ \text{ }\\ \cite{guberman2014}-$k_f$}} \\
 & & \parbox{9cm}{\centering $v\in\{1,3,4\}$ and \\ $(e'_1,e'_2)\in\{({}^4\text{S}_\text{u},{}^2\text{D}_\text{u}),({}^4\text{S}_\text{u},{}^2\text{P}_\text{u}),({}^2\text{D}_\text{u},{}^2\text{D}_\text{u}),({}^2\text{D}_\text{u},{}^2\text{P}_\text{u})\}$,\\ \text{ }} & \\
 & & \parbox{9cm}{\centering Remainder of $v$ with\\ $(e'_1,e'_2)\in\{({}^4\text{S}_\text{u},{}^2\text{D}_\text{u}),({}^4\text{S}_\text{u},{}^2\text{P}_\text{u}),({}^2\text{D}_\text{u},{}^2\text{D}_\text{u}),({}^2\text{D}_\text{u},{}^2\text{P}_\text{u})\}$:\\ VRP from case $v=4$} & \\
\midrule
I-m-e & $\ch{N_2}\left(\text{X}{}^1\Sigma_{\text{g}}^+,\,v\right)\ch{ + e^- <=> N_2^+}\left(e',v'\right)\ch{ + 2 e^-}$ & \parbox{9cm}{\centering $\forall\,v$ and $e'\in\{\text{X},\text{A},\text{B}\}$, ADV, VRP on $v'$ from case $\sum_{v'}$} & \cite{laricchiuta2006}-$\sigma_p$ (from \cite{Phys4Entry})\\ 
\greymidrule
\multirow{3}{*}{I-a-e} & \multirow{3}{*}{$\ch{N}\left(e\right)\ch{ + e- <=> N+}\left({}^3\text{P}\right)\ch{ + 2 e-}$} & $e={}^4\text{S}_\text{u}$ & \cite{brook1978}-$\sigma_p$\\
 & & $e\in\{{}^2\text{D}_\text{u},{}^2\text{P}_\text{u}\}$ & \cite{wang2014}-$\sigma_p$ \\
 & & Remainder of $e$ & \cite{panesi2009}-$k_f$ \\
\bottomrule
\end{tabular}
\label{tab:kin_e_synopsis}
\end{scriptsize}
\end{threeparttable}}
\endgroup

\begingroup
\centerline{\begin{threeparttable}
\setlength\tabcolsep{20pt} 
\renewcommand{\arraystretch}{1.5} 
\centering
\caption{Molecular spontaneous emission processes for which Einstein coefficients were obtained. The symbol after the reference in the column ``Reference'' represents the quantity which was extracted from it: Einstein coefficient (if $A$) or sum of the electronic-vibrational transition moments  (if $\sum R_e^2$).}
\begin{scriptsize}
\begin{tabular}{ccccc}
\toprule
Species & Electronic system & $e$ - $e'$ & $(v_\text{max},v'_\text{max})$ & Reference \\
\midrule
\multirow{10}{*}{\ch{N2}} & Vegard-Kaplan &  A${}^3\Sigma_\text{u}^+$ - X${}^1\Sigma_\text{g}^+$ & $(21,21)$ & \cite{quin2017} - $A$ (from \cite{quin2017data})\\ 
 & First positive & B${}^3\Pi_\text{g}$ - A${}^3\Sigma_\text{u}^+$ & $(21,21)$ & \cite{laux1992} - $A$\\
 & Wu-Benesch & W${}^3\Delta_\text{u}$ - B${}^3\Pi_\text{g}$ & $(21,17)$ & \cite{quin2017} - $A$ (from \cite{quin2017data})\\
 & IR afterglow & B$'{}^3\Sigma_\text{u}^-$ - B${}^3\Pi_\text{g}$ & $(21,21)$ & \cite{quin2017} - $A$ (from \cite{quin2017data})\\
 & Lyman-Birge-Hopfield & a${}^1\Pi_\text{g}$ - X${}^1\Sigma_\text{g}^+$ & $(21,21)$ & \cite{quin2017} - $A$ (from \cite{quin2017data})\\
 & Second positive & C${}^3\Pi_\text{u}$ - B${}^3\Pi_\text{g}$ & $(4,21)$ & \cite{laux1992} - $A$\\ 
 & Birge-Hopfield I &  b${}^1\Pi_\text{u}$ - X${}^1\Sigma_\text{g}^+$ & $(24,60)$ & \cite{liebhart2010} - $\sum R_e^2$\\
 & Worley-Jenkins &  c$_3{}^1\Pi_\text{u}$ - X${}^1\Sigma_\text{g}^+$ & $(11,60)$ & \cite{liebhart2010} - $\sum R_e^2$\\ 
 & Carroll-Yoshino &  c$_4'{}^1\Sigma_\text{g}^+$ - X${}^1\Sigma_\text{g}^+$ & $(11,60)$ & \cite{liebhart2010} - $\sum R_e^2$\\
 & Birge-Hopfield II &  b$'{}^1\Sigma_\text{u}^+$ - X${}^1\Sigma_\text{g}^+$ & $(46,60)$ & \cite{liebhart2010} - $\sum R_e^2$\\
 & Worley &  o$_3{}^1\Pi_\text{u}$ - X${}^1\Sigma_\text{g}^+$ & $(21,60)$ & \cite{liebhart2010} - $\sum R_e^2$\\
\hline
\multirow{3}{*}{\ch{N2+}} & Meinel & A${}^2\Pi_\text{u}$ - X${}^2\Sigma_\text{g}^+$ & $(27,27)$ & \cite{quin2017} - $A$ (from \cite{quin2017data})\\
 & First negative &  B${}^2\Sigma_\text{u}^+$ - X${}^2\Sigma_\text{g}^+$ & $(12,21)$ & \cite{laux1992} - $A$\\
 & Second negative & C${}^2\Sigma_\text{u}^+$ - X${}^2\Sigma_\text{g}^+$ & $(6,27)$ & \cite{quin2017} - $A$ (from \cite{quin2017data})\\
\bottomrule
\end{tabular}
\label{tab:rad_mol_synopsis}
\begin{tablenotes}
\end{tablenotes}
\end{scriptsize}
\end{threeparttable}}
\endgroup
\vspace{17.5pt}

\begingroup
\centerline{\begin{threeparttable}
\setlength\tabcolsep{52.5pt} 
\renewcommand{\arraystretch}{1.5} 
\centering
\caption{Atomic spontaneous emission processes for which Einstein coefficients were computed.}
\begin{scriptsize}
\begin{tabular}{ccc}
\toprule
Species & Number of processes & Reference \\
\midrule
\ch{N} & 279 (\tnote{a} ) & NIST\cite{NIST}\\
\ch{N+} & 276 (\tnote{a} ) & NIST\cite{NIST}\\
\bottomrule
\end{tabular}
\label{tab:rad_ato_synopsis}
\begin{tablenotes}
\item[a]{As a reminder to the reader: representative Einstein coefficients were computed considering the lumping procedure performed on the split electronic levels.}\\
\end{tablenotes}
\end{scriptsize}
\end{threeparttable}}
\endgroup

\pagebreak

\end{document}